%% file: BASS_XLF_BHM_ERDF_draft_overleaf.tex
\documentclass[twocolumn]{aastex631}
\pdfoutput=1
\usepackage[latin1]{inputenc}
\usepackage{amsmath}
\usepackage{amsfonts}
\usepackage{amssymb}
\usepackage{chngcntr}

\usepackage{graphicx}
\usepackage[varg]{txfonts}
\usepackage{mwe} 
\usepackage{url}
\usepackage{longtable}
\usepackage{float}

\newcommand{\magenta}[1]{{\color{black}{#1}}}
\newcommand{\blue}[1]{{\color{black}{#1}}}

\newcommand{\todo}{\ifmmode \text{\color{red}\Huge{\(\bullet\)}} \else {\color{red}{\Huge$\bullet$}}\fi}
\newcommand{\tido}{\ifmmode {{\color{red}\bullet}} \else {\color{red}$\bullet$}\fi}
\newcommand{\tomike}{\ifmmode \text{\color{red}\Huge{\(\bullet\)}MIKE} \else {\color{red}{\Huge$\bullet$}MIKE}\fi}

\newcommand{\Msun}{\ifmmode M_{\odot} \else $M_{\odot}$\fi}
\newcommand{\Msol}{\Msun}
\newcommand{\Vmax}{$V_\mathrm{max}$}
\newcommand{\Mstar}{\ifmmode \mathrm{M}^{*} \else $\mathrm{M}^{*}$ \fi}
\newcommand{\Mstarbh}{\ifmmode \mathrm{M}^{*}_{\rm BH} \else $\mathrm{M}^{*}_{\rm BH}$\fi}
\newcommand{\Mstargal}{\ifmmode \mathrm{M}^{*}_{\rm gal} \else $\mathrm{M}^{*}_{\rm gal}$\fi}

\newcommand{\ergs}	{\ifmmode {\text{erg\,s}}^{-1} \else erg s$^{-1}$\fi}
\newcommand{\kms}	{\ifmmode {\rm km\,s}^{-1} \else km\,s$^{-1}$\fi}
\newcommand{\kev}	{\ifmmode {\text{keV}} \else keV\fi}
\newcommand{\ergcms}	{\ifmmode {\rm erg\,cm}^{-2}\,{\rm s}^{-1} \else erg\,cm$^{-2}$\,s$^{-1}$\fi}
\newcommand{\cmii}	{\ifmmode {\rm cm}^{-2}    \else cm$^{-2}$\fi}

\newcommand{\Mgal}{\ifmmode M_{\rm gal} \else $M_{\rm gal}$\fi}
\newcommand{\Mbh   }{\ifmmode M_{\rm BH} \else $M_{\rm BH}$\fi}	
\newcommand{\logNH }{\ifmmode \log (N_{\rm H}/{\rm cm}^{-2}) \else $\log (N_{\rm H}/{\rm cm}^{-2})$\fi}	
\newcommand{\NH }{\ifmmode N_{\rm H} \else $N_{\rm H}$\fi}	
\newcommand{\logNHo}{\ifmmode \log N_{\rm H} \else $\log N_{\rm H}$\fi}	
\newcommand{\lamEdd }{\ifmmode \lambda_{\rm E} \else $\lambda_{\rm E}$\fi}
\newcommand{\lamEddstar }{\ifmmode \lambda_{\rm E}^{*} \else $\lambda_{\rm E}^{*}$\fi}
\newcommand{\LLedd }{\ifmmode L/L_{\rm Edd} \else $L/L_{\rm Edd}$\fi}
\newcommand{\lledd }{\lamEdd}
\newcommand{\Lbol }{\ifmmode L_{\rm bol} \else $L_{\rm bol}$\fi}
\newcommand{ \sigmas}{\ifmmode \sigma_{\star} \else $\sigma_{\star}$\fi}
\newcommand{  \sigs }{\sigmas}
\newcommand{  \lamLlam  }{\ifmmode \lamEdd L_{\lamEdd} \else $\lamEdd L_{\lamEdd}$\fi}
\newcommand{ \Lhard   }{\ifmmode L_{2-10\,\kev} \else $L_{2-10\,\kev}$\fi}
\newcommand{ \Luhard  }{\ifmmode L_{14-150\,\kev} \else $L_{14-150\,\kev}$\fi}
\newcommand{ \Lbat  }{\ifmmode L_{14-195\,\kev} \else $L_{14-195\,\kev}$\fi}
\newcommand{ \Lx  }{\Lbat}
\newcommand{  \Halpha   }{\ifmmode {\text{H} }\alpha \else H$\alpha$\fi}
\newcommand{  \Hbeta    }{\ifmmode {\text{H} }\beta \else H$\beta$\fi}

\newcommand{  \swift   }  {{Swift}}

\newcommand{\siglscatt}{\ifmmode{\sigma_{\log L, {\rm scatt}}} \else $\sigma_{\log L, {\rm scatt}}$\fi}

\newcommand{\zminsamplebhmf}{0.01}
\newcommand{\zmaxsamplebhmf}{0.3}
\newcommand{\mbhminsample}{6.5}
\newcommand{\mbhmaxsample}{10.5}
\newcommand{\lamminsample}{-3}
\newcommand{\lammaxsample}{1}
\newcommand{\binxlf}{0.3}
\newcommand{\binbhmf}{0.3}
\newcommand{\binerdf}{0.3}

\input{parameter_shortcuts}
\defcitealias{Schulze:2010aa}{SW10}
\defcitealias{Schulze:2015aa}{S15}

\shorttitle{BASS XXX: ERDF, BHMF \& XLF}
\shortauthors{Ananna et al.}
\begin{document}

\title{BASS XXX: Distribution Functions of DR2 Eddington-ratios, Black Hole Masses, and X-ray Luminosities}

\correspondingauthor{Tonima Tasnim Ananna}
\email{Tonima.Ananna@dartmouth.edu}


\author[0000-0001-8211-3807]{Tonima Tasnim Ananna}
\affiliation{Department of Physics and Astronomy, Dartmouth College, 6127 Wilder Laboratory, Hanover, NH 03755, USA}

\author[0000-0002-5489-4316]{Anna K. Weigel}
\affiliation{Modulos AG, Technoparkstrasse 1, CH-8005 Zurich, Switzerland}

\author[0000-0002-3683-7297]{Benny Trakhtenbrot}
\affiliation{School of Physics and Astronomy, Tel Aviv University, Tel Aviv 69978, Israel}

\author[0000-0002-7998-9581]{Michael J. Koss}
\affiliation{Eureka Scientific, 2452 Delmer Street Suite 100, Oakland, CA 94602-3017, USA}
\affiliation{Space Science Institute, 4750 Walnut Street, Suite 205, Boulder, Colorado 80301, USA}


\author[0000-0002-0745-9792]{C. Megan Urry}
\affiliation{Department of Physics and Yale Center for Astronomy \& Astrophysics, Yale University, PO BOX 201820, New Haven, CT 06520, USA}

\author[0000-0001-5231-2645]{Claudio Ricci}
\affiliation{N\'ucleo de Astronom\'ia de la Facultad de Ingenier\'ia, Universidad Diego Portales, Av. Ej\'ercito Libertador 441, Santiago 22, Chile}
\affiliation{Kavli Institute for Astronomy and Astrophysics, Peking University, Beijing 100871, People's Republic of China}

%
\author[0000-0003-1468-9526]{Ryan C. Hickox}
\affiliation{Department of Physics and Astronomy, Dartmouth College, 6127 Wilder Laboratory, Hanover, NH 03755, USA}

\author[0000-0001-7568-6412]{Ezequiel Treister}
\affiliation{Instituto de Astrof{\'i}sica, Facultad de F{\'i}sica, Pontificia Universidad Cat{\'o}lica de Chile, Casilla 306, Santiago 22, Chile}

\author[0000-0002-8686-8737]{Franz E. Bauer}
\affiliation{Instituto de Astrof{\'{\i}}sica and Centro de Astroingenier{\'{\i}}a, Facultad de F{\'{i}}sica, Pontificia Universidad Cat{\'{o}}lica de Chile, Casilla 306, Santiago 22, Chile}
\affiliation{Millennium Institute of Astrophysics, Nuncio Monse{\~{n}}or S{\'{o}}tero Sanz 100, Of 104, Providencia, Santiago, Chile} 
\affiliation{Space Science Institute, 4750 Walnut Street, Suite 205, Boulder, Colorado 80301}

\author[0000-0001-7821-6715]{Yoshihiro Ueda}
\affiliation{Department of Astronomy, Kyoto University, Kitashirakawa-Oiwake-cho, Sakyo-ku, Kyoto 606-8502, Japan}

\author[0000-0002-7962-5446]{Richard Mushotzky}
\affiliation{Department of Astronomy, University of Maryland, College Park, MD 20742, USA}

\author[0000-0001-5742-5980]{Federica Ricci}
\affiliation{Dipartimento di Fisica e Astronomia, Universit\'a di Bologna, via Piero Gobetti 93/2, I-40129 Bologna, Italy}
\affiliation{INAF - Osservatorio di Astrofisica e Scienza dello Spazio di Bologna, Via Gobetti, 93/3, I40129 Bologna, Italy}


\author[0000-0002-5037-951X]{Kyuseok Oh}
\affiliation{Korea Astronomy \& Space Science institute, 776, Daedeokdae-ro, Yuseong-gu, Daejeon 34055, Republic of Korea}
\affiliation{Department of Astronomy, Kyoto University, Kitashirakawa-Oiwake-cho, Sakyo-ku, Kyoto 606-8502, Japan}
\affiliation{JSPS Fellow}

\author[0000-0001-8450-7463]{Julian E. Mej\'ia-Restrepo}
\affiliation{European Southern Observatory, Casilla 19001, Santiago 19, Chile}

\author[0000-0002-8760-6157]{Jakob Den Brok}
\affiliation{Argelander-Institut fur Astronomie, Universit{\"a}t Bonn, Auf dem H{\"u}gel 71, D-53121 Bonn, Germany}

\author[0000-0003-2686-9241]{Daniel Stern}
\affiliation{Jet Propulsion Laboratory, California Institute of Technology, 4800 Oak Grove Drive, MS 169-224, Pasadena, CA 91109, USA}

\author[0000-0003-2284-8603]{Meredith C. Powell}
\affiliation{Kavli Institute of Particle Astrophysics and Cosmology, Stanford University, 452 Lomita Mall, Stanford, CA 94305, USA}

\author[0000-0002-9144-2255]{Turgay Caglar}
\affiliation{Leiden Observatory, PO Box 9513, 2300 RA, Leiden, The Netherlands}


\author[0000-0002-4377-903X]{Kohei Ichikawa}
\affiliation{Frontier Research Institute for Interdisciplinary Sciences, Tohoku University, Sendai 980-8578, Japan}

\author[0000-0003-4264-3509]{O. Ivy Wong}
\affiliation{CSIRO Astronomy \& Space Science, PO Box 1130, Bentley, WA 6102, Australia}
\affiliation{ICRAR-M468, University of Western Australia, Crawley, WA 6009, Australia}


\author{Fiona A. Harrison}
\affiliation{Cahill Center for Astronomy and Astrophysics, California Institute of Technology, Pasadena, CA 91125, USA}

\author[0000-0001-5464-0888]{Kevin Schawinski}
\affiliation{Modulos AG, Technoparkstrasse 1, CH-8005 Zurich, Switzerland}


\begin{abstract}
We determine the low-redshift X-ray luminosity function (XLF), active black hole mass function (BHMF), and Eddington-ratio distribution function (ERDF) for both unobscured (Type~1) and obscured (Type~2) active galactic nuclei (AGN) using the unprecedented spectroscopic completeness of the BAT AGN Spectroscopic Survey (BASS) data release 2. 
In addition to a straightforward 1/\Vmax\ approach, we also compute the intrinsic distributions, accounting for sample truncation by employing a forward modeling approach to recover the observed BHMF and ERDF. 
As previous BHMFs and ERDFs have been robustly determined only for samples of bright, broad-line (Type~1) AGNs and/or quasars, ours is the first directly observationally constrained BHMF and ERDF of Type~2 AGN. 
We find that after accounting for all observational biases, the intrinsic ERDF of Type~2 AGN is significantly skewed towards lower Eddington ratios than the intrinsic ERDF of Type~1 AGN. This result supports the radiation-regulated unification scenario, in which radiation pressure dictates the geometry of the dusty obscuring structure around an AGN. 
Calculating the ERDFs in two separate mass bins, we verify that the derived shape is consistent, validating the assumption that the ERDF \magenta{(shape)} is mass independent.  
We report local AGN duty cycle as a function of mass and Eddington ratio, by comparing the BASS active BHMF with local mass function for all SMBH. 
We also present the $\log N{-}\log S$ of Swift-BAT 70-month sources.
\end{abstract}



\section{Introduction}\label{sec:intro}


\magenta{Supermassive black holes (SMBHs) are found at the centers of nearly all massive galaxies, and are understood to co-evolve with their host galaxies (see \citealp{Kormendy:2013aa} for a review). Actively accreting SMBHs, identified by their high luminosities or rates of accretion, are known as active galactic nuclei (AGNs).} 
The space density of AGN as a function of luminosity---i.e., the AGN luminosity function (LF)---
represents a key statistical measure for the AGN population which allows us to constrain the abundance and growth history of SMBHs (e.g., \citealp{Soltan1982}). 

The redshift-resolved AGN LF and space density have had major impacts on our understanding of the evolving SMBH population. For example, it is used to determine the epoch of peak SMBH growth at around $z\sim 2$ \cite[e.g.,][]{Barger2001,Ueda:2003aa, Hasinger:2005aa, Croom:2009aa, Ueda:2014aa, ananna2020ab}, quite similar to the peak in cosmic star formation activity \cite[e.g.,][]{Lilly1996,Madau1998,Zheng2009,Madau2014,Aird2015,Caplar:2015aa}.
It is also clear that the space densities of low-luminosity AGN peak at lower redshifts compared to higher-luminosity systems (so-called ``downsizing'', see, e.g., \citealt{Barger2001,Ueda:2003aa,Miyaji:2015aa,BrandtAlexander2015_AARv, Ueda:2014aa, ananna2020ab}). 
At yet higher redshifts, the AGN LF can help constrain the contribution of accreting SMBHs to cosmic reionization \cite[e.g.,][]{Willott:2010ab, Kashikawa:2015aa, Giallongo2015, RicciF2017_UV_emissivity, Parsa:2018aa, Matsuoka2018, ananna2020ab}. 
Indeed, when used as a key ingredient in phenomenological population models, the evolving AGN LF is used to trace the growth of SMBHs throughout cosmic history, ultimately accounting for the local population of (relic) SMBHs, and even SMBH-host relations \cite[e.g.,][]{Soltan1982,Marconi:2004aa,Shankar:2009aa,Ueda:2014aa,Aird2015,buchner2015,Caplar2018,ananna2019}. The AGN LF is therefore a very useful statistical tool for understanding the AGN population and its evolution (see, e.g., \citealp{BrandtAlexander2015_AARv} for a review). 

The AGN LF alone, however, cannot constrain the crucial characteristics of the underlying SMBH population.
This is because the AGN (bolometric) luminosity is essentially the product of two more fundamental properties of a black hole--- its mass (\Mbh) and relative accretion rate, which we parameterize as the dimensionless Eddington-ratio ($\lamEdd  \equiv L_{\rm bol}/L_{\rm Edd}$), that is
\begin{equation}\label{eq:LMlamEdd_gen}
\Lbol \propto \Mbh \times (\Lbol/L_{\rm Edd}) = \Mbh\times\lamEdd \, .
\end{equation}
Therefore, only after measuring the underlying BH mass and Eddington-ratio for sizable, representative AGN samples can we decisively answer questions such as when was the epoch during which the most massive BHs ($\Mbh \gtrsim 10^9\,\Msun$) grew most of their mass. 
Several studies show that such high-mass BHs accreted at maximal Eddington rates, reaching the Eddington limit at $z\gtrsim5$ \cite[e.g.,][]{Willott:2010aa, Trakhtenbrot:2011aa, De-Rosa:2014aa}.
In the local Universe, on the other hand, it seems that lower mass ($\sim 10^6 - 10^8\,\Msun$) AGN with lower \lamEdd\ dominate space density distributions, even among the most luminous AGN  \cite[i.e. quasars; e.g.,][]{McLureDunlop2004,NetzerTrakht2007, Schulze:2010aa, Schulze:2015aa}. 
To obtain a complete census of the AGN population, we thus have to consider three key distribution functions: the AGN (bolometric) LF, the \magenta{\it active} black hole mass function \magenta{(BHMF)\footnote{Note that in this work we use BHMF to denote the mass function of the active SMBH population alone, unless stated otherwise, whereas the \textit{total} BHMF is the sum of active and inactive BHMFs}}, and the Eddington-ratio distribution function (ERDF), after correcting for obscuration,  uncertainties on the observationally-derived key quantities (\Mbh, \Lbol, and \lamEdd), and selection effects. 
These distributions are fundamentally interlinked through the ensemble version of Eq.~\ref{eq:LMlamEdd_gen}; that is, the bolometric LF can be expressed as the convolution of the BHMF and the ERDF.

Compared to the AGN LF, determining the BHMF and the ERDF is much more challenging. 
First, it requires reliable \Mbh\ and \lamEdd\ measurements for large, unbiased AGN samples. 
In practice, beyond the local Universe this is only possible for unobscured, broad-line AGNs, thanks to ``virial'' mass prescriptions calibrated against reverberation mapping experiments \cite[e.g.,][]{Shen2013_rev,Peterson2014_rev}.
Furthermore, certain selection effects have to be taken into account, which go beyond the more common biases affecting the LF (i.e., the Eddington bias; \citealp{eddington1913}).
For flux limited surveys, the dominant effect is a bias against low mass and low Eddington-ratio AGN. 
To address this bias, and others, the selection function of the sample at hand has to be well understood, and both the BHMF and ERDF have to be constructed (and/or fitted) simultaneously. 

With these challenges in mind, the BHMF (both active and total) and ERDF have previously been constrained from large surveys, both indirectly and directly. 
First, by assuming a universal relation between total stellar mass and BH mass \cite[see, e.g.,][]{Marconi:2004aa, Sani2011_MM,Kormendy:2013aa}, the shape of the \magenta{\textit{total}} BHMF can be associated to the galaxy stellar mass function and determined empirically. 
Since the AGN LF can be expressed as a convolution of the BHMF and the ERDF, once the shape of the BHMF is known (or assumed), then the LF can be used to constrain the ERDF indirectly \citep[e.g.,][]{Caplar:2015aa, Weigel:2017aa}. Other studies have used the ratio between bolometric AGN luminosity and \textit{stellar} mass as an indirect proxy for \lamEdd\ and thus the ERDF \cite[e.g.,][]{Aird:2018aa, Georgakakis:2017aa}.

In contrast to such indirect approaches, the \magenta{\textit{active}} BHMF and the ERDF can also be determined directly from observations AGN for which it is possible to get reliable SMBH mass measurements.
\citet{Greene:2007aa,Greene:2009aa} constrained a low-redshift \magenta{active} BHMF for broad-line AGNs drawn from the Sloan Digital Sky Survey (SDSS), focusing on relatively low-mass systems, and using the 1/\Vmax\ method to estimate the space densities \citep{Schmidt:1968aa}. 
On the other hand, \cite{Schulze:2010aa} determined the active BHMF and ERDF of highly luminous low-redshift quasars, drawn from the Hamburg-ESO survey. 
The 1/\Vmax\ method was used to determine the BHMF of 
$0.3{\lesssim} z {\lesssim}5$ quasars drawn from the SDSS/DR3 \citep{vestergaard2008} and L/BQS surveys \citep{Vestergaard2009}. 
The BHMF of quasars of comparably high luminosities and redshifts from the BQS and SDSS was also determined by \cite{Kelly2009} and \citet{Kelly:2013aa}, respectively, with the latter also constraining the ERDF. \citet{nobuta2012} reported the BHMF and ERDF of broad-line AGN at $z = 1.4$ selected from the Subaru \textit{XMM}-Newton Deep Survey (SXDS) field. 
\citet{Schulze:2015aa} constrained the BHMF and ERDF for AGNs in the $1< z < 2$ range, combining data from the SDSS, zCOSMOS and VVDS optical surveys. 
Assuming a luminosity-dependent fraction of obscured AGN, \citet{Schulze:2015aa} also indirectly deduced the BHMF and ERDF of obscured (Type~2) AGNs.

Several of these studies addressed some limitations of the 1/\Vmax\ method, and generally showed that the BHMF can be described by a (modified) Schechter-like functional form \citep{allerrichstone2002}, resembling the shape of the galaxy stellar mass function.
The ERDF, on the other hand, is often described using a broken-power-law shape \citep{Caplar:2015aa,Weigel:2017aa}, sharply decreasing towards high \lamEdd, and rarely exceeding the nominal Eddington limit.


%
High energy X-rays are understood to be 
more suited for probing large samples of obscured, narrow-line (Type~2) AGNs, than the optical and UV bands due to the penetrating power of X-ray photons through high column densities of gas and dust. 
X-rays are also considered a higher-purity tracer of AGN than the infrared (IR) band, as X-rays arising from SMBH accretion are less contaminated by stellar and gas emission originating from the host galaxy. 
Surveys such as SDSS are highly complete in optical bands, however they (as well as soft X-ray surveys) are naturally biased towards unobscured AGN [i.e., \logNH $< 22$]. 
This, combined with the inability to measure \Mbh\ and/or \lamEdd\ in narrow-line (mostly obscured) AGNs, means that essentially all existing literature presents BHMFs and ERDFs for only broad-line, unobscured (Type~1) AGNs. \citet{nobuta2012} reported BHMF/ERDF for an X-ray selected sample, but only for broad-line AGN. \citet{Aird2018_ERDF} constrained the distribution of \lamEdd, as probed indirectly by the X-ray luminosity to \textit{host} mass ratio, out to $z \simeq 4$, using a large (\textit{Chandra}) X-ray-selected sample, also concluding that the AGN population tends towards higher \lamEdd\ with increasing redshift, but not exceeding the Eddington limit.
However, that study did not correct for obscuration, as it was assumed that obscuration does not significantly affect a hard X-ray-selected (2--7 keV) sample.
Probing the key distributions of the AGN population using a large sample selected in the ultra-hard X-ray regime thus offers a crucial addition to our understanding of SMBH accretion and triggering, even in the local Universe.

The \textit{Swift/BAT AGN Spectroscopic Survey} (BASS, originally presented in \citealt{Koss:2017aa})
provides a large, highly complete sample of AGN selected in the ultra-hard X-ray band (14--195 keV), along with reliable measurements of key properties.  
This includes X-ray fluxes and column densities derived from detailed X-ray spectral analysis \citep{Ricci2017_Xray_cat}, and---crucially---optical counterpart matching, redshift measurements, and \Mbh\ and \lamEdd\ measurements, obtained through extensive optical spectroscopic observations and analysis. As shown in Figure~2 of \citet{koss2016b}, in terms of obscuration, BASS is the least biased of all X-ray surveys to date, and largely unaffected by obscuring column densities below \logNH $\simeq 23$.
The 2nd data release of BASS (BASS/DR2; \citealt{Koss_DR2_overview}) provides an essentially complete census of luminosity, BH mass, Eddington-ratio, and obscuration towards all AGNs in the 70-month catalog of the \swift/BAT all-sky survey, with over 800 unbeamed AGNs, mostly at $z\lesssim0.3$.
BASS/DR2 therefore allows us to determine the XLF, BHMF and ERDF of low-redshift AGNs in an unprecedented way. 
It also presents the first sample of ultra-hard X-ray selected AGN large enough to allow direct measurements of the BHMF and the ERDF for both Type~1 and Type~2 AGN. \magenta{Importantly, the BH masses for Type~1 and Type~2 AGN are derived through different, though consistent \& inter-calibrated methods: broad Balmer line measurements and ``virial'' prescriptions are used for the former, while stellar velocity dispersion (\sigs) measurements and the $\Mbh-\sigs$ relation is used for the latter.} 
Highly complete (\magenta{95\% mass measurement completeness at z $< 0.3$}) and with a well-understood selection function, BASS thus allows us to gain a new understanding of the local AGN population, while accounting for potential biases and statistical inference limitations \cite[e.g., see][]{Schulze:2010aa, Kelly:2013aa, Schulze:2015aa}.
With the intrinsic, bias-corrected XLF, BHMF, and ERDF at hand for both Type~1 and Type~2 AGN, we can also investigate trends with column density, AGN duty cycle, \Mbh\ and \lamEdd, as well as whether the AGN BHMF is consistent with galaxy-SMBH scaling relations.

We present our work as follows: in \S\ref{sec:sample_data} we describe the data and selection criteria, in \S\ref{sec:stat_method} we discuss the details of our method of calculating the BHMF, ERDF and the XLF, in \S\ref{sec:results} we present the results of our analysis and in \S\ref{sec:discussion} we discuss the physical implications of our results. Further details about our methods, such as estimating errors and testing our code with mock catalogs are described in the Appendices. 
A flat $\Lambda$CDM cosmology with $H_{\rm 0} = 70\,{\rm km\,s^{-1}\,Mpc^{-1}}$, $\Omega_{\rm M} = 0.3$, and $\Omega_{\rm \Lambda} = 0.7$ is assumed throughout the paper. 
All uncertainties reported in this paper are $\pm 1\sigma$ from the best-fit values.

\section{Sample, Data and Basic Measurements}\label{sec:sample_data}

\subsection{Sample selection}\label{sec:sample}

\subsubsection{Input catalog and sample}\label{sec:sample_input}

\begin{figure*}

    \centering
	\includegraphics[width=0.5\textwidth]{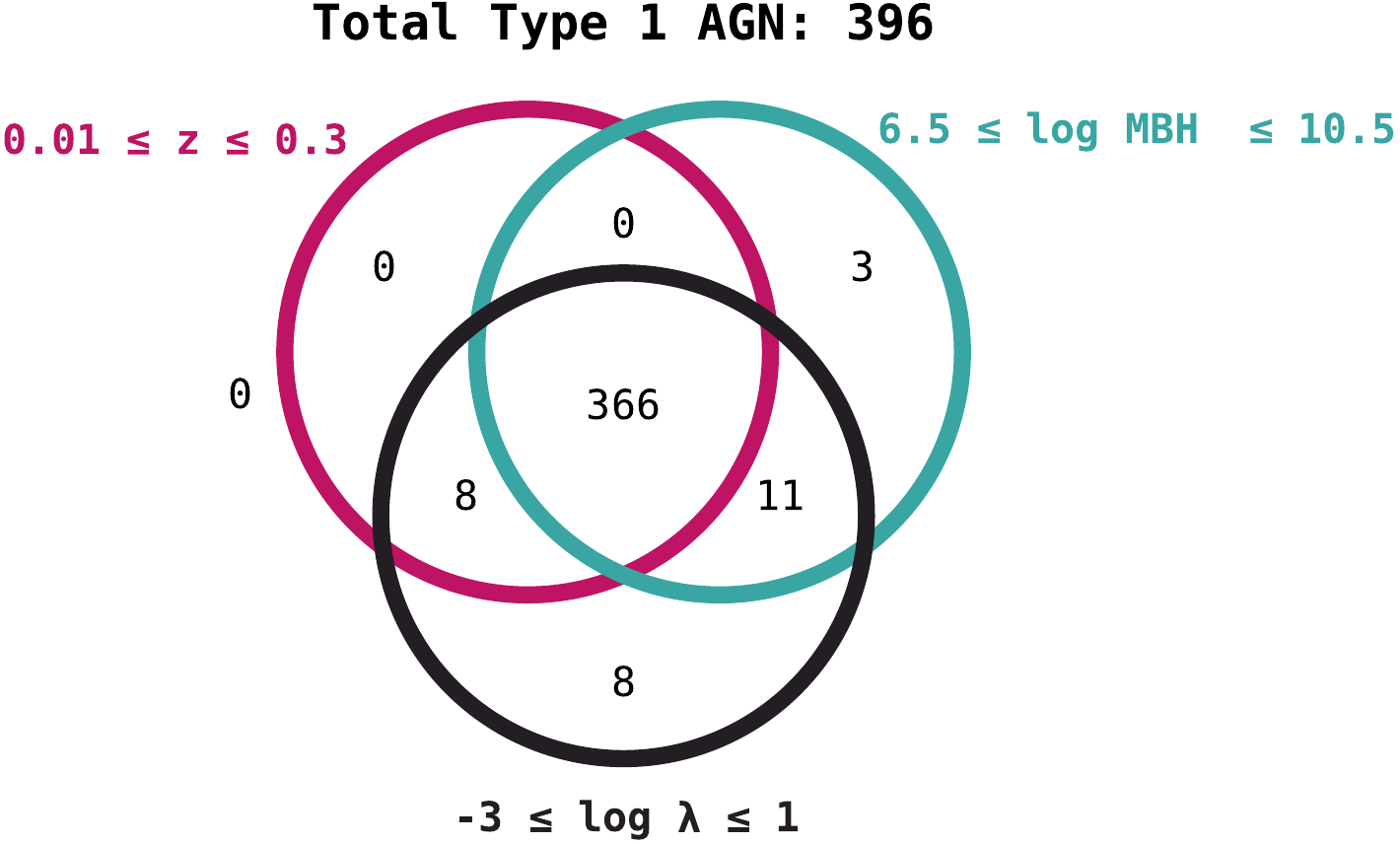}~
	\includegraphics[width=0.5\textwidth]{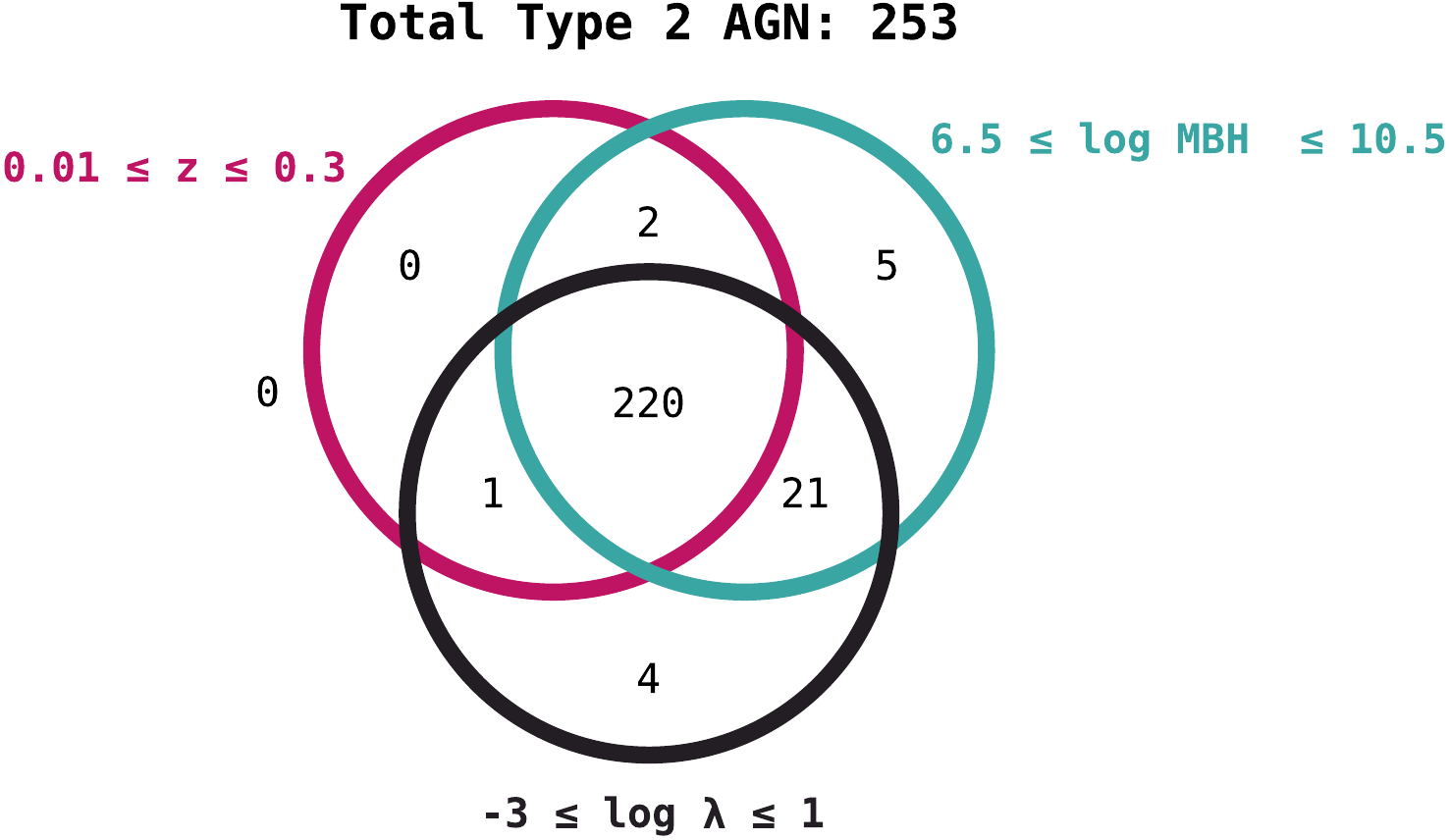}
	\caption{\label{fig:venn} Venn diagram of all \swift/BAT 70-month sources below $z =0.3$, after removing all beamed sources, Galactic-plane sources  ($|b| < 5^\circ$), as well as more subtle cuts that exclude a total of \ntotremoved\ 
	sources (see text for details). 
	The diagram shows how the final samples of \ntotallcutstypeone\ Type~1 objects and \ntotallcutstypetwo\ Type~2 objects were selected. All cuts applied to the parent sample are explicitly listed in Table~\ref{tab:bhmfpara}. 
	For each of the AGN sub-samples, the {\it red circle} denotes the set of objects that fall within $0.01 \leq  z \leq 0.3$, the {\it green circle} denotes the set that falls within $6.5 \leq \log(\Mbh/\Msun) \leq 10.5$ and the {\it black circle} denotes the set that falls within $-3 \leq \log \lamEdd \leq 1$ . 
	The numbers of sources in all sets and intersections of sets are also noted.
	}
\end{figure*}


\begin{figure*}[t]
    \centering
	\includegraphics[width=0.495\textwidth]{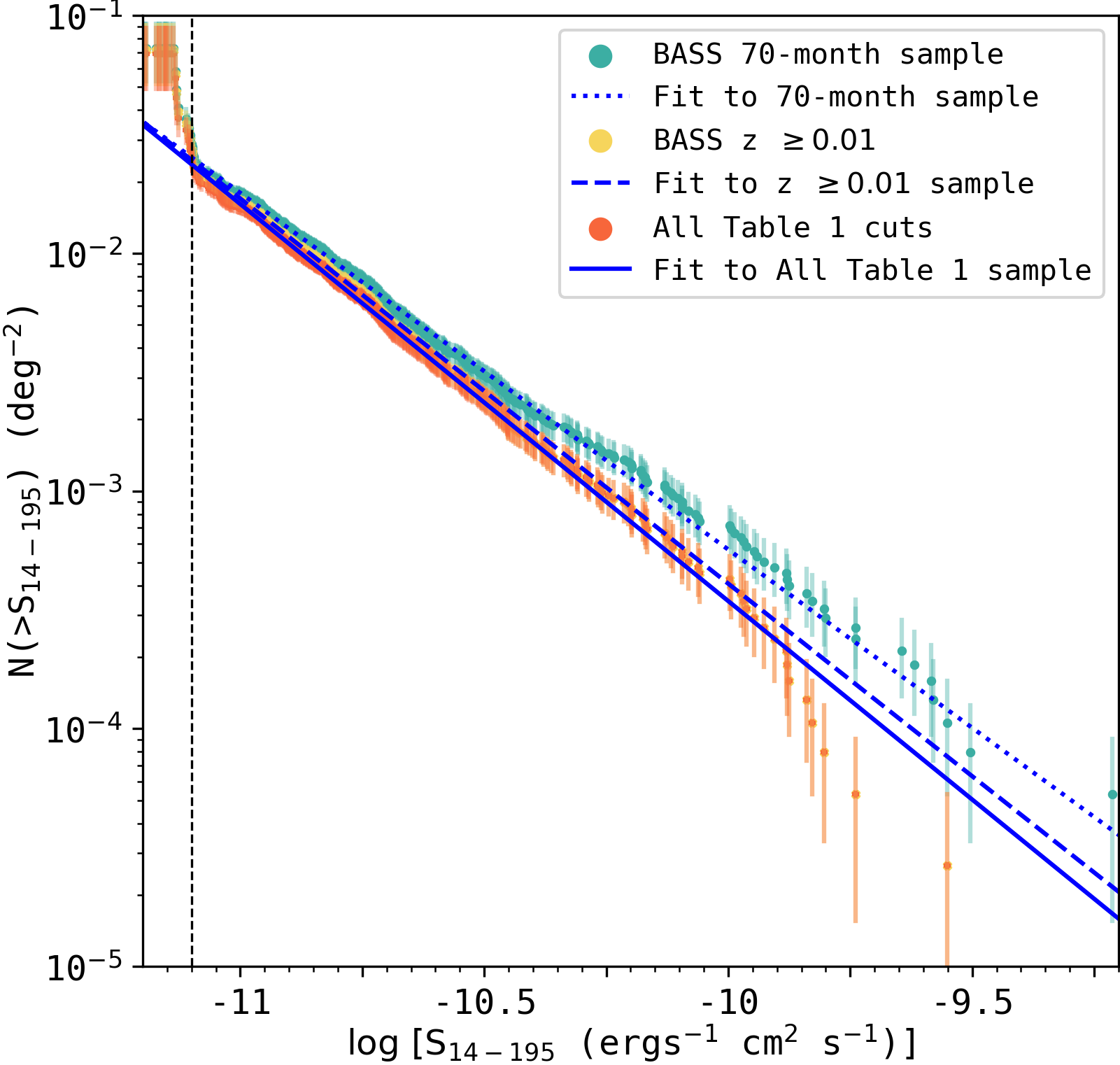}~
	\includegraphics[width=0.495\textwidth]{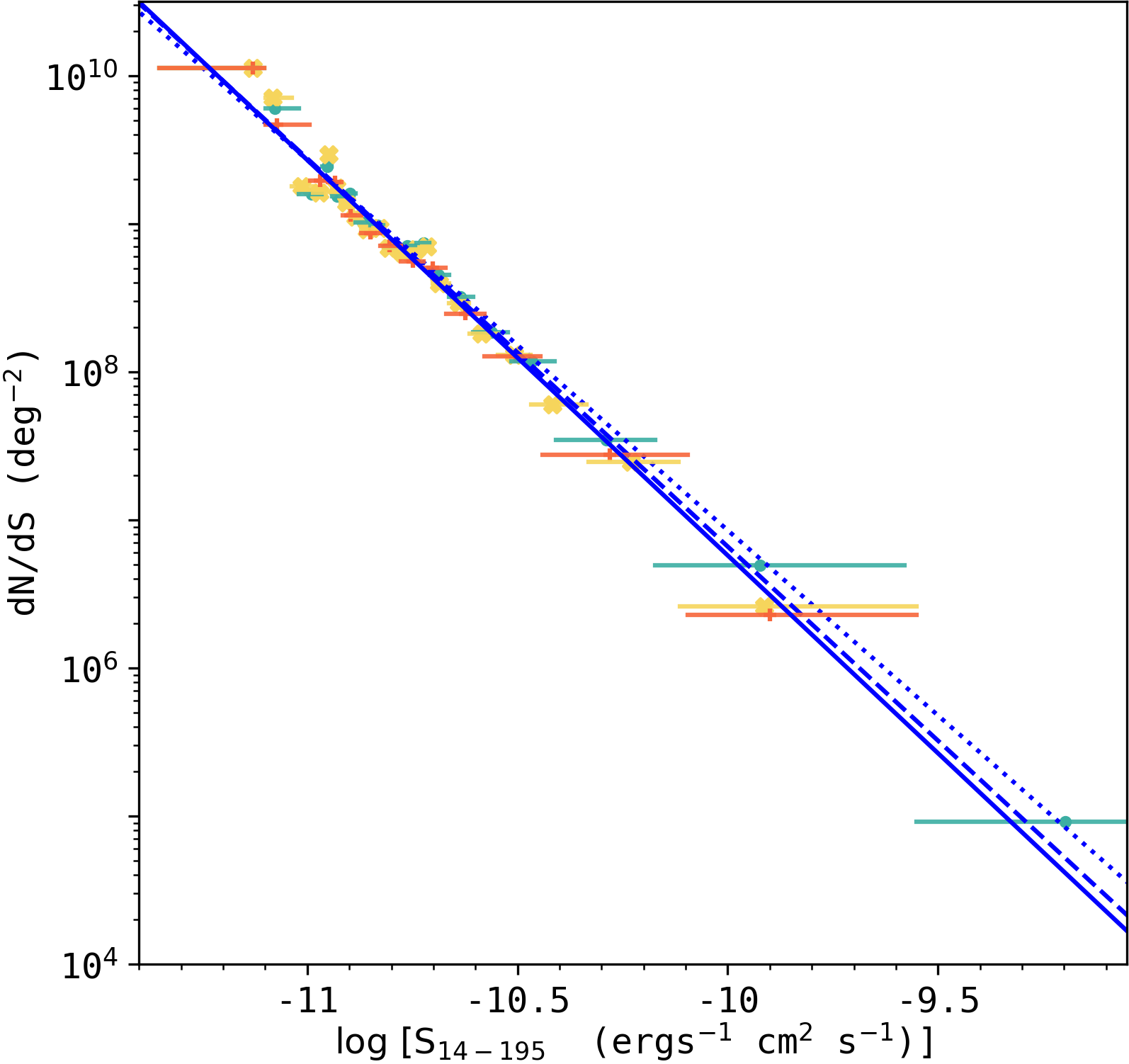}
	\caption{\label{fig:logN_logS} Swift-BAT cumulative number counts (\textit{left}) and differential number counts (\textit{right}) from 70-month survey with beamed sources, low flux dual sources, sources close to the Galactic plane and weak associations removed (\nnonblazarhighredshiftcut\ sources, at $z \leq 0.3$; {\it turquoise data points}), the same sample with an additional redshift cut applied ($z \geq 0.01$, \nnonblazarredshiftcut\ sources; {\it yellow data points}), and then with all Table~\ref{tab:bhmfpara} cuts applied (\ntotallcuts\ sources; {\it red data points}). 
	The latter sample is used in our BHMF and ERDF calculations, and as seen in this figure, it is representative of the redshift restricted sample. We produce fits to these data using Equation~\ref{eq:logNlogS}, and report the slope and normalization of the fitted line in \S\ref{sec:logNlogS}. 
	To produce the fits, we applied a flux cut at $F_{\rm 14-195,~obs} > 10^{-11.1}\,{\rm erg\,s^{-1}\,cm^{-2}}$ (shown in the left panel with \textit{black dashed line}), because the flux-area curve below that flux limit is not well constrained (\S\ref{sec:stat_method} and Fig.~\ref{fig:flux_area_curve}), artificially affecting the slope of the number counts (shown in the top left corner of the cumulative counts).}
\end{figure*}

We base our analysis on the second data release (DR2) of BASS, which is described in detail in \cite{Koss_DR2_overview}. 
BASS/DR2 combines extensive optical spectroscopic measurements, X-ray spectral analysis, and derived quantities, for essentially all \nxraydetected\ X-ray detected sources 
that are part of the 70-month Swift-BAT catalog \citep{Baumgartner:2013aa, Ricci2017_Xray_cat}. 
We further discuss aspects of this catalog and the BASS/DR2 parent sample that are crucial for our analysis, such as the flux limit (or rather, the flux-area curve), in subsequent Sections.

%
For the X-ray related properties of the vast majority of these AGN, we rely on the detailed spectral measurements described in \cite{Ricci2017_Xray_cat}. 
BASS/DR2 also includes 14 
AGN that were robustly detected as ultra-hard X-ray sources in the 70-month \swift/BAT catalog, but not identified as AGN in the \cite{Ricci2017_Xray_cat} compilation. 
The X-ray spectra of these source are analyzed as part of BASS/DR2, following the same procedures as in \cite{Ricci2017_Xray_cat}. 
A detailed account of these newly identified AGN, as well as on a few more minor changes to counterpart matches, can be found in  \cite{Koss_DR2_catalog}.

\subsubsection{Excluded sources}\label{sec:sample_exclusion}

First, we exclude all sources with Galactic latitudes $|b|<5^{\circ}$, as the reliability of cross-matching BAT sources with optical counterparts, as well as the completeness of the BASS optical spectroscopy efforts, drop significantly for sources in the Galactic plane. 
All our survey area and/or volume coverage calculations are adjusted to reflect the exclusion of this region of the sky. 

We also exclude from our analysis \nblazars\ beamed sources  (i.e., blazars --- BL-Lac-like or FSRQ sources).
More details about these sources are provided in \cite{Koss_DR2_overview}.
Here we briefly mention that this set constitutes of sources identified  through the \texttt{BZ\_flag} in \cite{Ricci2017_Xray_cat}, sources identified in a dedicated BAT-\textit{Fermi} analysis \citep{Paliya2019_BASS_blazars}, and a few additional sources for which extensive multi-wavelength data suggests that they are most likely beamed \cite[][]{Paiano2020_TeV_blazars,Marcotulli_BASS_blazars}. 
%
After making these two adjustments,  $\nnonblazar$ non-beamed, non-Galactic-plane AGNs remains in our sample.

We further exclude several dual AGN systems. 
These dual ultra-hard X-ray emitting systems are identified in the \swift/BAT 70-month catalog thanks to the \textit{combined} emission of the dual sources but are too faint to \textit{individually} fall above the flux limit (see, e.g., the earlier \swift/BAT study by \citealt{koss2012}).
Such sources should be removed from our analysis, as it is fundamentally based on a well-defined, flux-limited sample.
Details about each of these dual AGN systems are provided in \cite{Koss_DR2_overview}.
Of the 10 
sources in dual systems in BASS/DR2, only NGC~6240S \cite[BAT ID 841;][]{puccetti2016} and MCG+04-48-002 \cite[BAT ID 1077;][]{koss2016a} fall above our nominal the flux cut. 

An additional 26 sources that are detectable due to their flux being enhanced through blending with a nearby brighter BAT source are also excluded. 
These unassociated faint X-ray sources are described in detail in \citet{Ricci2017_Xray_cat} and \citealt{Koss_DR2_overview}. Together, the exclusion of faint/weak associations and of dual sources removes \nfaintdual\ objects from the sample, and leaves us with  of $\nnonblazarnonfaint$ AGN.
\subsubsection{Final, redshift-restricted AGN samples}\label{sec:sample_final}

After applying all the cuts mentioned so far, our base sample of non-beamed, non-Galactic-plane AGN includes \nnonblazarnonfaint\ 
sources, and is used for (parts) of our XLF analysis. 
We next introduce a number of additional criteria to retain only those AGN that lie in regions of parameter space in which the BAT and BASS selection functions are well understood, highly complete, and highly reproducible.

We first restrict our analysis to sources in the $\zminsamplebhmf \leq z \leq \zmaxsamplebhmf$ range. 
This excludes the \nlowzexclude\ 
nearest BASS/DR2 AGN, for which redshift-based distance determinations may be affected by peculiar velocities, and/or which may be outliers in terms of their location in the $L-z$ plane. 
Altogether, the $\zminsamplebhmf \leq z \leq \zmaxsamplebhmf$ sample contains \nnonblazarredshiftcut\ 
objects (after excluding \nhighzexclude\ more AGN at at $z > 0.3$). 

We next limit our sample to have reliable measurements of \Lbol, \Mbh, and \lamEdd, within ranges that are reasonable for persistent, radiatively-efficient accretion onto SMBHs and that can be probed within BASS/DR2 with a high degree of completeness.
These basic measurements are described in Section~\ref{sec:measurements} below, while the chosen ranges are listed in Table~\ref{tab:bhmfpara}.
Specifically, we include only those AGN with black hole masses in the range $\mbhminsample \leq \log (\Mbh / \Msol) \leq \mbhmaxsample$, and with Eddington ratios in the range $\lamminsample \leq \log \lamEdd \leq \lammaxsample$. 
%

Throughout our analysis, we further classify sources as being either broad-line (``Type~1'' hereafter) or narrow-line (``Type~2'' hereafter) AGN, based on the presence of {\it any} broad Balmer lines \cite[as described in][]{Koss_DR2_overview,Koss_DR2_catalog} and on how their most reliable measures of \Mbh\ and \lamEdd\ were derived (see Section~\ref{sec:measurements} below), which in turn has consequences for uncertainty and incompleteness estimation. 
Thus, our sample of Type~1 AGNs (\ntotallcutstypeone\ sources) --- sources that have at least one broad Balmer emission line --- also includes so-called ``Seyfert 1.9'' AGNs with broad \Halpha\ but no broad \Hbeta.
Our Type~2 AGNs (\ntotallcutstypetwo\ sources) have only narrow (Balmer) emission lines. 
Type~1 AGN are relatively unobscured [e.g., $\logNH \leq 22$] whereas Type~2 AGN tend to be more heavily obscured [$\logNH > 22$], although as shown in Section~\ref{sec:stat_method}, these \logNH\ limits do not apply strictly (see also \citealt{Oh_DR2_NLR}). 

\magenta{For our BHMF and ERDF analysis, we exclude the four Seyfert 1.9 sources that have mass estimates only from broad \Halpha\ lines, as such mass estimates were shown to be highly uncertain for heavily obscured Seyfert 1.9s in the companion BASS/DR2 paper by \cite{Mejia_DR2_BLR}. 
One of these four sources (ID 476) is the only AGN in our sample that has an estimated Eddington ratio formally greater than 10 ($\log\lamEdd \simeq 1.4$). 
Therefore, the \lamEdd\ ratio upper limit we impose does not exclude objects that would otherwise have been included in the analysis. Similarly, there are no sources in the BASS/DR2 sample with BH mass greater than the upper limit on \Mbh.
These two upper limits are important for computational reasons, and are discussed in \S~\ref{sec:bias}.}


The Venn diagrams in Figure~\ref{fig:venn} show the number of non-beamed BASS/DR2 AGN that meet each of our selection criteria (in $z$, \Mbh, and/or \lamEdd), as specified in Table~\ref{tab:bhmfpara}, for both Type~1 and Type~2 AGN. 
We also show the number of sources that fall \textit{outside} all these criteria. 
Our main BHMF and ERDF analysis is done using the \ntotallcuts\ AGN that meet all criteria (\ntotallcutstypeone\ Type~1 and \ntotallcutstypetwo\ Type~2 AGN), as shown in the central regions of the Venn diagrams.

\subsection{Data and Basic measurements}\label{sec:measurements}

The present analysis is based on four basic, interlinked measurements that are available for the BASS/DR2 AGN thanks to the extensive X-ray and optical spectroscopy that is in the heart of the BASS project: bolometric luminosities (\Lbol), black hole masses (\Mbh), Eddington-ratios (\lamEdd), and the line-of-sight hydrogen column densities (\NH).
The determination of these quantities from basic observables is described in detail in other BASS publications (see below). 
Here we provide only a brief summary of the aspects most relevant for the present analysis. 
One key consideration for deriving these quantities is to adopt a uniform approach to all BASS AGN, whenever possible (e.g., for \Lbol), and to revert to differential methodologies only when absolutely needed (e.g., for \Mbh\ determination).

First, as a primary probe of the AGN luminosity we use the integrated, intrinsic luminosity between 14--195 keV (\Lbat\ or $L_{\rm X}$ hereafter), as determined through an elaborate spectral fitting of all the available X-ray data for each BASS source, as described in detail in \cite{Ricci2017_Xray_cat}.
This X-ray spectral decomposition also yields the \NH\ measurements we use here, and we stress that the intrinsic \Lbat\ already accounts for the line-of-sight obscuration (as probed by \NH).
\magenta{The measurement uncertainties on the X-ray luminosities we use are rather small, not exceeding ${\sim}0.05-0.1$ dex for unobscured sources. Thus, whenever uncertainties on luminosities are invoked throughout our analysis, these relate to {\it bolometric} luminosities (unless otherwise noted), and are dominated by the systematics on the X-ray to bolometric luminosity conversion (see below).}

Bolometric luminosities, \Lbol, were then derived directly from \Lbat, by using a simple, universal scaling of 
\begin{equation}\label{eq:l14195_to_lbol}
    \Lbol = \kappa_{14{-}195\,\kev} \times L_{\rm 14-195\,\kev} \, .
\end{equation}
As explained in other key BASS publications \cite[e.g.,][]{Koss_DR2_catalog}, this bolometric correction of $\kappa_{14{-}195\,\kev}=7.4$ corresponds to a lower-energy bolometric correction of $\kappa_{2{-}10\,\kev}=20$, which is the average correction found for the BASS sample following the \Lhard-dependent prescription of \cite{Marconi:2004aa}, and further assumes the median X-ray power-law index found for the BASS sample, $\Gamma=1.8$ (e.g., \citealp{lanzuisi2013}). 
This bolometric correction is also in agreement with \citet{Vasudevan2009_BAT_9m}. 
Our choice to use \Lbat\ and not other (X-ray) luminosity probes is motivated by (1) the need to work as closely as possible with the \swift/BAT selection functions, (2) our desire to have the most reliable determinations of \Lbol\ for even the most obscured AGN [i.e., Compton-thick sources with $\logNH\gtrsim24$], and (3) our desire to be consistent with previous studies of the XLF of \swift/BAT-selected AGNs \cite[e.g.,][]{Sazonov:2007aa,Tueller:2008aa,Ajello:2012aa}. 
%
We acknowledge that several other, higher-order bolometric correction prescriptions were suggested and used in the literature, including corrections that depend on luminosity, \lamEdd, and/or other AGN properties \cite[e.g.,][and references therin]{Marconi:2004aa,Vasudevan2007_BC,Vasudevan2009_RM_SED_XMM,Jin2012,Lusso2012,Brightman2017,Netzer2019,Duras2020}.
\magenta{In the main part of the text,} we prefer to use a constant bolometric correction to simplify our already-complicated decomposition of the XLF, BHMF, and ERDF, and to be consistent with the rest of the BASS (DR2) analyses.
\magenta{However, we report how our conclusions change with a luminosity dependent bolometric correction in Appendix~\ref{sec:app_variable_kbol}.}

Black hole masses, \Mbh, are determined using two different approaches for AGN with or without broad Balmer line emission, and specifically broad \Halpha\ line emission, which allows for a certain \Mbh\ estimation procedure (see immediately below).
As noted above, for simplicity we refer to such sources simply as Type~1 and Type~2 sources, respectively, however we note that their detailed classification may be more nuanced (see \citealt{Koss_DR2_overview,Mejia_DR2_BLR} for a detailed discussion).

For broad line (Type~1) AGN, we rely on detailed spectral decomposition of the \Halpha\ spectral complex, and ``virial'' BH mass prescriptions which are calibrated against reverberation mapping experiments, as described in detail in \cite{Mejia_DR2_BLR}.
\magenta{Specifically, \citet{Mejia_DR2_BLR} followed} the prescription provided by \citet[Eq.~6]{GreeneHo2005}, but adjust the virial factor to $f=1$, yielding:
\begin{equation}\label{eqn:mbh_type1}
    \Mbh = 2.67 \times  10^6\, 
    \left(\frac{ L[{\rm b}\Halpha]}{10^{42}\,\ergs} \right)^{0.55} 
    \left(\frac{{\rm FWHM}[{\rm b}\Halpha]}{1000\,\kms}\right)^{2.06} 
    \Msun \, ,
\end{equation}
\magenta{where $L[{\rm b}\Halpha]$ and ${\rm FWHM}[{\rm b}\Halpha]$ are the luminosity and width of the broad part of the \Halpha\ emission line, respectively.}
For narrow line (Type~2) AGN, \citet{Koss:2017aa} and \citet{Koss_DR2_overview}  rely on measurements of the stellar velocity dispersion (\sigs) in the AGN host galaxies and the well-known $\Mbh-\sigs$ relation. 
Specifically, \sigs\ was measured from either the Ca\,\textsc{H+K}, Mg\,\textsc{i}, and/or the Ca\,\textsc{ii} triplet spectral complexes, as described in detail in \cite{Koss_DR2_sigs}. 
We then used the relation determined by \citet[Eq. 3]{Kormendy:2013aa}:
\begin{equation}\label{eqn:mbh_type2}
    \Mbh = 3.09\times 10^8 \left(\frac{\sigs}{200\,\kms}\right)^{4.38} \Msun .
\end{equation}
We stress that these two types of \Mbh\ prescriptions are consistently calibrated. Specifically, the broad-line \Mbh\ prescription in Eq.~\ref{eqn:mbh_type1} is derived by assuming that low-redshift, broad-line AGNs (particularly those with reverberation mapping measurements) lie on the same $\Mbh-\sigs$ relation as do narrow-line AGNs and inactive galaxies (that is - Eq.~\ref{eqn:mbh_type2}; see, e.g.,  \citealt{Onken2004} and \citealt{Woo2013_f_Msig} for detailed discussion).

The uncertainties in both types of \Mbh\ estimates are completely dominated by systematics and are of the order $0.3-0.5$ dex \citep{gultekin2009,shankar2019MNRAS,Shen2013_rev,Peterson2014_rev}. 
The lower end of this range is consistent with the scatter seen in the $\Mbh - \sigs$ (or $\Mbh-M_{\rm host}$) relation \cite[e.g.,][]{Sani:2011aa, Kormendy:2013aa}, while the upper end includes also the other key ingredients of ``virial'' \Mbh\ determinations in broad-line AGNs \cite[e.g.,][]{Shen2013_rev,Peterson2014_rev}. 
The uncertainties on \lamEdd\ are naturally dominated by the systematic uncertainties on \Mbh, and are thus also of the order $0.3-0.5$ dex, although the range and trends seen for bolometric corrections \cite[e.g.,][]{Marconi:2004aa,Vasudevan2009_RM_SED_XMM} suggest that uncertainties on \lamEdd\ may be yet higher. 
In comparison, our {\it measurement} uncertainties on $L({\rm b}\Halpha)$, FWHM(b\Halpha), and \sigs\ are typically of order 10\%, which would add up to $\sim$ 0.1 dex uncertainty for \Mbh\ estimates in broad-line AGN, and $<$ 0.2 dex for narrow-line AGN.
Our analysis takes into account the large (systematic) uncertainties on \Mbh\ determinations for all BASS AGN, as detailed in \S\ref{sec:stat_method}. 


Finally, \lamEdd\ is determined by combining the \Lbol\ and \Mbh\ estimates mentioned above, using the relation 
\begin{equation}\label{eq:bhmflbol}
\log \lamEdd = \log(\Lbol/\ergs) - \log(\Mbh/\Msun) - \lboltolam \, ,
\end{equation}
which is appropriate for solar metallicity gas. \magenta{As the uncertainty in both luminosity and mass must be taken into account to calculate \lamEdd, we present one case in our main analysis where we assume higher error in $\log~\lamEdd$ to take  the uncertainty in luminosity measurement/bolometric correction into account, and find that our results converge on the same solution. Bigger uncertainties in $\log~\lamEdd$ and results due to a variable bolometric correction is presented in the Appendix~\ref{sec:app_variable_kbol}.}

There are alternative galaxy-black hole scaling relationships suggested by \cite{bernardi2007}, \cite{Shankar:2017aa} and \cite{Shankar2020_NatAst}, that correct for selection biases that may affect the samples used for calibrating these mass measurement relationships. 
While recalculating masses computed using Equation~\ref{eqn:mbh_type2} (i.e., velocity dispersion) would be trivial, consistently recalibrating all the other masses, calculated using various other methods, to account for these selection effects is beyond the scope of the present analysis.

\subsection{Source number counts}\label{sec:logNlogS}

Figure~\ref{fig:logN_logS} shows the cumulative and differential source counts (number density per square degree and its differential form, respectively) for the various samples relevant for the present study, specifically:
(1) the input 70-month catalog ($z \leq 0.3$; \nnonblazarhighredshiftcut\ sources), 
(2) our redshift-restricted AGN sample ($0.01 \leq z \leq 0.3$; \nnonblazarredshiftcut\ sources), and
(3) our final BASS/DR2 BHMF/ERDF sample (i.e., with \Mbh\ and \lamEdd\ cuts; \ntotallcuts\ sources).
All three samples exclude the sources described in \S\ref{sec:sample_exclusion}. 
In all cases, the uncertainties are derived assuming that the source counts follow Poisson distributions. 
Figure~\ref{fig:logN_logS} also shows the best-fit curves corresponding to the various samples, 
\magenta{which use the following functional form for the differential number counts of the three samples:} 
\begin{equation}\label{eq:logNlogS}
    \frac{dN}{dS} = A \times (S/10^{-11})^{-\alpha} \, .
\end{equation}
\magenta{We limit our fits to the $F_{\rm 14-195,~obs} \geq 10^{-11.1}\,{\rm erg\,s^{-1}\,cm^{-2}}$ flux regime, since the flux-area curve for Swift-BAT 70-month survey (discussed in more detail in \S\ref{sec:stat_method}) is sparsely sampled at lower flux level, making it difficult to accurately calculate the surface density of faint BAT sources (see left panel of Fig.~\ref{fig:logN_logS}).
} 
\magenta{Our fits are derived using orthogonal distance regression (ODR), so to properly account for uncertainties on both axes (i.e., $\log S$ and $\log N$).
One limitation of the ODR method is the inherent assumption of symmetric uncertainties (which in our case are an average of the upper and lower errors). For the Poisson uncertainties relevant for our analysis, this introduces only a minor change compared with the real, asymmetric uncertainties (e.g., for a bin with 25 objects, the errors are ${+6.07}$ and ${-4.97}$; see \citealt{Gehrels:1986aa}). 
To verify that our results are not significantly affected by choice of fitting method (and associated treatment of uncertainties), we have carried out an additional maximum likelihood estimator (MLE) fitting, using a Fechner distribution (see \citealp{wallis2014} and references therein).
} 

The best-fit slopes derived using ODR are: $\alpha = 2.50 \pm 0.04$, $2.62 \pm  0.07$ 
and $2.67 \pm 0.05$, for the 70-month AGN, the redshift-restricted, and final BHMF/ERDF AGN samples, respectively. 
The best-fit normalizations are $\log A = 9.43 \pm  0.03$, $9.44 \pm  0.03$, and $9.43 \pm 0.02$, respectively. 
\magenta{The MLE-based best fitting results are in agreement with the ORD ones, within $\pm 1 \sigma$ errors.}
Our best-fit curve for the full sample is consistent with the expected slope for a fully uniform, Euclidean distribution ($\alpha=2.5$), and also with the results of several previous studies of ultra-hard X-ray selected AGN \cite[][]{Tueller:2008aa,Cusumano2010_Palermo_BAT_39m,Krivonos2010_INTEGRAL_7yr,Ajello:2012aa,Harrison2016_NuSTAR_surveys}. 

For the two limited samples, the slope is not perfectly Euclidean, but that is expected as objects have been removed from the total sample. 
We conclude that our BASS/DR2 AGN sample(s), which are used to determine the XLF, BHMF and ERDF, do not show any significant biases compared to the all-sky \swift/BAT 70-month catalog and/or other samples of ultra-hard X-ray selected AGN. We are thus confident that we can rely on the same selection criteria, and specifically sky coverage curves, as derived for the parent \swift/BAT 70-month  catalog.

\section{Statistical Inference Methods}\label{sec:stat_method}

\begin{deluxetable}{ll}
\label{tab:bhmfpara}
\tablecaption{Overview of sample selection and analysis parameters.$^{\rm a}$}
\tablewidth{0pt}
\tablehead{
\colhead{Quantity/variable} & \colhead{Symbol/value}}
\startdata
Minimum redshift considered & $z_{\rm min, s} = \zminsamplebhmf$\\
Maximum redshift considered & $z_{\rm max, s} = \zmaxsamplebhmf$\\
Minimum black hole mass considered & $\log (M_{\rm BH, min, s}/\Msun) = \mbhminsample$\\
Maximum black hole mass considered & $\log (M_{\rm BH, max, s}/\Msun) = \mbhmaxsample$\\
Minimum Eddington-ratio considered  & $\log \lambda_{\rm E, min, s} = \lamminsample$\\
Maximum Eddington-ratio considered  & $\log \lambda_{\rm E, max, s} = \lammaxsample$\\
Luminosity bin size for \Vmax\ method & $d\log L_{\rm X} = \binxlf$\\
BH mass bin size for \Vmax\ method    & $d\log (\Mbh/\Msol) = \binbhmf$\\
Edd. ratio bin size for \Vmax\ method & $d\log \lamEdd = \binerdf$\\
Assumed uncertainty on $\log \Mbh$     & $\sigma_{\Mbh} =$ 0.3, 0.5 dex\\
Assumed uncertainty on $\log \lamEdd$  & $\sigma_{\log \lamEdd} =$ 0.3$^{\rm b}$, 0.5 dex\\
Galactic plane exclusion$^{\rm c}$ &  $|b|\geq 5^\circ$\\
Other cuts & Beamed AGN are excluded \\
\enddata
\tablenotetext{a}{No explicit flux limits were imposed during our XLF/BHMF/ERDF analysis; instead, we used the full flux-area curve of the 70-month Swift-BAT survey \cite[][see main text]{Baumgartner:2013aa}.}
\magenta{\tablenotetext{b}{We also investigate the effect of an uncertainty of 0.2 dex in luminosity due to measurement uncertainty and the X-ray bolometric correction (Eq.~\ref{eq:l14195_to_lbol}).
Variable bolometric corrections are explored in  Appendix~\ref{sec:app_variable_kbol}.}} 
\tablenotetext{c}{Meant to guarantee a high completeness of optical counterpart identification and spectroscopic coverage.}
\end{deluxetable}

The main goal of the present study is to determine and interpret the intrinsic distributions of X-ray luminosity (\Lx), SMBH mass (\Mbh), and Eddington-ratio (\lamEdd)---namely the X-ray luminosity function (XLF), black hole mass function (BHMF), and Eddington-ratio distribution function (ERDF)---for AGN in the present-day Universe, in the most complete way possible. 
In what follows, we provide a detailed description of the statistical inference methods we use to derive these distributions from the basic measurements available for our sample of AGN drawn from the all-sky Swift-BAT survey and the BASS project.

The main obstacle in deriving the statistical properties of any sample of astrophysical sources is to account for the various factors of incompleteness and bias that are encoded in the observed sample in hand.
The first and most obvious source of incompleteness for a flux-limited survey such as that of Swift-BAT AGN is the Malmquist bias, where less luminous sources can only be detected within a small volume at low redshift, whereas higher-luminosity ones are detected even at high $z$. 
This leads to the severe underestimation of the space densities of the former and an overestimation of the space densities of the latter. 

More complex forms of bias are introduced once the incompleteness in terms of luminosity is translated into incompletenesses in the distributions of related quantities, such as masses or growth rates \cite[e.g.,][]{Marchesini:2009aa, Pozzetti:2010aa, Weigel:2016aa}. 
Specifically for the present study, the BASS AGN are selected based on their ultra-hard X-ray AGN luminosity. 
For the BHMF, this results in a bias against low mass black holes which will be too faint to lie above the flux limit, unless they have exceptionally high Eddington-ratios. Similarly, low Eddington-ratio AGN will not be part of the ERDF since they are too faint, even if they are massive. 
\citet[][hereafter \citetalias{Schulze:2010aa}]{Schulze:2010aa} refer to this incompleteness as ``sample censorship'', \magenta{however we refer to it as ``sample truncation'' hereafter.}

Another potential source of bias may arise from the need to measure the properties for which the statistics are to be surveyed. 
Specifically, our estimates of BH mass depend on robustly measuring the luminosities and widths of broad emission lines (Eq.~\ref{eqn:mbh_type1}) or the widths of stellar absorption features (Eq.~\ref{eqn:mbh_type2}).
This, in turn, requires well-calibrated, medium-to-high spectral resolution observations, and also carries significant (systematic) uncertainties.

Our statistical inference methodology accounts for all these possible biases.
We first constrain the XLF by using the classical 1/\Vmax\ method \citep{Schmidt:1968aa,avnibahcall1980}. We apply the same method to the \Mbh\ and \lamEdd\ measurements to gain an initial guesses for the BHMF and the ERDF. We assume functional forms and fit the resulting distributions. We then use a parametric maximum likelihood approach and our initial guess to simultaneously correct the BHMF and the ERDF for sample truncation. For the bias correction we follow the approach by \citetalias{Schulze:2010aa} and \citet[][hereafter \citetalias{Schulze:2015aa}]{Schulze:2015aa}. 
We test our approach by creating mock catalogs, determining the corresponding distributions, and comparing those to the assumed input. 

Throughout this work, we implicitly assume that the ERDF is \Mbh-independent. 
This assumption is further justified by the analysis we present, based on splitting our sample to two \Mbh\ regimes (Section~\ref{sec:results}).
We also do not impose any Eddington-ratio ($\lamEdd$) dependence of the distribution of absorbing columns (i.e., \NH).
Although a certain, complex, link between \lamEdd\ and \NH\ was suggested by several studies \cite[][and references therein]{Ricci2017_Nat}, our choice allows us to obtain independent evidence for such a link using our Type~1 (mostly unobscured) and Type~2 (mostly obscured) AGN samples, instead of assuming it {\it a priori}.
%


\begin{figure}[h]
\label{fig:flux_area_curve}
    \centering
	\includegraphics[width=0.47\textwidth]{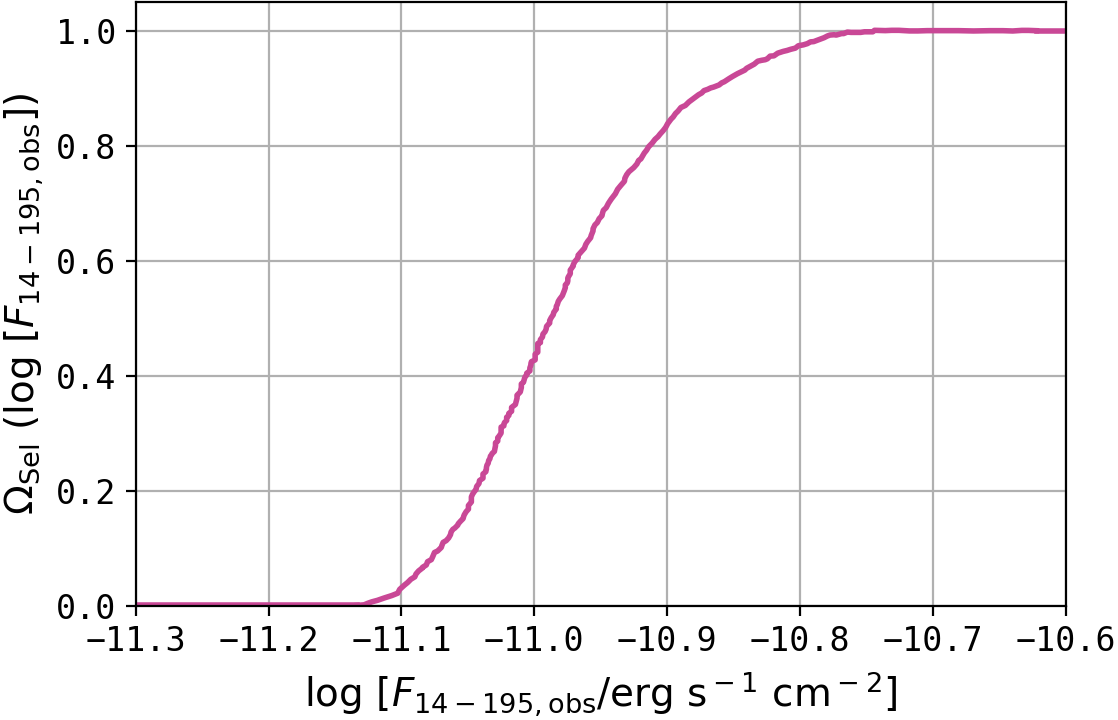}~
	\caption{Flux-area curve of the \swift/BAT 70-month all-sky survey. To compute the XLF and gain an initial guess for the BHMF and the ERDF, we use this flux-area curve rather than a constant flux limit. 
	For bright ultra-hard X-ray sources, $\log (F_{\rm 14{-}195, obs}/\ergcms) > -10.7$, the survey is complete over the entire sky. For fainter sources, the completeness decreases as a function of the observed 14$-$195 keV X-ray flux. 
	This Figure converts Figure 10 of \citet{Baumgartner:2013aa} cumulative flux-area curve from mCrab to \ergcms units.}
\end{figure}

\subsection{Survey specific considerations}\label{sec:flux_area_attenuation}

\begin{figure}[h]
    \centering
	\includegraphics[width=0.47\textwidth]{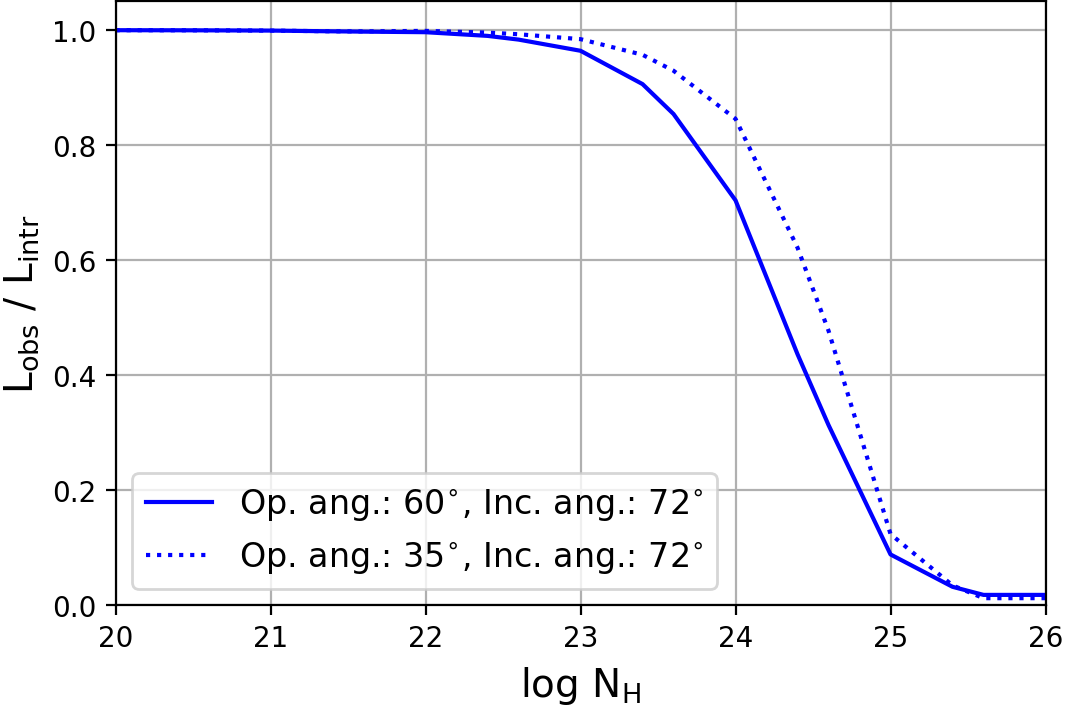}
	\caption{\label{fig:attenuation_curve} Hard X-ray (14$-$195 keV) attenuation curve for the BAT survey, assuming two torus opening angles: 60$^{\circ}$ (\textit{solid blue line}) and 35$^{\circ}$ (\textit{dotted blue line}). For extremely obscured sources,  with $\logNH > 25$, the fraction of observed luminosity relative to the intrinsic luminosity is less than 5\%, making them very difficult to detect even with the high energy X-ray window of BAT. This is an update to the \citet{Ricci:2015aa} attenuation curve with a newer torus model (see text for details).} 
\end{figure}


We take advantage of the well-constrained flux-area curve of the Swift-BAT survey \citep{Baumgartner:2013aa}. The flux-area curve $\Omega_{\rm sel}(\log F_{\rm X})$, shown in Figure \ref{fig:flux_area_curve}, accounts for the fact that the effective area of the BAT survey is a function of the observed $14- 195$ keV X-ray flux. 
For bright ultra-hard X-ray sources with $\log (F_{\rm 14{-}195, obs}/\ergcms) > -10.7$, the BAT survey is complete over the entire sky, i.e. $\Omega_{\rm sel} = 1$. Sources with $\log ({\rm F}_{\rm X, obs}/\ergcms) < -11.3$ are completely missed by the BAT 70-month survey \& catalog, and thus $\Omega_{\rm sel} = 0$. 
Our analysis makes explicit use of the complete flux-area curve, including the intermediate values for sources with $-11.3 \lesssim \log (F_{\rm 14{-}195, obs}/\ergcms) \lesssim -10.7$. 
Note that this is the cumulative flux-area curve over the entire sky, and the sensitivity is somewhat non-uniform (as shown in Figure~1 of \citealp{Baumgartner:2013aa}). Taking the slight positional differences in sensitivity is beyond the scope of this work. 

Additionally, we account for the fact that the XLF describes the distribution of AGN according to their \textit{intrinsic} ultra-hard X-ray luminosity, while their detectability is a function of their \textit{observed} (ultra-hard) X-ray flux, which depends on the amount of obscuration along the line of sight (in addition to distance, obviously). 
To correct the intrinsic luminosities for obscuration and to compute observed ultra-hard X-ray luminosities, we use the \swift/BAT attenuation curve shown in Figure \ref{fig:attenuation_curve}. 
This observed luminosity is a sum of transmission of the intrinsic power-law radiation and reflection from the accretion disk through the obscuring torus. 
This 14$-$195 keV attenuation curve is similar to the attenuation curve of  \citet{Ricci:2015aa}, which was calculated using the torus model of \cite{Brightman:2011aa}, based on the spectral models used in \citet{Ueda:2014aa}. 
We recalculated the attenuation curve using the \textsc{borus02} model  \citep{mislav2018}, which updated the \citet{Brightman:2011aa} torus, further assuming a photon index of $\Gamma = 1.8$, ${\rm E}_{\rm cutoff} = 200$ keV, a torus opening angle of $60^\circ$ and an inclination angle of $72^\circ$ \citep{Gilli2007,lanzuisi2013,Ueda:2014aa,Aird2015,ananna2019,ananna2020aa}. We have tested how our results vary with changes in torus opening angle and (line-of-sight) inclination angle. 
Following \citet{Ricci:2015aa}, we also consider another  attenuation curve for a model with opening angle of $35^{\circ}$ and an inclination angle of $72^{\circ}$ (shown in Figure~\ref{fig:attenuation_curve}). We discuss the impact of varying our template spectra (and other model dependent parts of our approach) on our final results in \S\ref{sec:pi}.

While the BAT 14$-$195 keV energy range is not strongly affected by attenuation up to $\logNH\simeq23.5$, Fig.~\ref{fig:attenuation_curve} shows that for Compton-thick sources [$\logNH\gtrsim24$] at least 30\% and up to $\sim$98\% of the intrinsic ultra-hard X-ray emission is lost [i.e. only the scattered component is detectable as \logNH\ approaches 26], and the luminosities of such sources may have been drastically underestimated. Our analysis makes explicit use of the complete attenuation curve, thus properly linking (limiting) observed fluxes and intrinsic luminosities, given the measured \NH\ of each AGN \citep{Ricci2017_Xray_cat,Koss_DR2_overview}.

We note that these links between observed fluxes, intrinsic luminosities and surveyed angles (and volumes) affect not only the XLF but also the BHMF and ERDF, as the three key properties (\Lx, \Mbh, \lamEdd) are closely related, through Eqs.~\ref{eq:l14195_to_lbol} and \ref{eq:bhmflbol}.

\subsection{The 1/\Vmax\ method and the XLF}\label{sec:Vmax}

The 1/\Vmax\ method \citep{Schmidt:1968aa} provides a way to correct for the incompleteness (i.e., lower luminosity/mass objects fall below survey sensitivity at larger distances). 
When counting the sources within a given luminosity (or mass etc.) bin, each source $i$ is weighted by $V_{\rm max,i}$ -- the maximum volume within which source $i$ could have been detected, given the survey properties. 
In the case of a luminosity function, low- and high-luminosity objects are weighted by small and large volumes, respectively. This increases and decreases their relative contribution to the respective space densities. 

To measure the distribution of AGN luminosities, black hole masses, and Eddington-ratios of the BASS sample, we bin in $\log \Mbh$, $\log \lamEdd$, and $\log L_{\rm X}$.\footnote{Note that for simplicity, we use $L_{\rm X}$ to denote \Lbat unless explicitly noted otherwise.} 
We then use the 1/\Vmax\ method to determine the corresponding space densities $\Phi_{\rm M}(\log \Mbh)$, $\xi(\log \lamEdd)$, and $\Phi_{\rm L}(\log L_{\rm X})$. 
We compute the \Vmax\ values corresponding to the intrinsic, ultra-hard X-ray luminosity of each source by considering the respective observed flux and the survey completeness, as detailed below.
As this 1/\Vmax-based calculation does not include a robust correction neither for sample truncation nor for uncertainties on the relevant key quantities, it only allows us to gain an initial guess for the BHMF and the ERDF. 

For the XLF, the space density in luminosity bin $j$ is given by the sum over all $N_{\rm bin}$ weighted objects within this bin:
\begin{equation}
\Phi_{\rm L, j} \, d\log L_{\rm X} = \sum_{i}^{N_{\rm bin}} \frac{1}{V_{\rm{max}, i}} .
\end{equation}
We compute $\Phi_{\rm M}(\log \Mbh)$ and $\xi(\log \lamEdd)$ similarly and use the bin sizes $d\log L_{\rm X}$, $d\log \Mbh$, and $d\log \lamEdd$ given in Table \ref{tab:bhmfpara}. 

To compute \Vmax\ values for each AGN, we express the completeness of the BAT survey as a function of the intrinsic ultra-hard X-ray luminosity, redshift, and column density: $\Omega_{\rm sel}(\log L_{\rm X}, N_{\rm H}, z)$. For each object $i$ at redshift $z_i$ with intrinsic luminosity $\log L_{\mathrm{X}, i}$ and column density $N_{\mathrm{H}, i}$, the maximum comoving volume $V_{\mathrm{max}, i}$ is then given by \cite[e.g.,][]{Hogg:1999aa, Hiroi:2012aa}:
\begin{equation}
\begin{aligned}
&V_{\rm{max}, i}(\log L_{\mathrm{X}, i}, N_{\mathrm{H}, i}, z_i)=\\
&4 \pi \times \frac{c}{H_0} \times \int_{z_{\rm min, s}}^{z_{\rm max, s}} \Omega_{\rm sel}(\log L_{\mathrm{X}, i}, N_{\mathrm{H}, i}, z)  \frac{D_{C}(z)^2}{E(z)} dz.
\end{aligned}
\end{equation}
In this expression, $D_C(z)$ corresponds to the comoving distance to redshift $z$, $E(z)$ is given by $\sqrt{\Omega_{\rm M} (1+z)^3 + \Omega_{\rm \Lambda}}$. 
The integration limits $z_{\rm min, s}$ and $z_{\rm max, s}$ correspond to the minimum and maximum redshifts of our BASS/DR2 (refined) sample, respectively (see Table \ref{tab:bhmfpara}). 
Note that, unlike what is done in some studies, our calculation does not explicitly introduce a rigid flux limit, which in turn would impose a maximal redshift up to which each source could be observed, $z_{\mathrm{max}, i}$. Instead, this information is encoded in the flux-area curve, which ultimately provides the same outcome as $\Omega_{\rm sel}=0$ when the source observed flux becomes too faint [i.e., $\log (F_{\rm 14{-}195, obs}/\ergcms) \lesssim -11.3$].

Note that the 1/\Vmax\ method is very sensitive to the flux-area curve, which is non-uniform (as described in \S\ref{sec:flux_area_attenuation}). 
As a specific example, the AGN in NGC~5283 (BAT ID 684), with $z  = 0.01036$ and $\log (F_{\rm 14{-}195, obs}/\ergcms) = -11.14$, falls within the criteria used to select the redshift-restricted sample (Table~\ref{tab:bhmfpara}), and it is close both to the low-$z$ cut we use and to the regime where area flux curve drops to zero. 
As a result, the \Vmax\ within which this object can be detected above $z > 0.1$ is very low, and this drives up the corresponding 1/\Vmax\ value and any related contribution to the population-wide distributions under study. 
However, the method used to calculate the bias-corrected, intrinsic BHMF and ERDF (described in \S\ref{sec:bias}) is {\it not} as dramatically affected by a single object as the direct 1/\Vmax\ approach. 
We show the effect of including this single object on 1/\Vmax\ values in Appendix~\ref{sec:app_tables}, whereas the 1/\Vmax\ values shown in plots in the main body of the text exclude this object. We include this object in the bias-corrected part of our analysis.

\magenta{There are several published statistical methods that address (some of) the limitations of the 1/\Vmax\ approach, such as the Lynden-Bell-Woodroofe-Wang estimator \citep{lyndenbell1971,woodroofe1985,Choloniewski1987,wang1989,efronpetrosian1992} which corrects for sample truncation. 
Unlike the 1/\Vmax\ method, this estimator does not depend on the assumed bin size and does not assume a constant space density over every given bin. Another flexible Bayesian parametric framework to estimate luminosity functions while correcting for sample truncation was suggested by \citet{kelly2008}. 
However, most previous (X)LF works used the 1/\Vmax\ approach.
Motivated by the desire for our XLF results to be directly comparable to previous studies, and by the fact that our core analysis methodology \textit{does} ultimately correct for both sample truncation and  uncertainties on key quantities (in all three distribution functions), 
we chose to present the 1/\Vmax\ results, if only as a first order (albeit somewhat biased) estimate of the XLF.}

To estimate the random uncertainties on $\Phi_{\rm L}(\log L_{\rm X})$, $\Phi_{\rm M}(\log \Mbh)$, and $\xi(\log \lamEdd)$, we follow the approach by \citet[][see also \citealt{Zhu:2009aa, Gilbank:2010aa}]{Weigel:2016aa}. \magenta{These errors are essentially Poisson errors on the number of sources in each bin, with effective weights applied based on the corresponding 1/\Vmax\ estimates. 
The exact prescriptions used to calculate these uncertainties are} provided in Appendix~\ref{sec:app_error_vmax}.

\subsection{Functional forms for XLF, BHMF and ERDF}\label{sec:funcitonal_forms}

After having determined the XLF and having determined an initial guess for the BHMF and the ERDF via the 1/\Vmax\ method, we assume specific functional forms for all three distributions and fit the corresponding (differential) space densities. 
For the XLF we assume a double power-law of the following form:
\begin{equation}\label{eq:bhmfbpl_xlf}
\Phi_{\rm L}(\log L_{\rm X}) = \frac{dN}{d\log L} = \Phi^{*}_{\rm L} \times \left[\left(\frac{L_{\rm X}}{L^{*}_{\rm X}}\right)^{\gamma_1} + \left(\frac{L_{\rm X}}{L^{*}_{\rm X}}\right)^{\gamma_2}\right]^{-1}\,.
\end{equation}
For the fitting procedure we parametrize the second power-law slope as $\gamma_2 = \gamma_1 + \epsilon_{\gamma}$ with $\epsilon > 0$ to avoid degeneracy between the two exponents. 
Given what is known about the rather universal nature of the AGN LF \cite[e.g.,][and references therein]{Shen2020_QLF}, in practice this means that our $\gamma_1$ and $\gamma_2$ correspond to what is often referred to as the faint- and bright-end LF slopes, respectively.

For the BHMF, $\Phi_{\rm M}(\log \Mbh) \equiv \frac{dN}{d\log \Mbh}$, 
we use a modified Schechter function, defined as:
\begin{equation}
\begin{aligned}\label{eq:bhmf_msch_def}
\Phi_{\rm M}(\log \Mbh) 
\propto \ln(10)  \Phi^{*} \left(\frac{\Mbh}{\Mstarbh}\right)^{\alpha + 1} \exp{\left[-\left(\frac{\Mbh}{\Mstarbh}\right)^{\beta}\right]}. 
\end{aligned}
\end{equation}
This choice is motivated by the general shape of the galaxy stellar mass function \cite[e.g.,][and references therein]{Baldry:2012aa,Weigel:2016aa,Davidzon2017} and the close relations between SMBH and galaxy mass \citep{Kormendy:2013aa}.

For the ERDF, we use a broken power-law which we define as:
\begin{equation}\label{eq:erdf_dbpl_def}
\xi(\log \lamEdd) = \frac{dN}{d\log\lamEdd} \propto \xi^{*} \times \left[\left(\frac{\lamEdd}{\lamEdd^*}\right)^{\delta_1} + \left(\frac{\lamEdd}{\lamEdd^*}\right)^{\delta_2}\right]^{-1} \, .
\end{equation}
This functional form is motivated by the fact that, qualitatively, the convolution of the BHMF with the ERDF should reproduce the LF, which directly links the bright-end slope of the XLF and the high-\lamEdd\ slope of the ERDF \cite[e.g.,][]{Caplar:2015aa,Weigel:2017aa}. 
Similar to what was done with the XLF, we parametrize $\delta_2$ as $\delta_1 + \epsilon_{\lambda}$ with $\epsilon_{\lambda} > 0$, thus linking $\delta_1$ and $\delta_2$ with the low- and high-\lamEdd\ ERDF slopes, respectively.

In the initial part of our analysis, we fit the 1/\Vmax\ measurements of the BHMF and ERDF with these functional forms simply to obtain initial guesses for the more elaborate recovery of the intrinsic BHMF and ERDF. 
These same functional forms, however, are also relevant in various other parts of our analysis and interpretation. 

To find the best-fitting parameters for the 1/\Vmax-based XLF, BHMF, and ERDF, we use the Markov chain Monte Carlo (MCMC) sampler \textsc{emcee} \citep{emcee}.  
In Appendix \ref{sec:app_fit_vmax}, we outline how we find the best-fitting functional form for the XLF. The BHMF and the ERDF are fit accordingly. 
We fit all three distributions independently. 
\magenta{Note that this MCMC analysis only fits the functions given in Equations~\ref{eq:bhmfbpl_xlf}, \ref{eq:bhmf_msch_def} and \ref{eq:erdf_dbpl_def} to 1/\Vmax\ estimates, and not directly to the observed astrophysical quantities (luminosities, masses, and/or Eddington ratios). 
In the next section, we describe a more elaborate method to correct for the limitations of 1/\Vmax\ approach (discussed in \S~\ref{sec:Vmax}), where we fit the aforementioned functions to the data directly.}

\subsection{Correcting the BHMF and the ERDF for sample truncation and for uncertainties}\label{sec:bias}

\subsubsection{The basic principle}\label{sec:biasintro}
To correct for sample truncation---i.e., the bias against low mass and low Eddington-ratio AGN due to the flux limited nature of the Swift-BAT and BASS samples---we follow the approach of \citetalias{Schulze:2010aa} and \citetalias{Schulze:2015aa}. The technique requires the assumption of (intrinsic) functional forms for the BHMF and the ERDF and thus should be considered as a parametric maximum likelihood approach.

Our aim is to constrain the intrinsic bivariate distribution function $\Psi(\log \Mbh, \log \lamEdd, z)$, which also provides the BHMF and ERDF, when integrated over $\log \lamEdd$ and $\log \Mbh$, respectively:
\begin{equation}\label{eq:bhmfpsi}
\begin{aligned}
\Phi_{\rm M}(\log \Mbh, z) =& \int_{\log \lambda_{\rm E, min, s}}^{\log \lambda_{\rm E, max, s}} \Psi(\log \Mbh, \log \lamEdd, z) d\log \lamEdd ,\\
\xi(\log \lamEdd, z) =& \int_{\log M_{\rm BH, min, s}}^{\log M_{\rm BH, max, s}} \Psi(\log \Mbh, \log \lamEdd, z) d\log \Mbh.
\end{aligned}
\end{equation}
The integration limits
$\log \lambda_{\rm E, min, s}$, $\log \lambda_{\rm E, max, s}$, $\log M_{\rm BH, min, s}$, and $\log M_{\rm BH, max, s}$ correspond to the minimum and maximum Eddington-ratios and black hole mass values that we are considering in our analysis (see Table~\ref{tab:bhmfpara}). 
The bivariate distribution $\Psi(\log \Mbh, \log \lamEdd)$ has the physical units of space density, namely h$^{\rm 3}$ Mpc$^{-3}$ dex$^{-2}$ (i.e., per dex in \Mbh\ and per dex in \lamEdd). 
As the redshift range is very small, we consider only a single redshift bin and do not model the redshift evolution of the XLF, the BHMF, and/or the ERDF; this is why our bivariate distribution function $\Psi(\log \Mbh, \log \lamEdd)$ does not depend directly on $z$. 
As noted above, we further assume that the ERDF is mass independent. 
Under all these assumptions, the calculation of the BHMF and the ERDF from $\Psi(\log \Mbh, \log \lamEdd)$ (the expressions in Eq.~\ref{eq:bhmfpsi}) reduces to:
\begin{equation}\label{eq:bhmfbi}
\begin{aligned}
\Phi_{\rm M}(\log \Mbh) = & \tilde{\Phi}(\log \Mbh) \times \int_{\log \lambda_{\rm E, min, s}}^{\log \lambda_{\rm E, max, s}} \tilde{\xi}(\log \lamEdd) d\log \lamEdd\\
\xi(\log \lamEdd) = & \tilde{\xi}(\log \lamEdd) \times \int_{\log M_{\rm BH, min, s}}^{\log M_{\rm BH, max, s}} \tilde{\Phi}(\log \Mbh) d\log \Mbh .
\end{aligned}
\end{equation}
{The assumed functional forms for $\tilde{\Phi}(\log \Mbh)$ and $\tilde{\xi}(\log \lamEdd)$ are given by the right sides of Equations~\ref{eq:bhmf_msch_def} and \ref{eq:erdf_dbpl_def}, respectively. These two functions are proportional to the actual BHMF and ERDF. To determine $\Phi_{\rm M}(\log \Mbh)$ and $\xi(\log \lamEdd)$, we have to define the \Mbh\ and \lamEdd\ range that is being considered and marginalize over the other distribution. 
As equation \ref{eq:bhmfbi} shows, $\tilde{\Phi}(\log \Mbh)$ and $\tilde{\xi}(\log \lamEdd)$ retain their assumed shape, however their normalizations are adjusted. In \S\ref{sec:norm}, we explain how these normalizations are determined after constraining the functional forms.}

To estimate $\Psi(\log \Mbh, \log \lamEdd)$, we compute the log-likelihood of observing our main BASS sample, which is given by

\begin{equation}\label{eq:likelihood}
    \ln \mathcal{L} =\sum_i^{N_{\rm obs}} \ln p_i(\log M_{\mathrm{BH}, i}, \log N_{\mathrm{H},i}, \log \lambda_{\mathrm{E}, i}, z_i) \, ,
\end{equation}
where $p_i$ is the probability of observing object $i$ with black hole mass $\log M_{\mathrm{BH}, i}$, Eddington-ratio $\log \lambda_{\mathrm{E}, i}$, and redshift $z_{i}$. 
The log-likelihood thus represents the sum of such (log) probabilities, over the total number of observed sources, $N_{\rm obs}$. 
For an assumed $\Psi(\log \Mbh, \log \lamEdd)$, $p_i$ is given by the expected number of objects with similar properties to object $i$, relative to the total number of sources which are predicted to be observed; 
$p_i$ thus encodes all the relevant selection effects, such as the flux limit of the survey, as we explain in detail in Section~\ref{sec:pi} immediately below. To estimate $\Psi(\log \Mbh, \log \lamEdd)$, and thus the intrinsic BHMF and ERDF, we maximize the likelihood $\mathcal{L}$, i.e., the probability of observing our ensemble of $N_{\rm obs}$ sources.

\subsubsection{The probability of observing a given source}
\label{sec:pi}

For a given bivariate distribution function $\Psi$, we express the probability of observing a specific object (index $i$) in the following way:
\blue{
\begin{equation}\label{eq:bhmfprob}
\begin{aligned}
p_i & (\log M_{\mathrm{BH}, i}, \log \lambda_{\mathrm{E}, i}, \log N_{\mathrm{H},i}, z_i)\\ 
& = \frac{N_i(\log M_{\mathrm{BH}, i}, \log \lambda_{\mathrm{E}, i}, \log N_{\mathrm{H},i}, z_i)}{N_{\rm tot}}\\
& = \frac{1}{N_{\rm tot}}\Psi(\log M_{\mathrm{BH}, i}, \log \lambda_{\mathrm{E}, i})\\
&~\Omega_{\rm sel}(\log M_{\mathrm{BH}, i}, \log \lambda_{\mathrm{E}, i}, \log N_{\mathrm{H},i}, z_i) \\
&~~p (\log N_{\mathrm{H}, i})~p(z_i)~\frac{dV_C(z_i)}{dz}
\end{aligned}
\end{equation}
Note that this expression by itself does \textit{not} correct for measurement uncertainties. 
We explain the mechanism that does ultimately account for these uncertainties in \S\ref{sec:Psi_obs} below.} 
\blue{For simplicity, we assume that $p(z_i)$ and $p (\log N_{\mathrm{H}, i})$ are independent of $\Psi$ in this expression. 
The specifics of how these functions are treated for our survey and redshift range are discussed in more detail below.} 

\begin{figure}
\begin{centering}
	\includegraphics[width=0.45\textwidth]{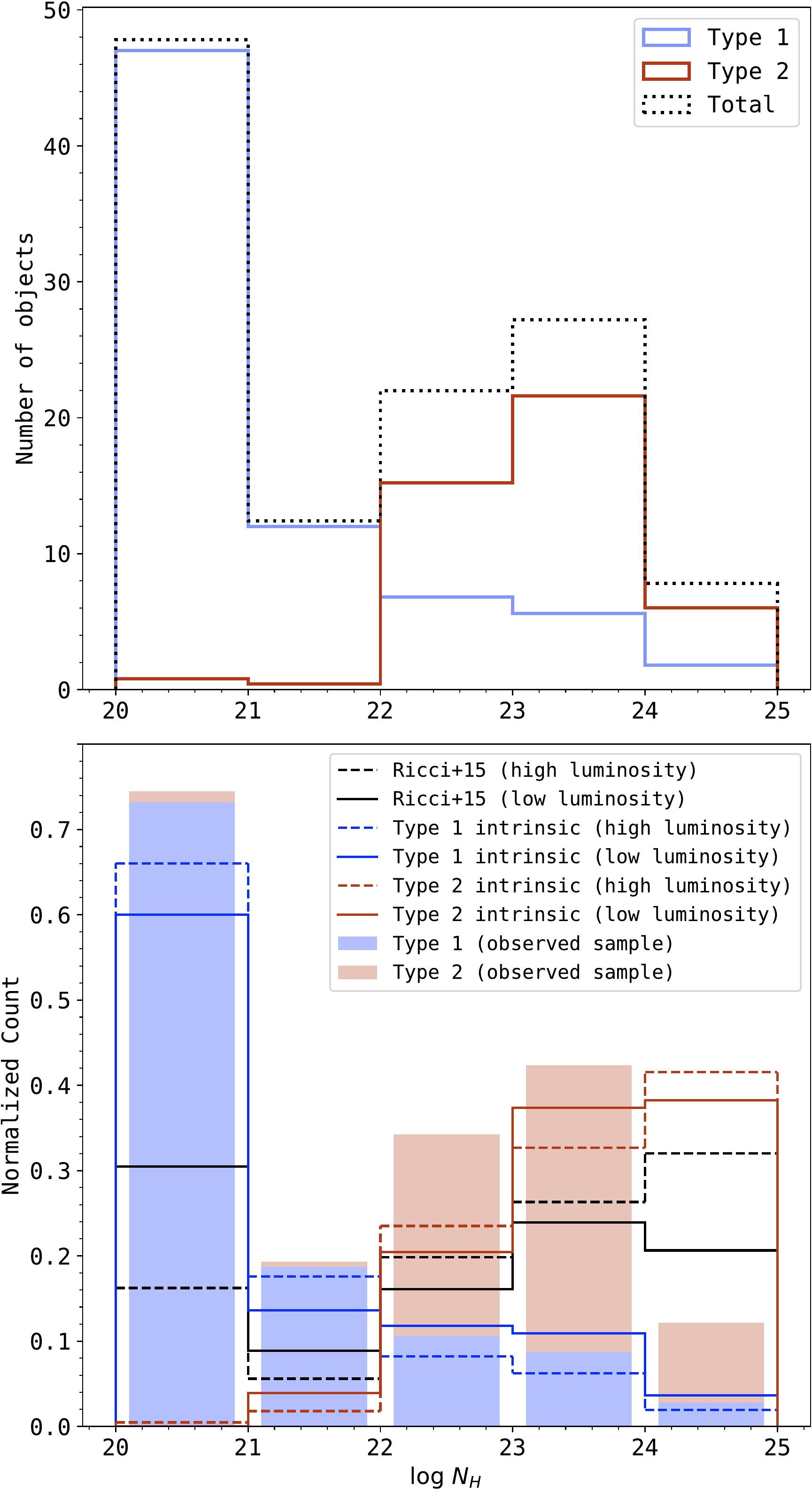}
	\caption{\label{fig:nh_dist} {\it Top panel:} Observed histograms of absorbing column density, $\log N_{\rm H}$, for Type~1 ({\it blue lines}) and Type~2 ({\it red lines}) AGN, and the overall distribution ({\it black dotted lines}). {\it Bottom panel:} Normalized 
	distributions observed for Type~1 ({\it blue stacked bars}) and Type~2 AGN ({\it red stacked bars}), along with the bias-corrected column density distribution of the BASS 70-month survey sample \citep{Ricci:2015aa}, calculated for 
	two luminosity bins: $\log (\Lbat / \ergs) \leq 43.7$ ({\it black solid lines}) and $\log (\Lbat / \ergs) > 43.7$ ({\it black dashed lines}). 
	For Type~1 AGN ({\it blue lines}), we assume the intrinsic and the observed column densities are the same, as Type~1 AGN tend to be relatively unobscured, then we calculate the Type~2 distribution ({\it red lines}) by subtracting the Type~1 distribution from the overall distribution in each luminosity bin.
	We use the intrinsic distributions when calculating the BHMF and ERDF for 
	the sample.
	}
\end{centering}
\end{figure}

Here, we describe each of the the terms used for the calculation of $p_i$:
\begin{itemize}
	
	\item $\Psi$, the intrinsic bivariate distribution function of BH mass and Eddington ratio: For a given set of $\log M_{\mathrm{BH}, i}$ and $\log \lambda_{\mathrm{E}, i}$ 
	values, the bivariate distribution function returns the space density of objects \blue{with those properties, i.e. the {\it intrinsic} number of such AGN per unit comoving volume, per dex of BH mass and Eddington ratio.}
	
	\item
	$dV_C/dz$, the comoving volume element: By multiplying the space density of objects with $\log M_{\mathrm{BH}, i}$ and $\log \lambda_{\mathrm{E}, i}$ with the comoving volume element at $z_i$, the space density is converted to an absolute number of sources. 
	The comoving volume element for the entire sky is defined as \cite[e.g.,][]{Hogg:1999aa}:
	\begin{equation}
	\frac{dV_{C}(z)}{dz} = 4\pi \frac{c}{H_0} \frac{D_{C}(z)^2}{E(z)} ,
	\end{equation}
	where $D_C$ and $E(z)$ were defined in Section~\ref{sec:Vmax}.
		
	\item
	$N_{\rm tot}$, the normalization: To obtain a probability, the number of sources with properties similar to object $i$ is normalized by the total number of sources that are expected to be part of the sample after selection effects have been taken into account. $N_{\rm tot}$ is given by an integral that accounts for all the quantities that affect the observability of our AGN within the survey:
	\blue{
	\begin{equation}\label{eq:bhmfNtot}
	\begin{aligned}
	N_{\rm tot} = \iiiint &\Omega_{\rm sel}(\log M_{\mathrm{BH}}, \log \lambda_{\mathrm{E}}, \log N_{\rm H}, z) \\
	&~\Psi(\log \Mbh, \log \lamEdd)~p (\log N_{\rm H})~p(z) \\ &~ \frac{dV_C(z)}{dz} d\log \Mbh ~ d\log N_{\rm H} ~ d\log \lamEdd ~ dz \, .
	\end{aligned}
	\end{equation}
	}
	Here, $d\log \Mbh$, $d\log \lamEdd$, and $dz$ are computed over the corresponding minimum and maximum values considered in the sample (see Table \ref{tab:bhmfpara}). 
	\item $p (\log N_{\rm H})$: A bias-corrected intrinsic absorption function needs to be assumed from which the \logNH\ values of the intrinsic AGN population are drawn. 
    As explained above, while the intrinsic luminosities of AGN are the product of their \Mbh\ and \lamEdd, any selection or distribution function that depends on their \textit{observed} fluxes would also have to depend on (the distribution of) line-of-sight obscuration, which we generally associate with circumnuclear (torodial) dusty gas. \citet{Ricci:2015aa} derived an \textit{intrinsic} $\log\NH$ distribution specifically for Swift-BAT detected AGN considering X-ray reflection from a torus. This model assumed an opening angle of 60$^{\circ}$ and considered two luminosity bins - above and below $\log (L_{\rm 14-195\,\kev}/\ergs)  = 43.7$. This intrinsic $\log\NH$ distribution is shown in Figure~\ref{fig:nh_dist}. 
    
    To test the impact of different models, we also calculate results for an absorption function with a torus opening angle of 35$^{\circ}$ (also from \citealt{Ricci:2015aa}), the results of which are reported in \S\ref{sec:results} and shown in Appendix~\ref{sec:app_tables}. 
    For consistency, the attenuation curve used when testing the effect of this absorption function also assumes an opening angle of 35$^{\circ}$ (see Figure~\ref{fig:attenuation_curve}).
    %
    While one may expect that our results would have changed significantly if the input absorption function and attenuation curve were substantially different, we find that this small change in the assumed opening angle did not alter our results substantially. 
    This indicates that our overall conclusions are robust against reasonably motivated changes to the model dependent components. 
    
    We argue that since the intrinsic absorption function derived in \cite[][in particular, with $60^\circ$]{Ricci:2015aa} is calculated specifically for the BAT sample, this is the appropriate function for our purposes. 
    We note that this absorption function is defined over the range $\logNH = 20-25$ in discrete bins of 1 dex width. 
    
    While calculating $N_{\rm tot}$ for all AGN, we assign \logNH\ to each object from the underlying distribution by drawing values from this absorption function, taking the luminosity dependence into account. 
    In this work, we chose to incorporate the luminosity dependent absorption function as provided by \citet{Ricci:2015aa}, and do not impose any direct \lamEdd\ dependence on the absorption function.
    While a \lamEdd\ dependence is supported by some studies  \cite[e.g.,][and references therein]{Ricci2017_Nat}, imposing such a dependence \textit{a-priori} would limit our ability to reveal differences in the BHMF and ERDF of obscured and unobscured sources.
    The distributions used in our calculations are shown in the bottom panel of Figure~\ref{fig:nh_dist}. 
    When we compute the BHMF and ERDF for Type~1 AGN, we assume the intrinsic absorption function to be identical to the observed one, which is justified given the negligible effect of absorption in such sources. 
    For Type~2 AGN, we subtract the distribution of Type~1 AGN from the bias-corrected overall distribution in each luminosity bin. These distributions are shown using red dashed and solid lines (for high and low luminosity bins, respectively) in the bottom panel of Figure~\ref{fig:nh_dist}. 
    \blue{Note that as the \citet{Ricci:2015aa} absorption function is luminosity dependent, the $p (\log \NH)$ used in this work is in fact $p (\log \NH \vert L_{\rm X} (\Mbh, \lamEdd))$. }
    
    \item 
    \magenta{$p(z)$: This is the redshift dependence term in the general expression, which in our case is considered constant (i.e., set to 1 in both numerator and denominator of $p_i$).}
    
    \item 
    \magenta{$\Omega_{\rm sel}$:
    In equation \ref{eq:bhmfprob}, the term $\Omega_{\rm sel}(\log M_{\mathrm{BH}, i}, \log \lambda_{\mathrm{E}, i}, \log N_{\rm H,i}, z_i)$ corresponds to the selection function of the survey. In its simplest form, $\Omega_{\rm sel}$ returns the value of the flux-area curve, $\Omega_{\rm sel}(\log F_{\rm X})$, for each source (see Section \ref{sec:Vmax}).
    To predict the X-ray flux $\log F_{\mathrm{X}, i}$ for source $i$ with BH mass $\log M_{\mathrm{BH}, i}$, an Eddington ratio $\log \lambda_{\mathrm{E}, i}$, a column density $\log N_{\mathrm{H}, i}$ and a redshift $z_i$, we perform the following sequence of calculations.
    
    First, we compute the corresponding bolometric luminosity, \Lbol, using Eq.~\ref{eq:bhmflbol} (in Appendix~\ref{sec:app_variable_kbol} we experiment with a variable, luminosity-dependent bolometric correction); 
    from \Lbol, we then calculate the intrinsic ultra-hard X-ray luminosity over the 14--195 keV range, \Lbat, using Eqn.~\ref{eq:l14195_to_lbol}; 
    from that we deduce the luminosity and flux as measured by BAT, taking into account the column density (using the curve shown in Fig.~\ref{fig:attenuation_curve}) and the luminosity distance to the source (using $z_i$).
    The calculated $\log F_{\mathrm{X}, i}$ is then used to obtain the selection function based on the curve shown in Fig.~\ref{fig:flux_area_curve}. 
    }
    
\end{itemize}

\subsubsection{The observed bivariate distribution function}
\label{sec:Psi_obs}

\magenta{
As presented in Equation~\ref{eq:bhmfprob}, $p_{i}$ is a four-dimensional normalized probability distribution function, and in order to properly account for the various selection functions in play, it needs to be convolved with the uncertainty in each of the four underlying parameters. 
Specifically, the selection effect corrected $p_i$ is as follows:
\begin{equation}\label{eq:bhmfprobconv}
\begin{aligned}
p_{\rm i, conv} & (\log M_{\mathrm{BH}, i}, \log \lambda_{\mathrm{E}, i}, \log N_{\mathrm{H},i}, z_i) = \frac{N_{\rm i, conv}}{N_{\rm tot}}\\ 
& = \frac{1}{N_{\rm tot}}\iiiint \Psi(\log M_{\mathrm{BH}}, \log \lambda_{\mathrm{E}}, \log N_{\mathrm{H}}, z) \\
&\ \times\Omega_{\rm sel}(\log M_{\mathrm{BH}}, \log \lambda_{\mathrm{E}}, \log N_{\rm H}, z)\\
&\ \times p (\log N_{\rm H})~p(z)~\frac{dV_C(z)}{dz}\\
&\ \times \omega(\log M_{\mathrm{BH}, i}, \log \lambda_{\mathrm{E}, i}, \log N_{\mathrm{H}, i}, z_i|\log \Mbh, \log \lamEdd, \log N_{\mathrm{H}}, z) \\
& ~~d\log\Mbh ~~ d\log\lamEdd ~~ d{\log \NH} ~~ dz \, .
\end{aligned}
\end{equation}
Here the denominator $N_{\rm tot}$ is the normalization constant of the original probability distribution $p_i$ and retains the value presented in Eq.~\ref{eq:bhmfNtot}, while $\omega$ is the four-dimensional Gaussian distribution function that reflects the uncertainties related to the $i$-th object, for all four directly or indirectly measured quantities. 
That is:
%
%
\begin{equation}
\label{eq:fourdomega}
\begin{aligned}
\omega & (\log M_{\mathrm{BH}, i}, \log \lambda_{\mathrm{E}, i}, \log N_{\mathrm{H},i}, z_i|\log \Mbh, \log \lambda_{\rm E}, \log N_{\mathrm{H}}, z)\\
& = \frac{1}{4\pi^2 \sigma_{\log \Mbh}~\sigma_{\log \lambda_{\rm E}}~\sigma_{\log N_{\mathrm{H}}}~\sigma_{z}} \times \\
& \exp \left( - \frac{(\log M_{\mathrm{BH}, i} - \log \Mbh)^2}{2\sigma_{\log \Mbh}^2} -  \frac{(\log \lambda_{\mathrm{E}, i} - \log \lambda_{\rm E})^2}{2\sigma_{\log \lambda_{\rm E}}^2} \right) \\
& \times \exp \left( - \frac{(\log N_{\mathrm{H}, i} - \log N_{\mathrm{H}})^2}{2\sigma_{\log N_{\mathrm{H}}}^2} -  \frac{(z_i - z)^2}{2\sigma_{z}^2} \right) \, ,
\end{aligned}
\end{equation}

where $\sigma_{\log \Mbh}$, $\sigma_{\log\lamEdd}$, $\sigma_{\log N_{\mathrm{H}}}$ and $\sigma_z$ correspond to the assumed uncertainties on $\log \Mbh$, $\log \lamEdd$, $\log \NH$ and $z$ (respectively; see Table~\ref{tab:bhmfpara}). 
Note that---at least for $\log~\Mbh$ and $\log~\lamEdd$---these are dominated by systematic uncertainties inherent to the BH mass estimation methods we rely on.

Since we have spectroscopic redshifts for all sources in our sample, and the obscuring column densities $\log\NH$ are derived from extensive spectral decomposition of multi-mission X-ray data (combining Swift-BAT and ancillary spectra at $E < 10$ keV), the convolution over $\log~\NH$ and $z$ (with $\sigma_{\log~\NH} = 0.25$ and $\sigma_z = 0.005$) does not change our results but makes the process significantly more computationally expensive. 
Thus, in practice we choose to only convolve over two dimensions:

\begin{equation}
\label{eq:twodomega}
\begin{aligned}
\omega & (\log M_{\mathrm{BH}, i}, \log \lambda_{\mathrm{E}, i}|\log \Mbh, \log \lambda_{\rm E})\\
& = \frac{1}{2\pi\sigma_{\log M_{\rm BH}} \sigma_{\log \lambda_{\rm E}}} \times \\
& \exp \left(- \frac{(\log M_{\mathrm{BH}, i} - \log \Mbh)^2}{2\sigma_{\log \Mbh}^2} -  \frac{(\log \lambda_{\mathrm{E}, i} - \log \lambda_{\rm E})^2}{2\sigma_{\log \lambda_{\rm E}}^2} \right) \, .
\end{aligned}
\end{equation}

We note that convolving over $z$ and/or $\log \NH$ may be, in principle, essential and beneficial for the analysis of samples that have photometric redshifts and/or more uncertain column density measurements.
} 


In our main analysis we have assumed $\sigma_{\log \Mbh} = \sigma_{\log \lamEdd} = 0.3$ dex (see Table \ref{tab:bhmfpara}).
The assumption of no uncertainties (i.e., $\sigma_{\log \Mbh} = \sigma_{\log \lamEdd} = 0$) is used only to test and demonstrate our analysis framework in Appendix~\ref{sec:app_test}.
A higher value of $\sigma_{\log \Mbh} = \sigma_{\log \lamEdd} = 0.5$ dex is required for Type~2 AGNs, to take into account rotation and aperture effects in the optical spectroscopy that is used to measure \sigs\ (see, e.g., \citealt{gultekin2009} and \citealt{shankar2019MNRAS}). 
Additionally, we report results for a scenario where $\sigma_{\log \Mbh} = 0.3$ dex, along with a total uncertainty of 0.2 dex in bolometric luminosity ($\siglscatt$), which reflects both measurement and systematic uncertainties (dominated by the latter).  
As $\log \lamEdd$ is calculated using two observed quantities (luminosity and mass), the uncertainty in luminosity measurement also contributes to the total uncertainty in $\log \lamEdd$.
Therefore, the uncertainty in $\log\lamEdd$ is $\sigma_{\log\lamEdd} \simeq 0.36$ (i.e. log uncertainties in mass and luminosity added in quadrature). 
\blue{We note that this approach to calculating $\sigma_{\log\lamEdd}$ is conservative, since the actual measurement errors on (X-ray) luminosities are much smaller than 0.2 dex, and since the intertwined nature of $L$, \Mbh, and \lamEdd\ means that any systematic errors on $\log\Mbh$ and $\log \lamEdd$ would be anti-correlated, rather than independent. 
As shown in \S~\ref{sec:results}, the (conservative) various levels of uncertainty we assume do not significantly affect our key results, which attests to the robustness of our conclusions.} 

In Appendix~\ref{sec:app_variable_kbol}, we report the results of an even more complex scenario, assuming luminosity dependent bolometric correction (from \citealt{Duras2020}), and an even larger error on $\log \lamEdd$. Our framework has also been tested using mock catalogs for this scenario, and the results are shown in Appendix~\ref{sec:app_test}. In the main part of the analysis, we present the results of the simpler scenario with constant bolometric correction.

We stress again that we have \textit{not} imposed any dependence of obscuration on either luminosity, \Mbh\ or \lamEdd\ in the main analysis. Thus, any difference between the XLF, BHMF, and/or ERDF derived for the obscured and unobscured sub-samples (Type~1 and 2 AGNs) would occur independent of our model assumptions. 
\magenta{In Appendix~\ref{sec:app_variable_kbol}, where we present results for a variable bolometric correction, some more complex assumptions are introduced. For example, Type~1 and Type~2 AGN have different bolometric corrections. This might artificially introduce difference in results between the two populations, therefore we keep such assumptions to a minimum in our main analysis.}

\subsubsection{Constraining $\Psi$}\label{sec:norm}

To determine the intrinsic bivariate distribution function, $\Psi(\log \Mbh, \log \lamEdd)$, we maximize the likelihood of observing our main input sample. 
Expressing the $\log$-likelihood (Eq.~\ref{eq:likelihood}) using  Eq.~\ref{eq:bhmfprob}:
\begin{equation}
\begin{aligned}
\ln \mathcal{L} =\sum_{i}^{N_{\rm obs}} \ln\left[N_{\rm i, conv}(\log M_{\mathrm{BH}, i}, \log \lambda_{\mathrm{E}, i}, \log N_{\mathrm{ H}, i}, z_i)\right] - \ln N_{\rm tot}
\end{aligned}
\end{equation}
Similar to the fitting procedure for the 1/\Vmax\ values, we use the MCMC \textsc{python} package \textsc{emcee} to maximize $\ln \mathcal{L}$ \citep{emcee}. 
The number of free parameters in these fits, which is six ($\gamma_1$, $\epsilon_{\lambda}$, $\lamEdd^*$, $\alpha$, $\beta$ and $\Mstarbh$), reflects the functional forms assumed for the BHMF and the ERDF. 
The initial guesses are based on the MCMC fits to the 1/\Vmax\ values, and 50 walkers are allowed to take 3,000 steps (the chains usually converge within 2,500 steps).

Equation \ref{eq:bhmfprobconv} shows that the normalization of the intrinsic bivariate distribution function, $\Psi^{*}$, does not affect the probability of observing source $i$ and thus also has no impact on the log-likelihood $\ln \mathcal{L}$. 
When applying the MCMC procedure we thus use constant normalizations for both the BHMF ($\Phi^{*}_{\rm init}$) and the ERDF ($\xi^{*}_{\rm init}$). After having determined the best-fitting parameters, we re-normalize $\Psi$ and determine $\Psi^{*}$ as follows:
\begin{equation}\label{eq:bhmfrescale}
\Psi^{*} = a_{\rm rescale} \times \frac{N_{\rm obs}}{N_{\rm tot, best-fit}(\Phi^{*}_{\rm init}, \xi^{*}_{\rm init})} \, .
\end{equation}
Here $N_{\rm obs}$ corresponds to the total number of objects observed in our sample, while $N_{\rm tot, best-fit}$ is determined by using Equation \ref{eq:bhmfNtot}, the best-fitting parameters for $\Psi$, and the initial normalizations $\Phi^{*}_{\rm init}$ and $\xi^{*}_{\rm init}$. 
The parameter $a_{\rm rescale}$ corresponds to an additional re-scaling factor which can be used to correct for only partially available data (i.e., $z$ and/or \Mbh\ measurements). 
%
As our sample is spectroscopically complete (excluding the non-beamed, non-Galactic sources, along with others discussed in \S~\ref{sec:sample_exclusion}), we assume $a_{\rm rescale} = 1$ for this work.

Finally, we use Eq.~\ref{eq:bhmfbi} to determine the bias-corrected intrinsic BHMF and ERDF:
\begin{equation}\label{eq:bhmfbinorm}
\begin{aligned}
&\Phi_{\rm M}(\log \Mbh) = \Psi^{*}\tilde{\Phi}(\log \Mbh, \Phi^{*}_{\rm init}) \\
&\int_{\log \lambda_{\rm E, min, s}}^{\log \lambda_{\rm E, max, s}} \tilde{\xi}(\log \lamEdd, \xi^{*}_{\rm init}) d\log \lamEdd\\
~&~\\
&\xi(\log \lamEdd) = \Psi^{*}\tilde{\xi}(\log \lamEdd, \xi^{*}_{\rm init}) \\
&\int_{\log M_{\rm BH, min, s}}^{\log M_{\rm BH, max, s}} \tilde{\Phi}(\log \Mbh, \Phi^{*}_{\rm init}) d\log \Mbh
\end{aligned}
\end{equation}
As $\Psi^{*}$ has the units of space density, namely h$^{\rm 3}$ Mpc$^{-3}$ dex$^{-2}$, so do $\Phi^{*}$ and $\xi^{*}$.

We tested our methodology end-to-end to verify that we can indeed uncover the intrinsic distributions of BHMF and ERDF using this approach. 
To this end, we created two mock catalogs for both Type~1 and Type~2 AGN and tested our method using three different levels of uncertainty on the (mock) $\log \Mbh$ and $\log\lamEdd$ values. 
These tests and mock catalogs are described in detail in Appendix~\ref{sec:app_test}.

\section{Results - the Intrinsic Distributions Governing the BASS Sample}\label{sec:results}

In this section we present our results for the XLF, the BHMF and the ERDF of the BASS/DR2 sample, including the distributions corresponding to the subsets of Type~1 and Type~2 AGN. 
We discuss some of the considerations made when determining the XLF, present the XLF for AGN in different $\log \NH$ bins (in addition to the Type~1 and Type~2 AGN XLFs), and compare to the results of previous studies of AGN selected in the ultra-hard X-ray regime. 
We then determine the bias-corrected BHMF and ERDF for BASS/DR2 AGN and again compare our results to those of previous studies. 
We finally 
demonstrate that the XLF that we derive from our fundamental bivariate, bias-corrected \Mbh- and \lledd-dependent distribution can reproduce the XLF that we measure directly from the observations. 

\magenta{
We note that the 1/\Vmax\ based XLF reported in this work is a way to understand the distribution of {\it observed} data with respect to AGN luminosity, rather than an involved derivation of the intrinsic XLF (such as the XLFs presented in \citealt{Gilli2007}, \citealt{Ueda:2014aa}, \citealt{Aird2015}, \citealt{buchner2015}, and \citealt{ananna2019}). 
Indeed, we report the 1/\Vmax\ based estimates (and fits) for all three distribution functions (BHMF, ERDF and XLF) as a first-order approximation, and as a way to make our results more directly and readily comparable with the observed distributions reported in previous studies \cite[e.g.,][]{Greene:2009aa, Ajello:2012aa,Schulze:2010aa,Schulze:2015aa}.} 

\subsection{The XLF of low-redshift AGN}

To determine the AGN XLF of ultra-hard X-ray selected AGN, we use the 1/\Vmax\ approach. As discussed in Section \ref{sec:Vmax}, we bin our sources according to their intrinsic $14{-}195$ keV luminosities. When determining the corresponding space densities, we take the flux-area curve and attenuation by high column densities into account (Fig.~\ref{fig:flux_area_curve}, \ref{fig:attenuation_curve}). Once we have determined $\Phi_{\rm L}$, we fit the XLF with a double power-law (Eq.~\ref{eq:bhmfbpl_xlf}). The 1/\Vmax\ space densities are given in Appendix~\ref{sec:app_tables} (specifically Table~\ref{tab:xlfvmaxallsamples}), and the best-fitting double power-law parameters of the various sub-samples we consider are given in Table~\ref{tab:bhmfxlffit}. 
The analysis steps are further discussed in what follows.

\begin{figure*}
\begin{centering}
	\includegraphics[width=\textwidth]{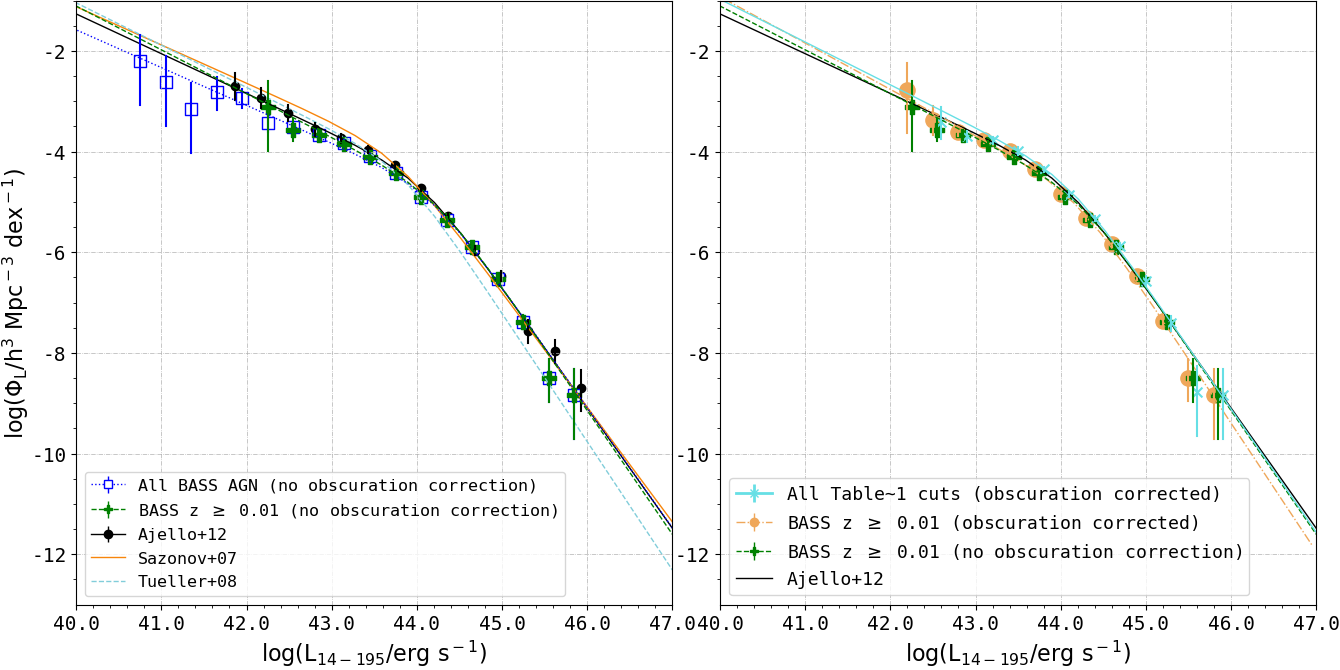}
	\caption{Effect of the low redshift cut and obscuration on the BASS XLF. \textit{Left panel:} The BASS XLF determined without a low-$z$ cut ({\it blue squares}; $\nnonblazarhighredshiftcut$ sources), 
    and the corresponding fit ({\it dotted blue line}). 
    {\it Green plus signs} show the XLF after excluding nearby sources with $z < \zminsamplebhmf$ (leaving $\nnonblazarlowredshiftcut$ sources in the sample), with the {\it green dashed line} fit to these $\Phi_{\rm L}$ values. For comparison we show the XLFs by \cite{Sazonov:2007aa}, \cite{Tueller:2008aa}, and \cite{Ajello:2012aa}. 
    \textit{Right panel:} The obscuring column density $N_{\rm H}$ has a small but systematic effect on the XLF. {\it Green plus signs} and {\it green dashed line} show the XLF computed by not taking obscuration into account when computing the 1/\Vmax\ values, while the {\it orange-filled circles} (shifted left by 0.05 dex for clarity) illustrate obscuration-corrected $\Phi_{\rm L}$ values. The $N_{\rm H}$ correction consistently increases the space densities in most luminosity bins. The {\it turquoise crosses} (shifted right by 0.05 dex for clarity) and the {\it turquoise solid line} show the XLF after applying all the cuts in Table~\ref{tab:bhmfpara} (\ntotallcuts\ sources).}
	\label{fig:xlfobs}
\end{centering}
\end{figure*}

In Figure~\ref{fig:xlfobs} we present the XLF determined for BASS/DR2 AGN, regardless of their classification (i.e., both Type~1 and Type~2 sources). 
First, we use the redshifts, intrinsic 14--195 keV X-ray luminosities (\Lbat), and column densities (\NH) of all \nnonblazarhighredshiftcut\ {non-beamed, high Galactic latitude} BASS/DR2 AGN at $z \leq 0.3$; their XLF is shown in the left panel with blue open squares and dotted line. 
Second, the left panel of Fig.~\ref{fig:xlfobs} shows the XLF of the \nnonblazarredshiftcut\ sources that meet our redshift restrictions ($\zminsamplebhmf \leq z \leq \zmaxsamplebhmf$), \textit{without} any corrections due to obscuration applied to their selection function. 
The only noticeable difference between the XLF of all BASS/DR2 AGN and the redshift-restricted sub-sample is in the low-$L$ end.
The redshift-restricted sub-sample lacks the \nlowlumagn\ lowest-$L$ AGN (i.e., in the $\log [L_{\rm X}/\ergs] < 42.25$ bins), which are also some of the lowest redshift AGN in BASS/DR2 (i.e., $z<\zminsamplebhmf$). At higher luminosities, the number of AGN in each luminosity bin for the second sample is non-zero, and usually equal to the number of AGN in the supersample. 
As shown in Figure~\ref{fig:xlfobs} and Table~\ref{tab:xlfvmaxallsamples}, in a few bins, the redshift-restricted sample has slightly higher space densities than the supersample, even when the latter has more sources in those bins. This is possible because the lower redshift cut leads to smaller \Vmax\ (and therefore higher 1/\Vmax) values in some cases. 

We analyse the XLF of our samples further in the \textit{right} panel of Figure~\ref{fig:xlfobs}. 
In addition to the XLF of the redshift-restricted sub-sample, we show how this XLF changes when obscuration corrections are applied to the selection function. The obscuration correction results in slightly higher values of 1/\Vmax\ at all luminosities (as the \Vmax\ value of the obscured object is smaller). The 1/\Vmax\ values and the corresponding errors are given in Tables~\ref{tab:bhmfxlfvmaxxlf} and \ref{tab:xlfvmaxallsamples}. The best-fitting parameters for each XLF are shown in Table \ref{tab:bhmfxlffit}. 

The correction due to obscuration is generally very small, which is expected as the radiation in the 14$-$195 keV regime only starts to get attenuated beyond $\logNH \gtrsim 23$ in the local universe (as shown in Figure~\ref{fig:attenuation_curve}). Therefore, taking the effect of obscuration into account for ultra-hard X-rays may not lead to significant changes unless the sample at hand preferentially selects heavily obscured sources. 
For example, the lowest luminosity bin of the redshift restricted sample contains a single object with $\logNH = 23.15$, and therefore the effect of obscuration in that bin is noticeable, as shown in the right panel of Figure~\ref{fig:xlfobs}. 
As our sample includes both unobscured and heavily obscured sources, we apply an obscuration correction to all other 1/\Vmax\ calculations except for the two cases shown in Fig.~\ref{fig:xlfobs}. In the Figure, we show the 1/\Vmax\ values and XLF fit to the sample used for the BHMF/ERDF analysis (after applying all cuts from Table~\ref{tab:bhmfpara}).

Figure~\ref{fig:xlfobs} also shows previous determinations of the ultra-hard XLF of low-redshift AGN by \cite{Sazonov:2007aa}, \cite{Tueller:2008aa}, and \cite{Ajello:2012aa}. We also show the binned 1/\Vmax\ XLF measurements of the latter study (black symbols). 
To convert these previous results to the 14--195 keV range we use here, we have assumed $\Gamma=1.8$. For the results by \cite{Ajello:2012aa}, this implies $L_{14{-}195\,\rm{keV}}  = 2.31 \times L_{15 - 55\,\rm{keV}}$. To convert from $L_{20 - 40\,\rm{keV}}$ and $L_{17{-}60\,\rm{keV}}$ to $L_{14{-}195\,\rm{keV}}$ we used multiplicative factors of $4.34$ and $2.33$, respectively. 

The left panel in Figure~\ref{fig:xlfobs} shows that our parent sample of (unbeamed) BASS/DR2 AGN reaches down to hard X-ray luminosities that are $\approx1$ dex fainter than the faintest luminosities covered by \cite{Ajello:2012aa}, which relied on the 60-month \swift/BAT all-sky catalog. 
As a natural result of the lower redshift cut at $z = 0.01$, our redshift-restricted sample ($\zminsamplebhmf \leq z \leq \zmaxsamplebhmf$) only goes down to $\log (L_{\rm X}/\ergs) = 42.48$, which is higher than the lowest luminosities reached by \cite{Ajello:2012aa}. 
 %
 
Overall, Figure \ref{fig:xlfobs} shows that the 1/\Vmax\ based $\Phi_{\rm L}$ values for our BASS/DR2 AGNs are in excellent agreement with previous results.

\begin{deluxetable*}{lllll}
\label{tab:bhmfxlffit}
\tablecaption{Best-fitting XLF parameters$^{\rm a}$.}
\tablewidth{0pt}
\tablehead{
\colhead{Selection} & \colhead{$\log (L^{*}_{\rm X}/\rm{erg\ s^{-1}})$} & \colhead{$\log (\Phi_{\rm L}^{*}/\rm{h^{3}\ Mpc^{-3}})$} & \colhead{$\gamma_1$} & \colhead{$\epsilon_{\gamma}$} }
\startdata
		{{BASS AGN at $z \leq 0.3$ ($\nnonblazarhighredshiftcut$; no obs corr)}}& {\xlfvmaxallnoobscorrL} & {\xlfvmaxallnoobscorrphi} & {\xlfvmaxallnoobscorrgamma} & {\xlfvmaxallnoobscorreps} \\
		{$0.01 \leq z \leq 0.3$ ($\nnonblazarredshiftcut$; no obscuration correction)} & {\xlfvmaxlowzcutnoobscorrL} & {\xlfvmaxlowzcutnoobscorrphi} & {\xlfvmaxlowzcutnoobscorrgamma} & {\xlfvmaxlowzcutnoobscorreps} \\
		{$0.01 \leq z \leq 0.3$ (obscuration corrected)} & {\xlfvmaxlowzcutobscorrL} & {\xlfvmaxlowzcutobscorrphi} & {\xlfvmaxlowzcutobscorrgamma} & {\xlfvmaxlowzcutobscorreps} \\
		{$0.01 \leq z \leq 0.3$, $\log~{\rm N}_{\rm H} < 22$} & {\xlfvmaxunabsL} & {\xlfvmaxunabsphi} & {\xlfvmaxunabsgamma} & {\xlfvmaxunabseps} \\
		{$0.01 \leq z \leq 0.3$, $22 \leq \log~{\rm N}_{\rm H} < 24$} & {\xlfvmaxctnL} & {\xlfvmaxctnphi} & {\xlfvmaxctngamma} & {\xlfvmaxctneps} \\
		{$0.01 \leq z \leq 0.3$, $\log~{\rm N}_{\rm H} \geq 24$} & {\xlfvmaxctkL} & {\xlfvmaxctkphi} & {\xlfvmaxctkgamma} & {\xlfvmaxctkeps}\\
		{Type~1 AGN only (All Table~\ref{tab:bhmfpara} cuts)} & {\xlfvmaxtypeoneL} & {\xlfvmaxtypeonephi} & {\xlfvmaxtypeonegamma} & {\xlfvmaxtypeoneeps} \\
		{Type~2 AGN only (All Table~\ref{tab:bhmfpara} cuts)} & {\xlfvmaxtypetwoL} & {\xlfvmaxtypetwophi} & {\xlfvmaxtypetwogamma} & {\xlfvmaxtypetwoeps} \\
		{\bf All Table~\ref{tab:bhmfpara} Selection AGN (\ntotallcuts)} & {\xlfvmaxtypeallL} & {\xlfvmaxtypeallphi} & {\xlfvmaxtypeallgamma} & {\xlfvmaxtypealleps} \\
\enddata
\tablenotetext{a}{For the XLF, we assume a double power-law shape (see equation \ref{eq:bhmfbpl_xlf}). All samples described here exclude sources falling within the Galactic plane (-5 $<$ latitude $<$ 5), weak X-ray associations, dual sources falling below the flux limit of the survey, beamed sources and sources with $z > 0.3$.}
\end{deluxetable*}

\begin{figure*}
\begin{centering}
	\includegraphics[width=0.9\textwidth]{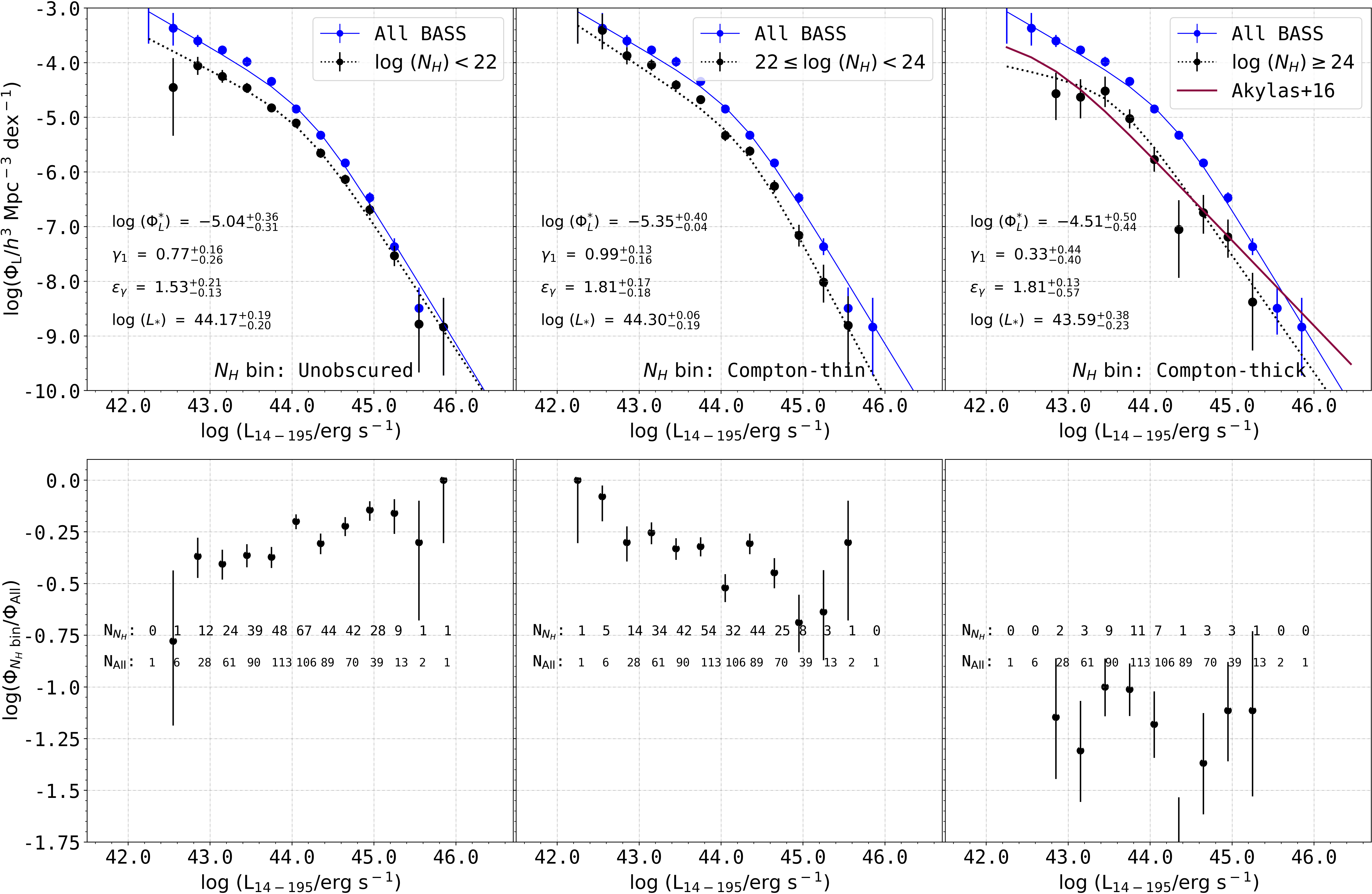}
	\caption{XLFs for BASS AGN in different regimes of \NH. 
	\textit{Top panels:} From left to right, \textit{black points, black lines} show the XLF for unobscured \magenta{Compton-thin} ($\NH  < 10^{22}\ \rm{cm}^{-2}$), \magenta{obscured} Compton-thin ($10^{22}\, \rm{cm}^{-2} \leq \NH < 10^{24}\ \rm{cm}^{-2}$), and Compton-thick ($N_{\rm H} > 10^{24}\ \rm{cm}^{-2}$) AGN. 
	The XLF for the entire BASS sample is shown by {\it blue filled circles} and 
	{\it blue solid lines}. 
	Also shown in the top right panel is the XLF of BAT Compton-thick AGN from \cite{Akylas:2016aa}, which includes objects at $z < 0.01$, below the limit used in our analysis.
	\textit{Bottom panels:} Ratios between the \NH\ sub-sample XLFs and the XLF of all AGN in each $\log L_{\rm X}$ bin ({\it black data points}), along with the associated Wilson score intervals ({\it black vertical lines}). 
	The numbers refer to the objects in each bin, with upper and lower rows corresponding to 
	the sub-sample and the full sample, respectively).
	\blue{These XLFs and ratios should be treated as representing the AGNs observed within BASS/DR2, in the corresponding absorption bins, rather than intrinsic XLFs (which would also account for AGNs completely missed by our survey; see main text).}	
	%
	} 
	\label{fig:xlfznh}
\end{centering}
\end{figure*}

\subsection{\magenta{The XLF of AGN with Various Levels of Absorption}}\label{ssec:xlf_lognh}

The size of the BASS sample allows us to determine the XLF for various subsets of AGN. 
In Figure~\ref{fig:xlfznh}, we split the AGN by $\log \NH$, and calculate the binned XLF separately for unabsorbed \magenta{Compton-thin} [$\log (\NH/\cmii) < 22$], \magenta{absorbed} Compton-thin [$22 \leq \log (\NH/\cmii) < 24$] and Compton-thick [$\log (\NH/\cmii) \geq 24$] sources. 
For comparison, the blue solid lines on the top panels of Figure~\ref{fig:xlfznh} illustrate the XLF for our redshift-restricted unbeamed AGN sample.
For reference, in the top right panel of Figure~\ref{fig:xlfznh}, we also show the XLF of Compton-thick AGN reported by \cite{Akylas:2016aa}. 
They used 53 Compton-thick AGN selected from the the same parent catalog as the one we use (Swift-BAT 70-month catalog), and determined a relatively low ``break'' luminosity in the 14$-$195 keV range, $\log (L^{*}_{\rm X}/\ergs)\simeq42.8$, compared to our value of \xlfvmaxctkL. 
Note that we also have 53 Compton-thick objects in our parent sample, 11 of which are at $z < 0.01$, one falls above $z > 0.3$ and another one which is a faint dual source. 
Our individual 1/\Vmax\ $\Phi_{\rm L}$ values are generally consistent with the fit by \cite[within errors]{Akylas:2016aa} in all but the highest luminosity bin, although at $\log(\Lbat/\ergs) > 44$, the \cite{Akylas:2016aa} Compton-thick XLF lies above our data points. 
This may be caused by the different volumes considered, and/or by the fact that \citet{Akylas:2016aa} uses a Poissonian maximum likelihood estimator that considers individual sources rather than a fit to 1/\Vmax\ \cite[see also][]{loredo2004}.

To further demonstrate the fractional densities of the $\log \NH$ sub-samples, the bottom panels of Figure~\ref{fig:xlfznh} show the ratios between the space densities of each 
sub-sample compared to the entire sample, as a function of \Lbat. 
These ratios are computed using the 1/\Vmax\ $\Phi_{\rm L}$ values, and the errors on the ratio are calculated using Wilson Score Interval method \citep{wilson1927}, which provides binomial confidence intervals without any assumption about the symmetry of the error bars. 

\magenta{
We stress again that the XLFs and the related ratios shown in Fig.~\ref{fig:xlfznh}\blue{, are estimated using the 1/\Vmax\ method and thus for the observed sample of sources only.
While these XLFs account for the effects of absorption on the sources observed in our survey, they do {\it not} account for absorbed populations that are completely missing from the sample due to selection biases 
(which are accounted for in more sophisticated studies, e.g.,} \citealp{Ueda:2014aa} and \citealp{ananna2019}). 
Deriving such intrinsic \NH-dependent XLFs and ratios requires a much more involved approach than the 1/\Vmax\ estimates we use here, which is beyond the scope of the present work.
} 

We discuss the insights from Figure~\ref{fig:xlfznh} in more detail in \S\ref{sec:discussion}.

\subsection{BHMF and ERDF of local AGN} 

\begin{figure*}
    
	\includegraphics[width=0.5\textwidth]{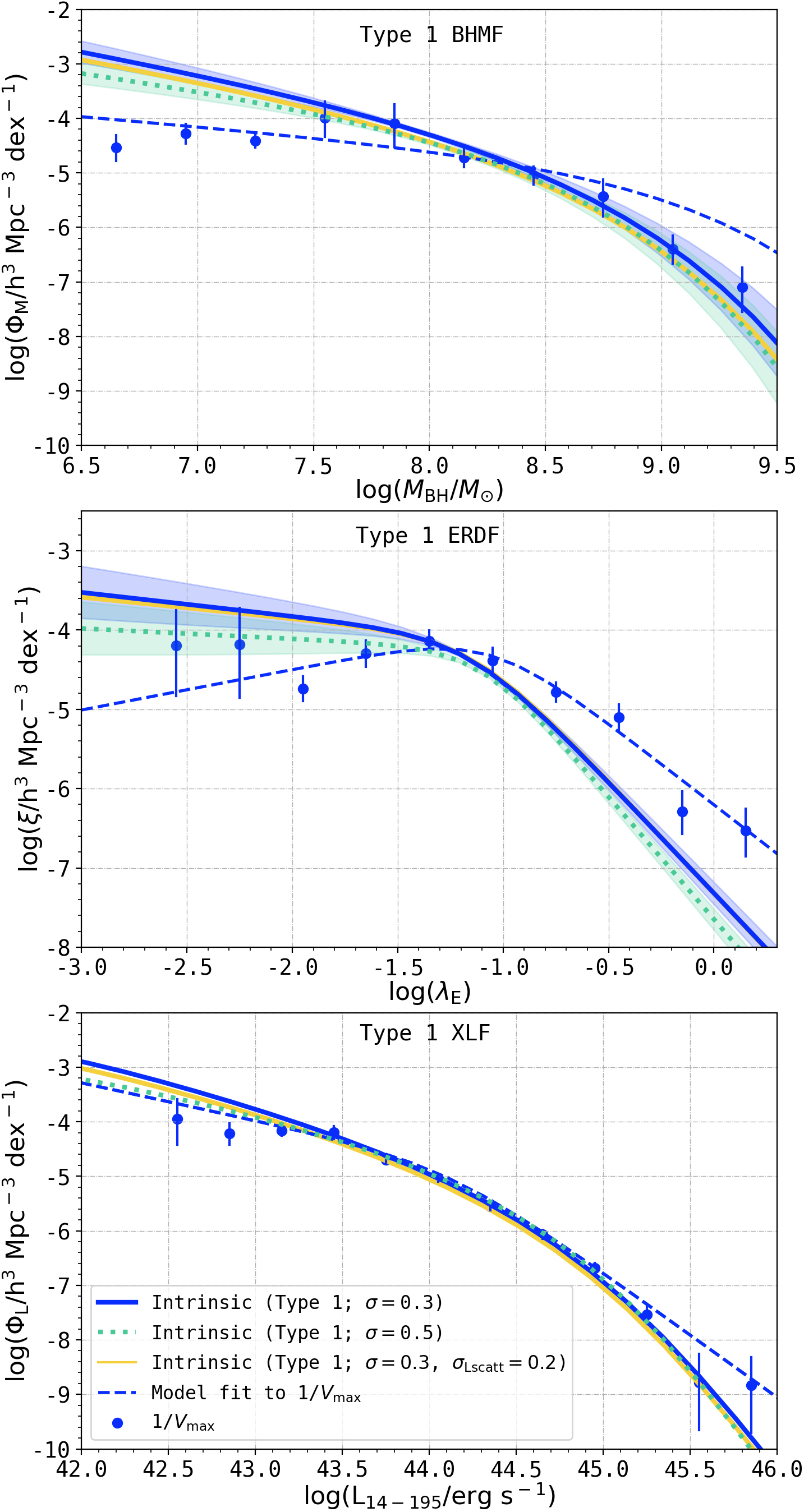}
	\includegraphics[width=.5\textwidth]{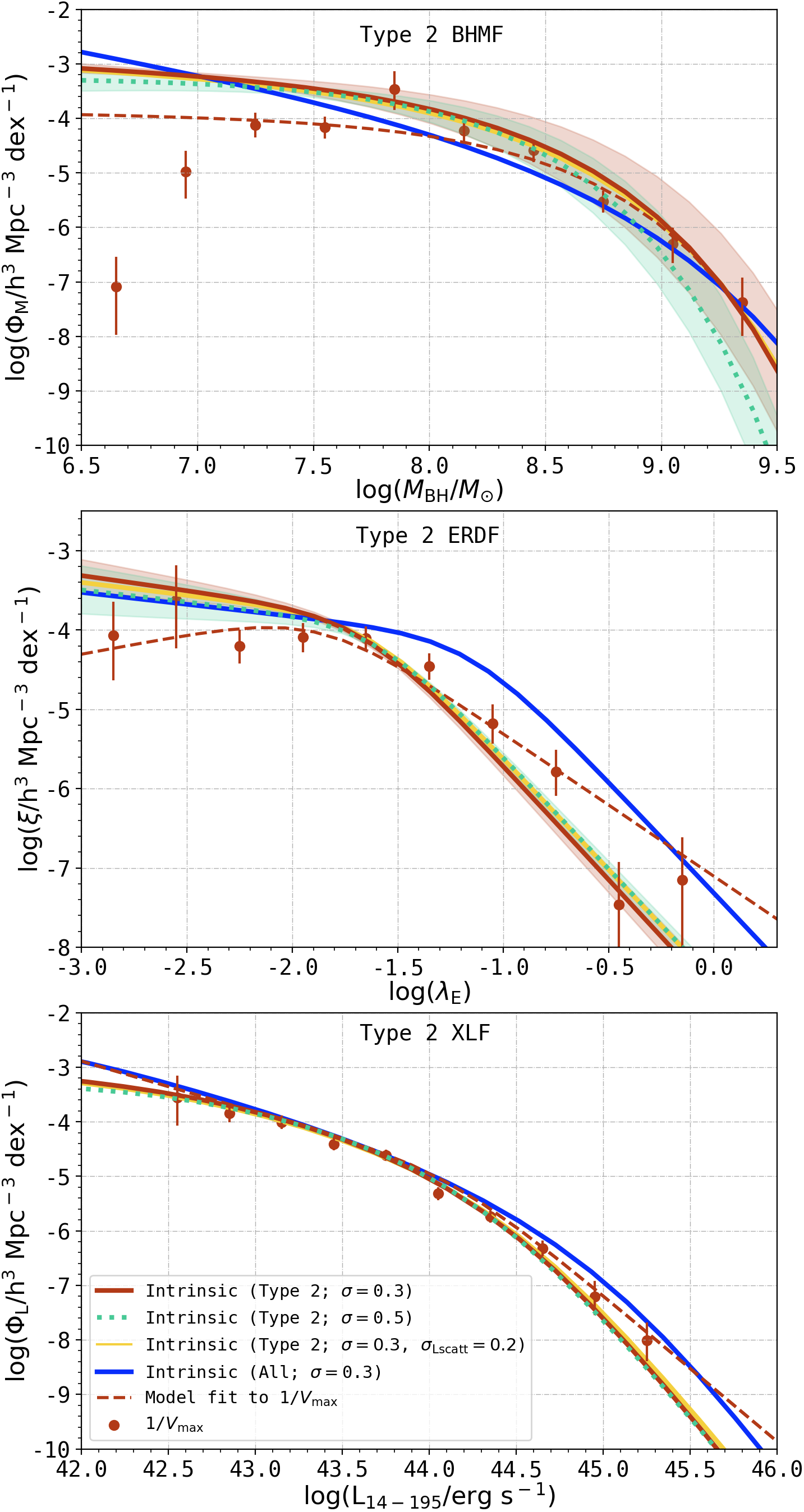}
	\caption{\label{fig:bhmferdfdis} BH mass, Eddington ratio, and luminosity distribution functions of BASS/DR2 
	Type~1 ({\it left} panels) and Type~2 ({\it right} panels) AGNs.
	For each type of AGNs, combining the BHMF ({\it top} row) and ERDF ({\it middle} row) reproduces the observed XLF ({\it bottom} row).
	{\it Data points} come from the 1/\Vmax\ analysis and {\it dashed lines} represent fits to these data points. 
	The solid lines and shaded areas show the final, bias-corrected intrinsic distribution functions. 
	\magenta{Lines with different colors trace the intrinsic distributions assuming various levels of uncertainty (see legend for details). 
	Specifically, the blue (left) and red (right) \textit{solid lines} assume uncertainties of $\sigma_{\log \Mbh} = \sigma_{\log \lamEdd} = 0.3$; the orange solid lines assume $\sigma_{\log \Mbh} = 0.3$ and $\siglscatt = 0.2$ (or $\sigma_{\log \lamEdd} = 0.36$); and the green dotted lines assume $\sigma_{\log \Mbh} = \sigma_{\log \lamEdd} = 0.5$ dex.}
	The intrinsic distributions of Type~1 AGNs (assuming $\sigma = 0.3$) are shown also in the {\it right} panels, to highlight differences between Type~1 and 2 AGNs.
	Note that the 1/\Vmax\ points shown for each subset of AGNs were calculated considering only the AGNs of the respective type (i.e., Type~1 or 2 AGNs). 
	The 1/\Vmax\ values are reported in Tables~\ref{tab:bhmfxlfvmaxxlf}, \ref{tab:bhmfvmax} and \ref{tab:erdfvmax}.
	} 
\end{figure*}

\begin{figure}

	\includegraphics[width=0.5\textwidth]{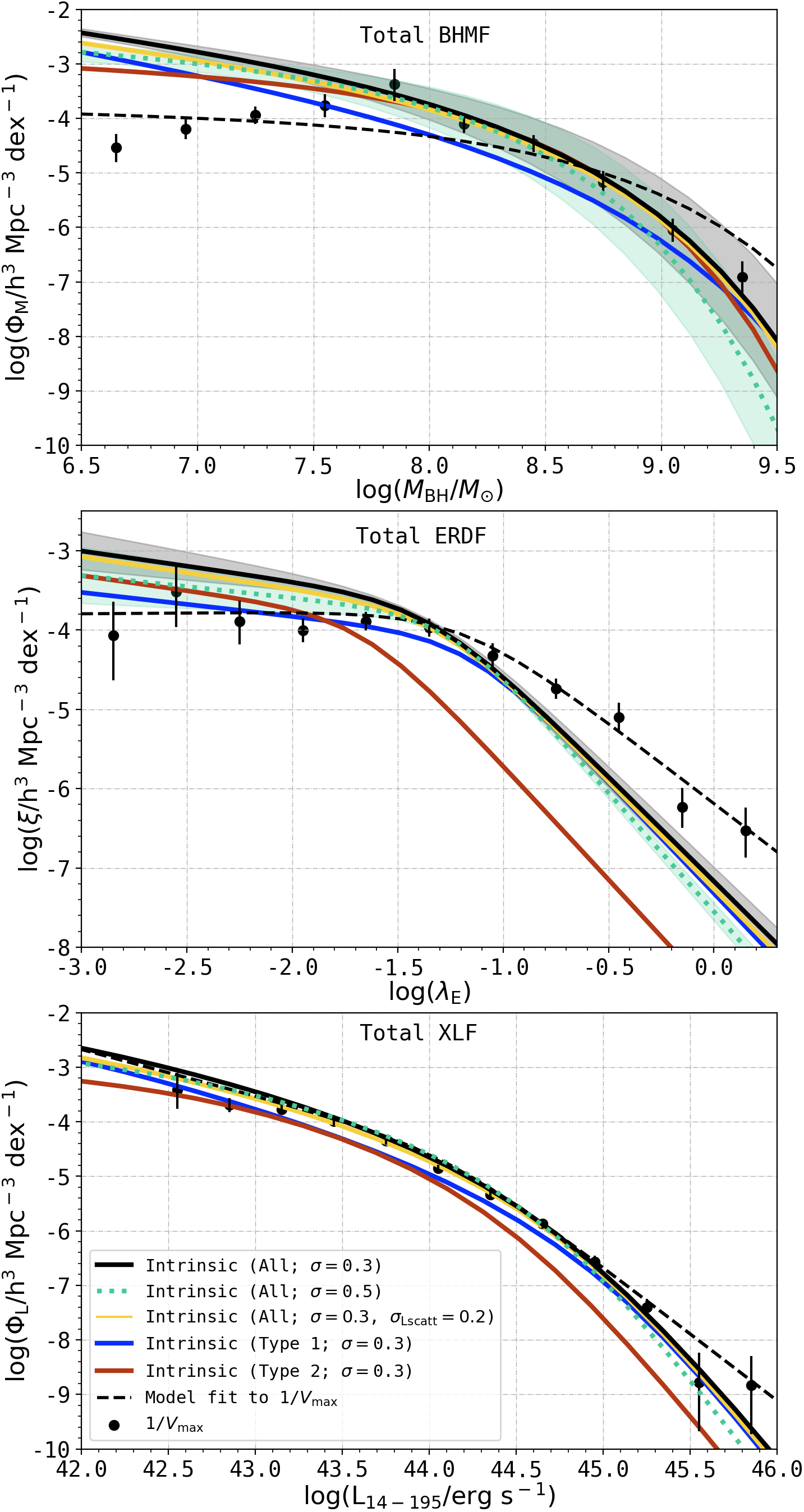}
	\caption{\label{fig:bhmferdfdis_all} Distribution functions for all BASS/DR2 AGNs that fall within the criteria specified in Table~\ref{tab:bhmfpara} (\ntotallcuts\ sources). 
	As in Fig.~\ref{fig:bhmferdfdis}, the BHMFs, ERDFs and reconstructed XLFs  are shown in the {\it top}, {\it middle} and {\it bottom} panels, respectively. 
	We show the intrinsic distributions derived for all AGNs assuming uncertainties of either $\sigma = 0.3$ dex or $\sigma = 0.5$ dex in $\log \Mbh$ and $\log \lamEdd$ ({\it black solid} and {\it green dotted} lines, respectively). 
	\magenta{The \textit{orange solid lines} show intrinsic functions, derived assuming $\sigma_{\log\Mbh} = 0.3$ and $\siglscatt = 0.2$ (or $\sigma_{\log \lamEdd} = 0.36$).}
	The {\it black data points} in the bottom panel show the direct 1/\Vmax\ XLF estimates, and the {\it black dashed line} is the fit to those points. 
	{\it Solid black lines} show the final, bias-corrected, intrinsic distribution functions for the complete AGN sample.
	For comparison, we also show the distribution functions of Type~1 and Type~2 AGNs ({\it blue} and {\it red} lines, respectively) from Figure~\ref{fig:bhmferdfdis}.
	} 
\end{figure}

To determine the BHMF and ERDF for local AGN we use the \Mbh\ and \lamEdd\ measurements of \ntotallcuts\ ultra-hard X-ray selected BASS/DR2 AGN, as described in \S\ref{sec:measurements}. 
Since the \Mbh\ (and thus \lamEdd) of Type~1 (broad line) and Type~2 (narrow line) BASS/DR2 AGN were determined through two different approaches, with different systematics and potentially different selection effects in play, we first treat these two subsets separately, and only then address the BHMF and ERDF of the total BASS/DR2 AGN sample.

To gain an initial guess for the two distribution functions we use the 1/\Vmax\ approach, assume functional forms, and fit the individual $\Phi$ and $\xi$ values independently. For the BHMF and the ERDF we assume a modified Schechter function and a double power-law, respectively (see Eqs.~\ref{eq:bhmf_msch_def} and \ref{eq:erdf_dbpl_def} in \S\ref{sec:funcitonal_forms}). To correct for sample truncation, i.e., the bias against low mass and low Eddington-ratio AGN, we use the parametric maximum likelihood approach outlined in \S\ref{sec:bias}. 
The 1/\Vmax\ values for both the BHMF and ERDF are given in Appendix~\ref{sec:app_tables}.

Figure \ref{fig:bhmferdfdis} shows the BHMF (top panels) and ERDF (center panels) for Type~1 and Type~2 BASS AGN (left and right columns, respectively). 
The two bottom panels in Figure~\ref{fig:bhmferdfdis} show the XLFs \textit{reproduced} from these BHMFs and ERDFs, as explained below. Figure~\ref{fig:bhmferdfdis_all} presents the BHMF, ERDF and XLF for the overall sample (Type~1 + Type~2) in a similar format.
We recall that the normalizations are kept constant during the bias correction, and the method of computing them is described in \S\ref{sec:biasintro}. We are thus left with six free parameters: the break and two slopes of the BHMF ($M^*_{\rm BH}$, $\alpha$, $\beta$), and the break and two slopes of the ERDF ($\lamEdd^*$, $\delta_1$, $\delta_2$). 
We note that the high-\lamEdd\ slope, $\delta_2$, is parametrized as $\delta_1 + \epsilon_{\lambda}$ with $\epsilon_{\lambda} > 0$. 
We vary $\delta_1$ and $\epsilon_{\lambda}$ in the MCMC. We also recall that for the bias correction we assume uncertainties of 0.3 dex and 0.5 dex 
on both $\log \Mbh$ and $\log \lamEdd$ (see Equations~\ref{eq:bhmfprobconv} and \ref{eq:twodomega}), \magenta{as well as an additional scenario with $\sigma_{\log \Mbh} = 0.3$ and $\siglscatt = 0.3$ (i.e., $\sigma_{\log \lamEdd} = 0.36$). The results from all these different $\sigma$ values reassuringly converges on the same solution, as shown in the Figure. In Appendix~\ref{sec:app_variable_kbol} we present BHMF/ERDF calculated assuming a luminosity dependent bolometric correction from \citet{Duras2020}. Note that we prefer the constant bolometric correction because it minimizes the number of assumptions we have to make
, and because there is some conflict between different prescriptions of luminosity dependent bolometric corrections (shown in Figure~6 of \citealt{Duras2020}), and it is unclear which prescription is the most accurate.
}
%
%

The contour plots presenting the likelihoods resulting from our MCMC analysis are shown in Figure~\ref{fig:chain_overlap} in Appendix~\ref{sec:app_tables}. 
The best-fitting BHMF and ERDF parameters for all three samples are given in Tables~\ref{tab:bhmfbhmferdffitBHMF} and ~\ref{tab:bhmfbhmferdffitERDF}, respectively. Note that we also report the results assuming an attenuation curve and absorption function calculated using a torus opening angle of 35$^{\circ}$ (as discussed in \S\ref{sec:pi}) in these tables. As shown in the Figure~\ref{fig:bhmferdfdisoa35}, our final results for both torus geometries are consistent with each other.

\begin{deluxetable*}{lclll}
\label{tab:bhmfbhmferdffitBHMF}
\tablecaption{Sample truncation corrected BHMF$^{\rm a}$, and fit to /\Vmax\ values for the BHMF for Type~1, Type~2 AGN, and both samples together$^{\rm b}$.}
\tablewidth{0pt}
\tablehead{
\colhead{} & \colhead{$\log (M_{\rm BH}^{*}/\Msun)$} & \colhead{$\log (\Psi^{*}/\rm{h^{3}\ Mpc^{-3}})$} & \colhead{$\alpha$} & \colhead{$\beta$} }
\startdata
        All & & & & \\
		Intrinsic ($\sigma = 0.3$) & \bhmflogmstar & \bhmflogphistar & \bhmfalpha & \bhmfbeta \\
		Intrinsic ($\sigma = 0.3$; $\siglscatt = 0.2$) & \bhmflogmstarsigLtwo & \bhmflogphistarsigLtwo & \bhmfalphasigLtwo & \bhmfbetasigLtwo \\
		Intrinsic ($\sigma = 0.5$) & \bhmflogmstarsigfive & \bhmflogphistarsigfive & \bhmfalphasigfive & \bhmfbetasigfive \\
		Intrinsic ($\sigma = 0.3$; OA = 35$^{\circ}$) & \bhmflogmstaroathreefive & \bhmflogphistaroathreefive & \bhmfalphaoathreefive & \bhmfbetaoathreefive \\
		1/\Vmax\  & \bhmfvmaxalllogMbh & \bhmfvmaxallphi & \bhmfvmaxallalpha & \bhmfvmaxallbeta\\ 
		\hline
		Type~1 & & & & \\
		Intrinsic ($\sigma = 0.3$) & \bhmflogmstartypeone & \bhmflogphistartypeone & \bhmfalphatypeone & \bhmfbetatypeone \\
		Intrinsic ($\sigma = 0.3$; $\siglscatt = 0.2$) & \bhmflogmstartypeonesigLtwo & \bhmflogphistartypeonesigLtwo & \bhmfalphatypeonesigLtwo & \bhmfbetatypeonesigLtwo \\
		Intrinsic ($\sigma = 0.5$) & \bhmflogmstartypeonesigfive & \bhmflogphistartypeonesigfive & \bhmfalphatypeonesigfive & \bhmfbetatypeonesigfive \\
		1/\Vmax\ & {\bhmfvmaxtyponelogMbh} & {\bhmfvmaxtyponephi} & {\bhmfvmaxtyponealpha} & {\bhmfvmaxtyponebeta}\\
		\hline
		Type~2 & & & & \\
		Intrinsic ($\sigma = 0.3$) & \bhmflogmstartypetwo & \bhmflogphistartypetwo & \bhmfalphatypetwo & \bhmfbetatypetwo \\
		Intrinsic ($\sigma = 0.3$; $\siglscatt = 0.2$) & \bhmflogmstartypetwosigLtwo & \bhmflogphistartypetwosigLtwo & \bhmfalphatypetwosigLtwo & \bhmfbetatypetwosigLtwo \\
		Intrinsic ($\sigma = 0.5$) & \bhmflogmstartypetwosigfive & \bhmflogphistartypetwosigfive & \bhmfalphatypetwosigfive & \bhmfbetatypetwosigfive \\
		Intrinsic ($\sigma = 0.3$; OA = 35$^{\circ}$) & \bhmflogmstartypetwooathreefive & \bhmflogphistartypetwooathreefive & \bhmfalphatypetwooathreefive & \bhmfbetatypetwooathreefive \\
		1/\Vmax\ & {\bhmfvmaxtyptwologMbh} & {\bhmfvmaxtyptwophi} & {\bhmfvmaxtyptwoalpha} & {\bhmfvmaxtyptwobeta}\\
\enddata
\tablenotetext{a}{We assume a modified Schechter function (see equation \ref{eq:bhmf_msch_def}) for the BHMF.}
\tablenotetext{b}{\magenta{We use a constant bolometric correction (Equation~\ref{eq:l14195_to_lbol}) to compute these results.}}
\end{deluxetable*}

\begin{deluxetable*}{lclll}
\label{tab:bhmfbhmferdffitERDF}
\tablecaption{Sample truncation-corrected ERDF$^{\rm a}$, and fit to /\Vmax\ values for the ERDF for Type~1 AGN, Type~2 AGN and the full AGN sample$^{\rm b}$.}
\tablewidth{0pt}
\tablehead{
\colhead{} & \colhead{$\log \lamEdd^{*}$} & \colhead{$\log (\xi^{*}/\rm{h^3\ Mpc^{-3}})$} & \colhead{$\delta_1$} & \colhead{$\epsilon_{\lambda}$} }
\startdata
		All & & & & \\
		Intrinsic ($\sigma = 0.3$) & {\erdfloglamstar} & {\erdflogxistar} & {\erdfdeltaa} & {\erdfepislonlam} \\
		Intrinsic ($\sigma = 0.3$; $\siglscatt = 0.2$) & {\erdfloglamstarsigLtwo} & {\erdflogxistarsigLtwo} & {\erdfdeltaasigLtwo} & {\erdfepislonlamsigLtwo} \\
		Intrinsic ($\sigma = 0.3$; OA = 35$^{\circ}$) & {\erdfloglamstaroathreefive} & {\erdflogxistaroathreefive} & {\erdfdeltaaoathreefive} & {\erdfepislonlamoathreefive} \\
		Intrinsic ($\sigma = 0.5$) & {\erdfloglamstarsigfive} & {\erdflogxistarsigfive} & {\erdfdeltaasigfive} & {\erdfepislonlamsigfive} \\
		1/\Vmax\ & {\erdfvmaxalllam} & {\erdfvmaxallphi} & {\erdfvmaxallgamma} & {\erdfvmaxalleps} \\
		\hline
		Type~1 & & & & \\
		Intrinsic ($\sigma = 0.3$) & {\erdfloglamstartypeone} & {\erdflogxistartypeone} & {\erdfdeltaatypeone} & {\erdfepislonlamtypeone} \\
		Intrinsic ($\sigma = 0.3$; $\siglscatt = 0.2$) & {\erdfloglamstartypeonesigLtwo} & {\erdflogxistartypeonesigLtwo} & {\erdfdeltaatypeonesigLtwo} & {\erdfepislonlamtypeonesigLtwo} \\
		Intrinsic ($\sigma = 0.5$) & {\erdfloglamstartypeonesigfive} & {\erdflogxistartypeonesigfive} & {\erdfdeltaatypeonesigfive} & {\erdfepislonlamtypeonesigfive} \\
		1/\Vmax & {\erdfvmaxtyponelam} & {\erdfvmaxtyponephi} & {\erdfvmaxtyponegamma} & {\erdfvmaxtyponeeps}  \\
		\hline
		Type~2 & & & & \\
		Intrinsic ($\sigma = 0.3$) & {\erdfloglamstartypetwo} & {\erdflogxistartypetwo} & {\erdfdeltaatypetwo} & {\erdfepislonlamtypetwo} \\
		Intrinsic ($\sigma = 0.3$; $\siglscatt = 0.2$) & {\erdfloglamstartypetwosigLtwo} & {\erdflogxistartypetwosigLtwo} & {\erdfdeltaatypetwosigLtwo} & {\erdfepislonlamtypetwosigLtwo} \\
		Intrinsic ($\sigma = 0.3$; OA = 35$^{\circ}$) & {\erdfloglamstartypetwooathreefive} & {\erdflogxistartypetwooathreefive} & {\erdfdeltaatypetwooathreefive} & {\erdfepislonlamtypetwooathreefive} \\
		Intrinsic ($\sigma = 0.5$) & {\erdfloglamstartypetwosigfive} & {\erdflogxistartypetwosigfive} & {\erdfdeltaatypetwosigfive} & {\erdfepislonlamtypetwosigfive} \\
		1/\Vmax\ & {\erdfvmaxtyptwolam} & {\erdfvmaxtyptwophi} & {\erdfvmaxtyptwogamma} & {\erdfvmaxtyptwoeps} \\
\enddata
\tablenotetext{a}{We assume a double power-law shape for the ERDF (see equation \ref{eq:erdf_dbpl_def}).}
\tablenotetext{b}{\magenta{We use a constant bolometric correction (Equation~\ref{eq:l14195_to_lbol}) to compute these results.}}
\end{deluxetable*}

As discussed in \S\ref{sec:intro} and shown in detail in Appendix~\ref{sec:app_test}, the bolometric luminosity function corresponds to the convolution of the BHMF and the ERDF. 
By ``reversing'' the bolometric correction, the XLF can thus be predicted from the best-fit BHMF and ERDF. We test if the bias-corrected BHMF and ERDF for each subset of AGNs (i.e., Type~1, Type~2, and the overall sample) allow us to predict the corresponding observed XLF. 
Note that for the convolution we use the normalized ERDF (see Eqn.\,\ref{eq:bhmfconvolution}). The normalization of the predicted XLF is thus driven by the normalization of the BHMF. 

Figures~\ref{fig:bhmferdfdis} and \ref{fig:bhmferdfdis_all} illustrate the importance of bias correction. For the BHMF, relying solely on the model fit to 1/\Vmax\ values would have led us to underestimate the space density of low mass AGNs at the lowest mass bin ($\log [\Mbh/\Msol] = 6.65$) by $\gtrapprox 1$ 
dex for both Type~1 and Type~2 sources. 
%
%
At low \Mbh\ and/or \lamEdd, the 1/\Vmax\ method underestimates the intrinsic space density of AGNs due to the survey incompleteness.  
At high \Mbh\ and/or \lamEdd, the 1/\Vmax\ method {\it overestimates} the intrinsic AGN space densities, due to the uncertainties associated with the key AGN parameters. 
As shown by the mock catalogs (Figures~\ref{fig:mock_catalog_type1} and \ref{fig:mock_catalog_type2}), and Figure 17 of \citetalias{Schulze:2015aa}, as measurement uncertainty ($\sigma_{\log \Mbh}$, $\sigma_{\log \lamEdd}$) increases, this overestimation increases at the high \Mbh\ and/or high \lamEdd\ end. 
\magenta{This effect is a manifestation of the so-called Eddington bias \citep{eddington1913}:} the uncertainty causes objects from lower \Mbh\ (or \lamEdd) bins to scatter into higher \Mbh\ (or \lamEdd) bins, and vice versa. 
Intrinsically, there are always fewer objects with higher \Mbh\ (or \lamEdd), therefore scattering from the lower to higher bins causes significant overestimation of space densities in the higher bins. 
The bottom panels in Figures \ref{fig:bhmferdfdis} and \ref{fig:bhmferdfdis_all} demonstrate that our convolution-based reconstruction of the XLF matches what we measure directly from observations.
%

Figure~\ref{fig:bhmferdfdis} shows that the normalization of the bias-corrected BHMF of Type~2 AGN is higher than that of Type~1 AGN at \magenta{most masses. The bias-corrected ERDF of Type~2 AGN has higher space densities at $\log \lamEdd\ \leq -1.7$
.} Beyond this point, the ERDF of Type~2 AGN drops off rapidly below the ERDF of Type~1 AGN. This is because the break in ERDF of Type~2 AGN is at $\log$ \lamEdd = \erdfloglamstartypetwo, which is significantly below the break of Type~1 AGN at $\log$ \lamEdd = \erdfloglamstartypeone. Figure~\ref{fig:bhmferdfdis_all} shows the Type~1 and Type~2 BHMF, ERDF and XLF along with the overall sample results. 

\begin{figure*}
\begin{centering}
	\includegraphics[width=0.8\textwidth]{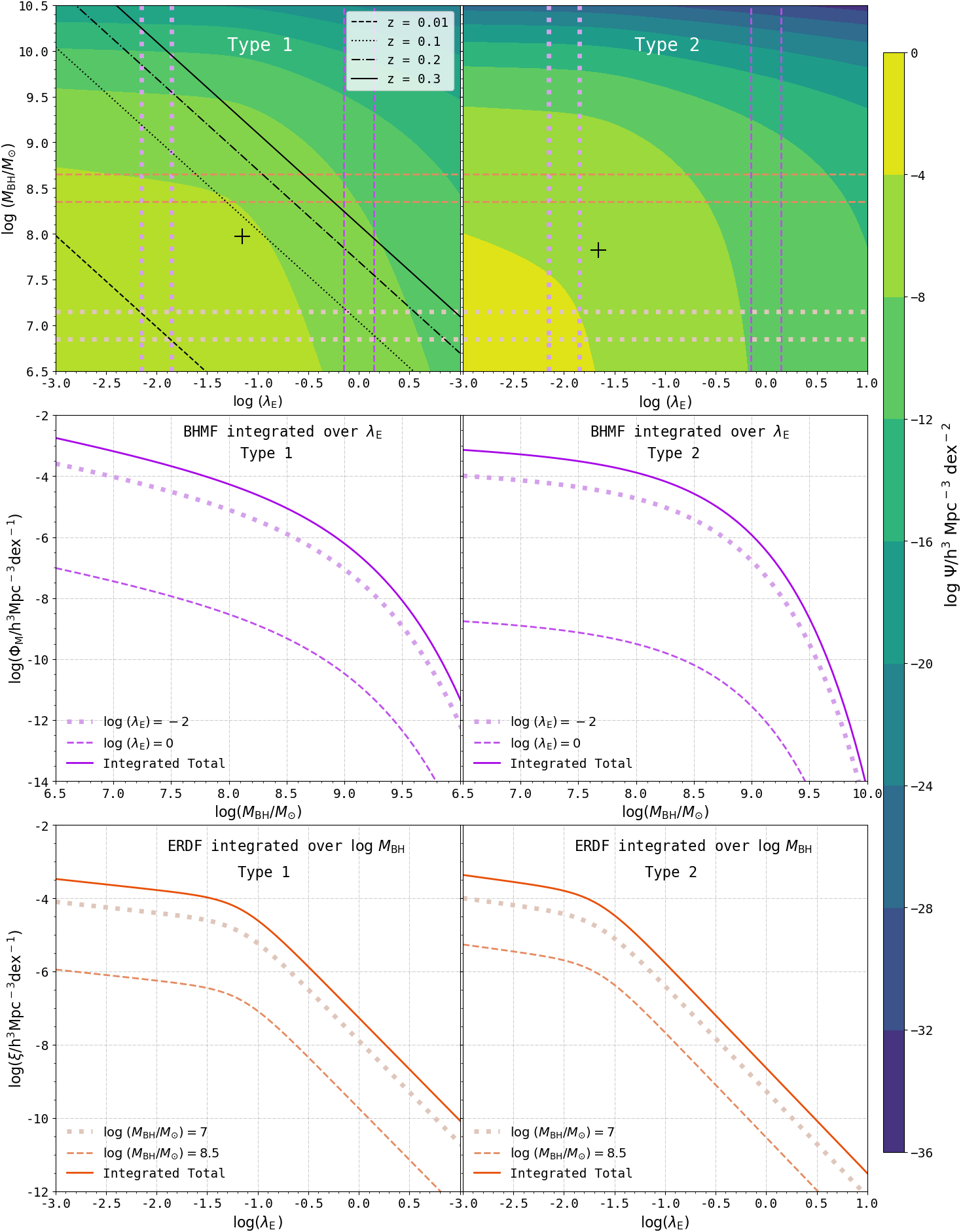}
	\caption{\label{fig:bivariate_contourplot} Contour plot of the best-fit bivariate distribution functions ({\it top panels}), BHMF marginalized over bins of $\log \lamEdd$ ({\it middle panels}), 
	and ERDF marginalized over bins of $\log \Mbh$ ({\it bottom panels}). 
	Type~1 AGN are shown on the left, Type~2 AGN on the right. 
	In the {\it top panels}, contours highlight levels of constant $\Psi$ values, with log-uniform spacing (see color bar); {\it black crosses} mark the positions of $[\log (\lamEdd^*)$, $\log (M^*_{\rm BH}/\Msun)]$; and {\it black lines} highlight the flux limits at different redshifts. 
	In the middle panels, we show the integrals for $\log \lamEdd = -2 \pm 0.15 $ ({\it dotted pink lines}), $\log \lamEdd = 0 \pm 0.15$ ({\it dashed pink lines}), and all $\log \lamEdd$, ({\it solid pink lines}).
	In the bottom panels, we show the integrals for $\log (\Mbh/\Msol) = 7 \pm 0.15$ ({\it dotted orange lines}), $\log (\Mbh/\Msol) = 8.5 \pm 0.15$ ({\it dashed orange lines}), and all $\log (\Mbh/\Msol)$. 
	Each $\log \lamEdd$ and $\log (\Mbh/\Msol)$ bin over which a function was marginalized is shown in the {\it top panels} in the corresponding color and line style (e.g., the {\it purple dashed lines} show the $\log \lamEdd$ range over which the BHMF is marginalized, and the result of the marginalization is shown with {\it dashed purple lines} in the {\it middle panels}). 
	For the purpose of this plot we assume a constant flux limit of $\log (F_{14-195\,\kev}/\ergs) = -11.1$ (which corresponds to a sky coverage completeness of just 2--3\%; see Fig.~\ref{fig:flux_area_curve}). The lines of constant flux limit in \textit{top left panel} are also lines of constant luminosity, and integrating along these lines produces the local XLF. This figure demonstrates that the bivariate distribution function fully captures the statistical properties of AGN.
     }
\end{centering}
\end{figure*}

Figure~\ref{fig:bivariate_contourplot} illustrates how we can use the bivariate distribution (i.e., $\Psi$) to reproduce the (univariate) BHMF and ERDF, following Eq.~\ref{eq:bhmfbinorm}. 
In the top panels we show how the bivariate distribution varies as a function of $\log \Mbh$ and $\log \lamEdd$ for Type~1 (left panels) and Type~2 (right panels) AGNs. 
We also indicate intervals over which we integrate to produce the BHMFs and ERDFs shown in the lower panels. 
In the middle panels, we show the reconstructed BHMFs for two bins of $\log$ \lamEdd\ (bin width 0.3 dex), as well as the integrated BHMF (over all \lamEdd\ we consider). 
Similarly, the bottom panels show the reconstructed ERDFs for two $\log$ \Mbh\ bins (bin width of 0.3 dex), as well as the integrated ERDF over all \Mbh. 
We note that, in a graphical sense, the reconstructed XLFs would correspond to integrating the bivariate distribution along anti-diagonal stripes (see top left panel in Fig.~\ref{fig:bivariate_contourplot}), illustrating that the bivariate distribution function fully captures the statistical properties of AGN. 

\begin{figure*}[h]
\begin{centering}
	\includegraphics[width=\textwidth]{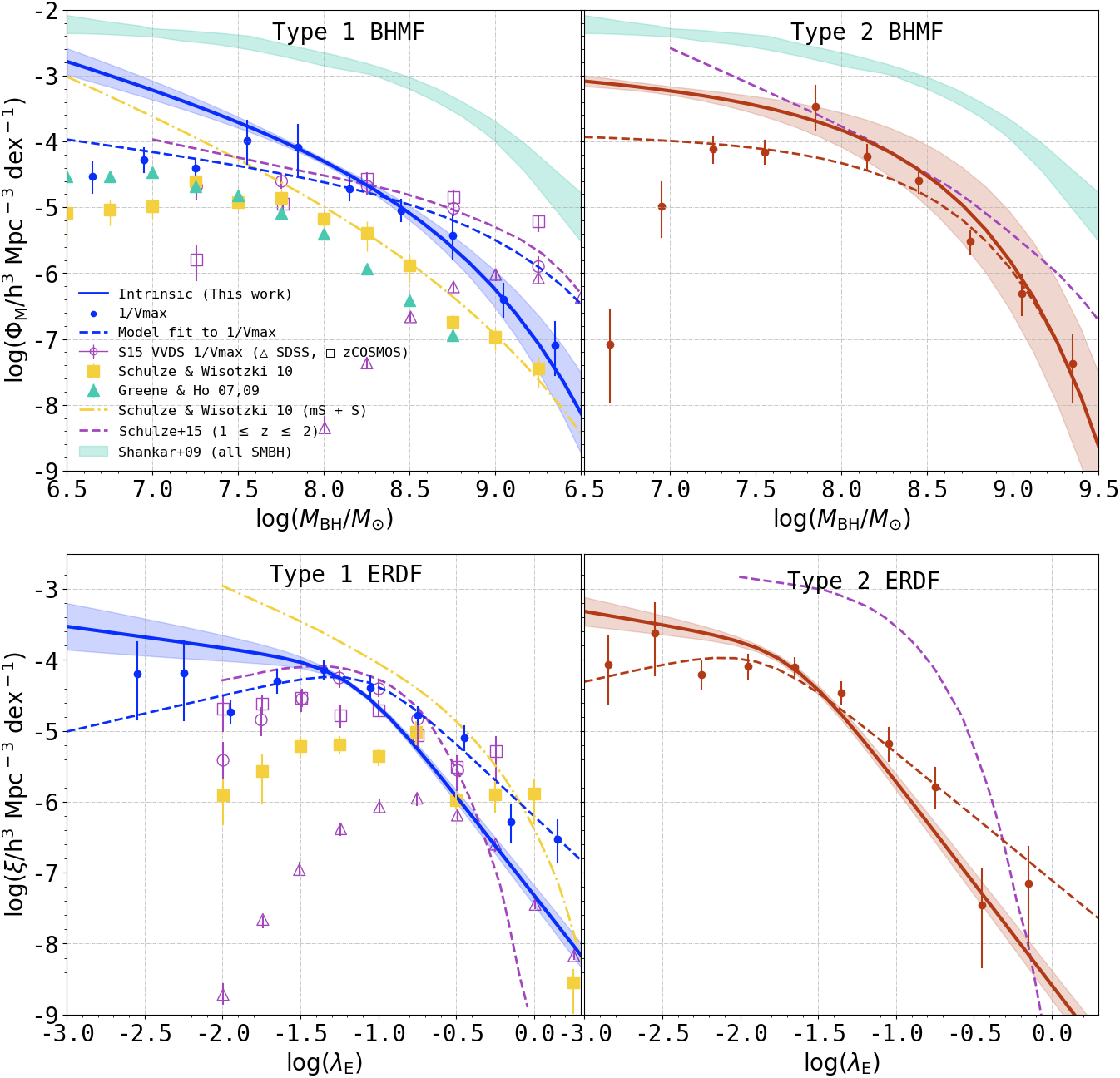}
	\caption{\label{fig:bhmferdfcomparison}Comparison of the BASS BHMF and ERDF for Type~1 AGN ({\it left panels}) and Type~2 AGN ({\it right panels}) to previous studies. 
	The {\it blue/red solid lines} show the {\it intrinsic} BASS BHMF and ERDF. \textit{Top panels:} The BHMF of Type~1 and Type~2 AGN. The \textit{green triangles} shows the bias-uncorrected 1/\Vmax\ results from \citet{Greene:2007aa,Greene:2009aa}, calculated using a sample of local broad-line AGN from SDSS. The \textit{yellow squares} and \textit{dashdotted lines} show the 1/\Vmax\ and incompleteness corrected (but not uncertainty corrected) BHMF of \citet{Schulze:2010aa}, calculated using local broad-line AGN from the Hamburg/ESO survey. The fully bias-corrected Type~1 BHMF and predicted Type~2 BHMF from \citet{Schulze:2015aa}, evaluated at $1 <$ z $< 2$, are shown with {\it purple dashed lines}. The 1/\Vmax\ points from \citet{Schulze:2015aa}, constrained using VVDS, SDSS and zCOSMOS data sets, are also shown using \textit{open purple circles, open purple triangles} and \textit{open purple squares}, respectively. For reference, we also plot the total BHMF (including inactive black holes) from \citet{Shankar:2009aa} ({\it turquoise shaded region}).
	\textit{Bottom panels:} The ERDF of Type~1 and Type~2 AGN. The data points and lines from \citet{Schulze:2010aa} and \citet{Schulze:2015aa} follow the same scheme as in the top panels. }
\end{centering}
\end{figure*}

Note that there is some degeneracy between several pairs of parameters describing the fitting functions, as shown in our MCMC chain contour plots in Appendix~\ref{sec:app_tables}  
(Figure~\ref{fig:chain_overlap}). \magenta{For all three AGN populations (Type~1, 2, and overall), the parameters of the BHMF show significant correlation. The pairs of parameters $(\alpha, \beta)$ and $(\alpha, \log M_{\rm BH}^*)$ seem to be anti-correlated, while the pair $(\beta, \log M_{\rm BH}^*)$ is positively correlated. \magenta{These trends are rather expected: while fitting the same intrinsic population, if we fix the break in BHMF at a higher mass, the slope at low mass ($\alpha$) has to be shallower and the slope at high mass ($\beta$) has to be steeper, to compensate for the higher mass break and produce a good fit.} For the ERDF, the slopes ($\delta_1$--$\epsilon_{\lambda}$) are negatively correlated for all three samples. The break of the ERDF ($\log \lambda_*$) is weakly positively correlated with $\delta_1$ and shows no significant correlation with $\epsilon_{\lambda}$.} The Figure also shows that even when these functions converge to the same distribution for one parameter, it may occupy distinctly different locations in the six-dimensional parameter space. These degeneracies should be kept in mind when one tries to directly compare individual fitting parameters within our own analysis (e.g., between Type~1 and 2 AGN), or when comparing our best-fit parameters to those found in other studies. In what follows, instead of comparing the values of individual parameters, we often refer to similarities (or lack thereof) in the shapes of certain fitting functions shown in the Figure. 

In Figure \ref{fig:bhmferdfcomparison} we compare the BASS BHMF and ERDF for Type~1 AGN ({\it left panels}) and Type~2 AGN ({\it right panels}) to previously published determinations of these distributions. In the top panels we show the 1/\Vmax\ $\Phi$ values and the bias-corrected BHMF in blue (\textit{left panels}; Type~1) and red (\textit{right panels}; Type~2). 
In the \textit{top left panels}, the green data points show the $\Phi$ values determined by \cite{Greene:2007aa,Greene:2009aa}, based on a large sample of SDSS AGNs with broad \Halpha\ lines, which is not corrected for sample truncation. 
The yellow data points and dash-dotted line illustrate the 1/\Vmax\ $\Phi$ values and incompleteness-corrected distribution, respectively, determined by \citetalias{Schulze:2010aa} based on the quasar sample of the Hamburg/ESO survey (HES; \citealt{Wisotzki:2000aa}). Note that \citetalias{Schulze:2010aa} does not correct for measurement uncertainties. 
The purple dashed lines in Fig.~\ref{fig:bhmferdfcomparison} show the BHMF and ERDF determined by \citet{Schulze:2015aa}, based on a joint analysis of $1 <  z < 2$ AGNs from the SDSS, zCOSMOS and VVDS samples (\citealt{Schneider2010_SDSS_DR7QSO}, \citealt{Lilly2009_zCOSb} and \citealt{Gavignaud2008_VVDS}, respectively).
Note that the \citet{Schulze:2015aa} BHMF and ERDF for Type~2 AGN are \textit{predicted} from the corresponding distributions for Type~1 AGN, by assuming a luminosity-dependent fraction of obscured systems, which is in turn determined from X-ray surveys. 
The higher space densities of the \citet{Schulze:2015aa} BHMFs and ERDFs are expected given the well-known redshift evolution of AGN abundance \cite[e.g.,][]{Aird2015,Caplar:2015aa,Aird:2018aa,Shen2020_QLF}. 
The turquoise shaded region in the \textit{top panels} of Fig.~\ref{fig:bhmferdfcomparison} also shows 
the \textit{total} local black hole mass function---that is, including both active and inactive SMBHs determined by \citet{Shankar:2009aa}. The total BHMF naturally occupy a much higher space density than our BASS AGN-only BHMF. 
The recent study by \citet{Shankar2020_NatAst} used a different set of $\Mbh-\sigma$ relations (discussed in \S\ref{sec:measurements}) to derive an updated total BHMF. 
However, as our analysis relies on the more widely used $\Mbh-\sigma$ relation of \cite{Kormendy:2013aa}, we do not compare our results with the \citet{Shankar2020_NatAst} BHMF.
%

%

%

\subsection{Mass Independence of the ERDF \magenta{Shape}}\label{sec:mass_independence}

\begin{figure}[h]
	\centering
	\includegraphics[width=0.55\textwidth]{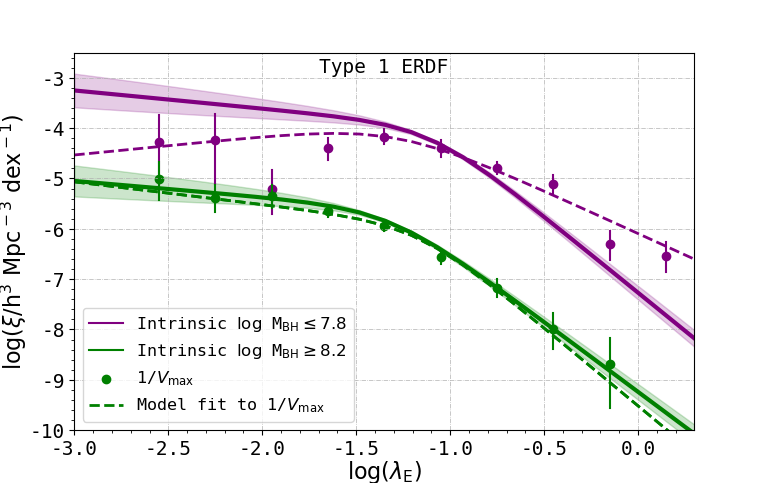}
	\includegraphics[width=0.55\textwidth]{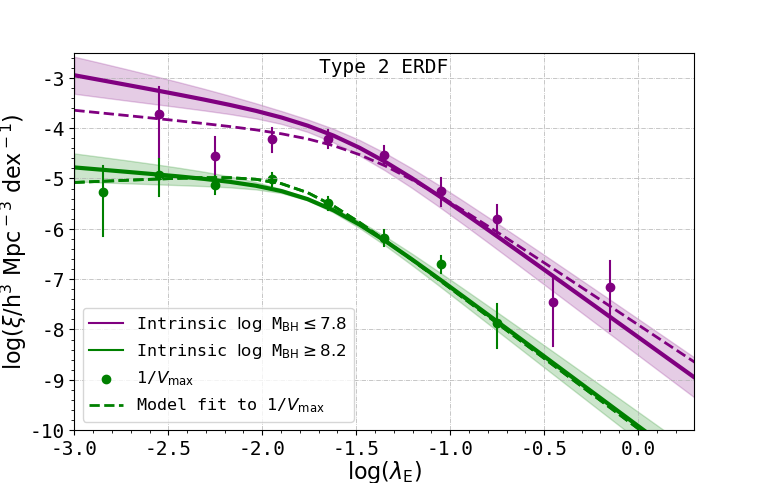}
	\caption{\label{fig:mass_indepen} ERDF of Type~1 ({\it top panel}) and Type~2 ({\it bottom panel}) AGN divided into two mass bins: $\log$ \Mbh $\leq 7.8$ ({\it purple lines and points}) and $\log$ \Mbh $\geq 8.2$ ({\it green lines and points}). The \textit{data points} represent 1/\Vmax\ in each \lamEdd\ bin, the \textit{dashed lines} represent fit to the 1/\Vmax\ points, and the \textit{solid lines} represent the bias-corrected, intrinsic distribution functions. 
	Dividing the sample in this way shows that, for both Type~1 and Type~2 AGN, the shape of the ERDF is independent of mass.}
\end{figure}


We used our BASS/DR2 sample to directly verify that our assumption of a mass-independent ERDF is a reasonable assumption, at least for the data in hand. 
To this end, we divided each of the Type~1 and Type~2 AGN sub-samples into two broad mass bins, $\log (\Mbh/\Msol) \leq 7.8$ and $\log (\Mbh/\Msol) \geq 8.2$. 
The 0.4 dex wide gap in $\log\Mbh$ was imposed to minimize ``mixing'' between the mass bins due to uncertainties on \Mbh\ estimation. 
We have then derived the 1/\Vmax\ measurements (which are susceptible to selection biases, as discussed in detail in \S\ref{sec:stat_method}), the associated functional fits of these measurements, and the bias-corrected, intrinsic functional forms, for each of the four sub-samples (Type~1s and 2s, low- and high-mass), following the same methodology as used for our main analysis.

The results of this analysis are shown in Figure~\ref{fig:mass_indepen}.
We find that the \magenta{\textit{shapes}} of the bias-corrected, intrinsic distributions of the low- and high-mass bins\magenta{---that is, power-law exponents and the the location of the break---}are in excellent agreement (for each of the AGN sub-types), as can also be seen from the best-fit parameters (Table~\ref{tab:bhmfbhmferdffitERDFmassdep}).
The low-mass subsets naturally have higher number densities (i.e., normalizations), as expected, given the generally decreasing nature of the BHMF with increasing \Mbh. 
This analysis indicates that the \magenta{shape of the} ERDF is indeed mass independent, for both obscured and unobscured AGN. 

\section{Discussion}\label{sec:discussion}

We have derived the XLF, BHMF and ERDF of a large, complete sample of ultra-hard X-ray-selected, low-redshift AGNs from the BASS/DR2 dataset. 
Our analysis included the direct 1/\Vmax\ approach as well as an elaborate inference scheme that allowed us to recover the intrinsic distribution functions, accounting for numerous generic, AGN-related, and survey-specific potential biases. 
In both cases, we derived functional fits of the distribution functions, considering the combined BASS/DR2 sample of (non-beamed) AGNs, as well as sub-samples based on the AGN optical spectral classes (Type~1s and 2s) and line-of-sight obscuration (parametrized by their \NH). 
We finally demonstrated that the intrinsic BHMF and ERDF can be combined to reproduce the XLF, and that the assumption of \magenta{mass-independence of the shape of the ERDF} is justified, at least for our sample.

Before moving to a higher-level discussion, we briefly list the main results of our main analyses:

\begin{itemize}

    \item \magenta{Our main result is the intrinsic BHMF and ERDF for Type 1 and Type 2 AGNs, presented in Figure~\ref{fig:bhmferdfdis}, and Tables~\ref{tab:bhmfbhmferdffitBHMF} and \ref{tab:bhmfbhmferdffitERDF}. 
    The method used to derive these functions overcomes the limitations of the 1/\Vmax\ approach (e.g., the Eddington bias; \citealp{eddington1913}) and incompleteness at low luminosities, masses, and/or accretion rates.}

    \item 
    We show the source number counts for all the AGN samples drawn from the \swift/BAT 70-month catalog in Figure~\ref{fig:logN_logS}, 
    and provide the best-fit power-law slopes and normalizations of the differential number counts for all samples under study in \S\ref{sec:logNlogS}.
    
    \item 
    We present the overall XLFs in Figure~\ref{fig:xlfobs}.
    The key observed quantities for these XLFs (i.e., 1/\Vmax\ based space densities) are given in Tables~\ref{tab:bhmfxlfvmaxxlf} and \ref{tab:xlfvmaxallsamples}, while their best-fit, broken-power-law parameters are given in Table~\ref{tab:bhmfxlffit}.
    
    \item 
    We present the XLFs of unabsorbed objects, Compton-thin objects and Compton-thick objects separately in Figure~\ref{fig:xlfznh}. 
    The corresponding 1/\Vmax\ values and best-fits parameters are given in Tables~\ref{tab:xlfvmaxabsbins} and \ref{tab:bhmfxlffit}, respectively.
    
    \item 
    The 1/\Vmax\ fits to the BHMF, ERDF, and XLF for Type~1, Type~2, and combined AGN samples are given in Tables~\ref{tab:bhmfvmax}, \ref{tab:erdfvmax}, and \ref{tab:bhmfxlfvmaxxlf}, respectively. 
    The parameter fits to the {\it observed}, 1/\Vmax\ based measurements for all three distributions, as well the {\it intrinsic} BHMF and ERDF are given in Tables~\ref{tab:bhmfbhmferdffitBHMF}, \ref{tab:bhmfbhmferdffitERDF}, and \ref{tab:bhmfxlffit} (again, respectively). 
    The top (BHMF), middle (ERDF), and bottom (XLF) panels in Figures~\ref{fig:bhmferdfdis} and ~\ref{fig:bhmferdfdis_all} show the observed quantities and the (intrinsic) best-fit distribution functions.
    
\end{itemize}

We next discuss several higher-level topics pertaining to the population of accreting SMBHs in the local Universe, based on the results of our main analysis. 
In particular, we 
(1) discuss the relation between accretion power and (circumnuclear) obscuration; 
(2) compare the intrinsic BHMF and ERDF to earlier studies and discuss possible ways to interpret their shapes; 
and (3) discuss the AGN duty cycle.

\subsection{AGN Obscuration and Demographics}\label{sec:discussion_agn_obsc}

Our analysis of the BASS/DR2 sample offers several insights concerning the role of obscuration in the distributions describing the (low-redshift) AGN population. First, we have fully considered the effect of line-of-sight obscuration on the derived intrinsic luminosity of every AGN in our sample, and thus on the derived XLF (as well as the BHMF and ERDF).
This is motivated by the significant attenuation expected for highly obscured sources [$\logNH>23$], even in the ultra-hard 14$-$195 keV band (i.e., Fig.~\ref{fig:attenuation_curve}). 
%

Second, the BASS sample allowed us to construct and explore the XLF for AGN of several \NH\ regimes.
The top panels of Fig.~\ref{fig:xlfznh} show the XLF in each of the three \NH\ bins, while the bottom panels show the fractions of objects in each \NH\ bin relative to all AGN (in a luminosity-resolved way; see details in \S\ref{ssec:xlf_lognh}). 
We find that 
(1) the fraction of unabsorbed AGN \textit{increases} with luminosity; 
(2) the fraction of Compton-thin sources \textit{decreases} with luminosity; 
and (3) the fraction of Compton-thick objects remains roughly \textit{constant} with luminosity. 

The observation that unobscured sources dominate the high-$L$ end of the AGN LF is often interpreted as evidence for the so-called ``receding torus model'' \cite[e.g.,][]{Lawrence:1991aa,simpson2005}. 
In this model the covering factor of the dusty torus, which obscures the nuclear region, decreases with increasing AGN luminosity---a trend that is observed in many (X-ray) AGN samples and surveys \cite[e.g.,][]{lawrenceelvis1982,steffen2003,Barger2005,simpson2005,Hasinger:2005aa,lafranca2005,treisterurry2006,Maiolino2007,brusa2010,burlon2011}. 
As the opening angle increases, the probability of detecting an obscured source decreases. 
Some studies suggested that unobscured AGN with higher bolometric luminosities may have less (dusty) torus material, based on the ratio of mid-IR to bolometric ratio, again supporting the receding torus model (e.g., \citealp{Treister2008}). 
Some alternative explanations have also been suggested. 
For example, \citet{akylas2008} showed that the photoionization of the obscuring screen around AGNs roughly reproduces the relationship between the fraction of obscured AGN and X-ray luminosity, as observed by \textit{XMM}-Newton and \textit{Chandra}. 
\citet{Honig:2007aa} suggested that the Eddington limit on a clumpy torus may also cause the obscured fraction to decrease with luminosity. 
It is important to note that the apparent decrease in the fraction of obscured AGN at high luminosities could be partially due to selection effects, as suggested by e.g. \cite{treisterurry2006}---further emphasizing the need for large, highly-complete samples, drawn from homogeneous input catalogs (such as BASS).

Many of these studies often made the (pragmatic) assumption that Compton-thin AGN trace {\it all} obscured objects (e.g., \citealt{Ueda:2014aa}; see also \citealt{HickoxAlexander2018_rev}). 
Specifically, several analyses of AGN LFs have assumed that the abundance of AGNs stays constant throughout the $\logNH = 24-26$ regime, thus predicting significant space densities of $\logNH > 25$ AGNs \cite[e.g.,][]{Ueda:2014aa,Aird2015,buchner2015,ananna2019}. 
In contrast, only a handful of objects (three) with $\logNH\ \geq 25$ are observed in the (70-month) all sky \swift/BAT survey, conducted in the ultra-hard 14-195 keV band \citep{Ricci:2015aa,Ricci2017_Nat}. 
\magenta{To see how our BASS-based results relate to this issue, we consider the \citet{ananna2019} XLF, which suggests that Compton-thick AGN would comprise $\approx$  50\% of all AGN in the local Universe. 
If we apply the appropriate BAT flux limits to that XLF (i.e., the 70-month survey) and limit the XLF to $\logNH < 25$, we find that $<10$\% of the {\it observable} sample would in fact be Compton-thick, which is consistent with the results shown in Figure~\ref{fig:xlfznh}.} 
\magenta{
We also note that the \cite{Ricci:2015aa} study, where the intrinsic \NH\ distribution shown in Fig.~\ref{fig:nh_dist} was derived, also presents the {\it observed} \NH\ distribution of the \textit{Swift}-BAT AGN sample used in that study (their Figure 4, bottom panel). The observed Compton-thick fraction in that study is ${\lesssim}10$\%, again consistent with the result we show in Fig.~\ref{fig:xlfznh} here (note that the \citealp{Ricci:2015aa} analysis also includes AGN at $z < 0.01$).}

As the attenuation curve in Figure~\ref{fig:attenuation_curve} shows, the observed luminosity of an object with $\logNH \geq 25$ is $\leq 5$\% of its intrinsic luminosity in the 14--195 keV band, therefore we have to probe very faint fluxes to be able to detect such heavily obscured AGN. \citet{vito2018}, \citet{yan2019} and \citet{carroll2021} find evidence for IR-selected luminous AGNs that completely lack X-ray detection, even in the \textit{NuSTAR} 3-79 keV band.\footnote{\citet{carroll2021} showed that X-ray emission from these sources can be determined via stacking analysis of \textit{Chandra} data, indicating that these objects do emit X-rays. It is possible that the heavy obscuration around them extinguishes most of it.} 
It is therefore possible that even high-energy X-ray observations such as the \swift/BAT survey do not individually identify the most obscured Compton-thick objects, and thus the fractions reported here for the high-\NH\ sub-sample should be treated with some caution.  

Considering the physical driver for the trends linking (Compton-thin) obscuration and accretion power, our results in Fig.~\ref{fig:xlfznh} seem at face value to be consistent with the receding torus scenario.
However, note there is \magenta{no clear drop in the (relative) space density of Compton-thick AGN with increasingly high luminosities (i.e., no downward trend in the bottom right panel of Figure~\ref{fig:xlfznh}). 
Indeed, the small sample size and the correspondingly large errors mean we cannot robustly rule out the possibility that the luminosity dependence of the Compton-thick space densities is consistent with that of Compton-thin sources.
With this caveat in mind, if the Compton-thick fraction indeed remains (roughly) constant with luminosity, it may lead to some important insights regarding the distribution of circumnuclear matter in AGNs.}

\cite{Fabian2006}, \cite{Fabian2008}, and \cite{Fabian2009_feedback}, have suggested that the radiation pressure exerted on the dusty torus gas is crucial for understanding links between AGN accretion power and obscuration. 
This was corroborated by the BASS/DR1-based study by \cite{Ricci2017_Nat}, which tied radiation pressure, gravity and orientation angle together and suggested an explanation for the relation between fraction of obscured sources and luminosity. 
%
In this scenario, the fraction of obscured (Compton-thin) sources fundamentally depends on the Eddington-ratio, which dictates the effective radiation pressure on the dusty torus gas. As the luminosity exceeds the effective Eddington limit for dusty gas, the torus material is pushed away from the central engines, thus significantly decreasing the abundance of high-\lamEdd, high-\NH\ sources. The observed luminosity dependence is then simply a consequence of the \lamEdd\ dependence, dictated by the (limited) range of \Mbh\ probed in AGN surveys. 
Most importantly, \cite{Ricci2017_Nat} showed that the fraction of Compton-thick AGN is independent of luminosity, indicating that these clouds are apparently unaffected by radiation pressure, or that the effective \lamEdd\ threshold for Compton-thick clouds is too high and is seldom exceeded.\footnote{The \NH\ dependence of this threshold value is shown in Figure~3 of \citealp{Ricci2017_Nat}.}

Our results provide further support for this radiation pressure-driven scenario. 
The ERDFs we constructed for Type~1 and 2 AGN (Fig.~\ref{fig:bhmferdfdis}) clearly show that obscured AGNs (Type~2 sources) are indeed increasingly rare beyond $\log\lamEdd^* \simeq -1.7$, which is remarkably consistent with the effective Eddington limit for dusty gas [$\log\lamEdd \approx -1.7$ for $\logNH=22$; see \citealp{Ricci2017_Nat} and references therein]. 
The downturn in the space densities of unobscured (Type~1) AGN occur at higher accretion rates, $\log\lamEdd^* = $ \erdfloglamstartypeone, and may instead be more linked to the physics of accretion disks and/or of the (circumnuclear-scale) fueling mechanisms.  
We therefore propose that the much higher space densities of unobscured AGN (relative to obscured AGN) at high \lamEdd\ could be naturally explained through the effect of radiation pressure on dusty obscuring material.

Indeed, looking more closely at the different shapes of the ERDFs for Type~1 and Type~2 AGN, another line of interpretation suggests itself, 
relating small-scale physics to large-scale population statistics. 
First, high accretion rates (i.e., high \lamEdd) can be triggered by major galaxy mergers, and 
theoretical models \cite[e.g.,][]{Hopkins2006_mergers,Hopkins2006_QLF_model,Blecha2018} and observations \citep{Treister2012,Glikman2012,Banerji2015,Glikman2015,Ricci2017_mergers,Glikman2018,Koss2018_Nat_mergers_AO,Banerji2021} suggest that luminous, merger-triggered AGN start from a highly obscured state (i.e., Type~2) but eventually blow away the obscuring material to become unobscured AGN (i.e., Type~1). 
The ratio of Type~2 to Type~1 AGN number densities at high \lamEdd\ could therefore reflect the short duration of the obscured phase to a much longer, unobscured phase. 
In contrast, at low \lamEdd, where accretion is less violent or disruptive, the ratio of Type~2 to Type~1 AGN likely reflects the geometry of circumnuclear obscuration, as in the traditional unification scheme that is well established locally \citep{barthel1989,antonucci1993,urry1995}. 
At intermediate \lamEdd, a transition occurs, where a mix of AGN types co-exist, and where the break in the ERDF of Type~2 AGN likely reflects the onset of significant radiative feedback, as explained above.
%
If the picture of obscured AGN transitioning into unobscured AGN during a merger-driven accretion episode is correct, then we would expect that at higher redshifts, more high \lamEdd\ obscured AGN would be transitioning into unobscured AGN, since both merger rates and gas fractions generally increase with redshift. 
Therefore, we predict that the break in the Type 2 AGN ERDF at higher redshift should occur at higher \lamEdd\, \cite[e.g.,][]{Jun21}.
Quantitative exploration of these ideas will be deferred to a future work.

We caution that our sample still has relatively few high-luminosity obscured sources. 
Still, the very existence of such sources indicates that the receding torus scenario is at the very least incomplete \cite[see also][]{Baer2019_BASS_hiL}, and our extensive bias corrections suggest that the dearth of high-\lamEdd\ obscured AGNs is real, and cannot be easily explained by obvious observational biases.

\subsection{Comparison of the Intrinsic BHMF and ERDF to Previous Studies}\label{ssec:comp_bhmf_erdf_discussion}

In Figure~\ref{fig:bhmferdfdis}, the similarity in shape of the BHMF for Type~1 and Type~2 AGN, and the difference in the shape of the ERDF, could \magenta{potentially imply that the observed difference between the two populations is mainly due to the different distributions in Eddington ratio. It could also highlight the possibility for a fundamental difference between the two AGN populations.}
\magenta{As explained in the preceding Section, this difference may be related to 
AGN fueling mechanisms, including galaxy mergers and large-scale environments (e.g., as seen in the BASS-based clustering analysis of \citealt{powell2018}); 
to smaller-scale feedback mechanisms, including the  radiation-regulated unification scheme (as supported by the BASS/DR1-based analysis of \citealt{Ricci2017_Nat}); 
or perhaps some other mechanisms.}
In any case, any comparison of our results to other studies should take into account these differences between Type~1 and 2 AGNs.

In Figure~\ref{fig:bhmferdfcomparison} we compare the BHMFs and ERDFs for BASS Type~1 and Type~2 AGNs with other BHMFs and ERDFs reported in the literature. 
We compare our BHMF for the Type~1 AGN sub-sample to the 1/\Vmax\ based BHMF of local broad-line, SDSS AGNs, derived by \citet{Greene:2007aa} and \citet{Greene:2009aa}. We find that the BASS observed space densities are higher, by $\geq 0.5 $ dex, in the range $\log (\Mbh/\Msun) > 7.0$, which may have been caused by differences in the selection method. We note that the \citet{Greene:2007aa} and \citet{Greene:2009aa} studies focused on a broad (optical) selection function, which included host-dominated continuum sources with broad \Halpha\ lines, and not just quasar-like AGNs. 
Our higher space densities thus indicate that ultra-hard X-ray selection of AGNs provides larger, likely more complete samples of even (high luminosity) broad-line AGNs, compared to SDSS. 
Some of this discrepancy may be related to the {\it bright} flux limit imposed as part of SDSS spectroscopy ($i>15$ mag), which BASS does not impose.
Specifically, the fact that our 1/\Vmax\ space densities extend to higher masses 
imply that the X-ray selection is probing a different \Mbh\ (and possibly \lamEdd) range. 

\citetalias{Schulze:2010aa} assumed a modified Schechter function-shaped BHMF and a Schechter function-shaped ERDF, evaluated over essentially the same redshift interval as the present work ($z < 0.3$). While \citetalias{Schulze:2010aa} does correct for incompleteness at the low \Mbh/\lamEdd\ range, unlike 1/\Vmax\ method, it does not account for measurement uncertainty. For the BHMF, \citetalias{Schulze:2010aa} report the following best-fitting parameters: $\log (M^*_{\rm BH}/\Msun) = 8.11$, $\log (\Phi^*/\rm{h^3\ Mpc^{-3}}) = -5.10$, $\alpha = -2.11$, and $\beta=0.5$. The 1/\Vmax\ values of the BHMF of \citetalias{Schulze:2010aa} is $\simeq 0.5$ dex lower at $\log (\Mbh/\Msol) \leq 9$ than this work. These differences in 1/\Vmax\ between \citetalias{Schulze:2010aa} and this work are also reflected in the discrepancy between the intrinsic BHMFs, particularly at intermediate masses. 
As both \citetalias{Schulze:2010aa} and the present work focus on the same redshift regime 
the differences between the two observed populations  
likely arise from the distinct selection by the optical and ultra-hard X-ray bands.  
%

The differences in the two samples become clearer when we look at the ERDF. There is a noticeable discrepancy between the \citetalias{Schulze:2010aa} and our Type~1 results --- in both the observed data points and the intrinsic functions. 
Again, the discrepancy in the observed samples might be due to different selection methods. 
Optical surveys are more likely to be biased towards high luminosity AGNs because in order for an object to be identified as an AGN, it has to dominate over host galaxy emission. Therefore, this sample would be skewed towards high luminosity (and \lamEdd).  Additionally, our sample of Type~1 AGN includes all objects with broad H$\alpha$ lines, which includes intermediate types (such as 1.5, 1.9). These intermediate types tend to be somewhat obscured, even in the Compton-thin regime, which means that if the radiation-regulated unification model is indeed the dominant mechanism, a sample of luminous quasars (such as that of \citetalias{Schulze:2010aa}) would be skewed towards higher \lamEdd, compared to the more complete BASS sample. 
As the \citetalias{Schulze:2010aa} study does not account for measurement uncertainty, that could also lead to part of the discrepancy. 
\citetalias{Schulze:2015aa} presented the BHMF and ERDF for Type~1 AGN in the redshift range $1 < z <2$. 
Additionally, \citetalias{Schulze:2015aa} predicted the BHMF and ERDF of Type~2 AGN in the same redshift range by using a luminosity-dependent obscured fraction function (from \citealt{merloni2014}). 
By comparing our local BASS results with the  $1 < z <2$  \citetalias{Schulze:2015aa} results, it appears that at higher redshifts, there are more high mass AGN of both types, whereas in the local Universe the lower mass AGNs become more abundant. 
This suggests that many high-mass SMBHs have become inactive between $1 \lesssim z  \lesssim 2$ and $ z \lesssim 0.3$, or that the average accretion rate has decreased over time (e.g.,  due to fewer mergers, or because interstellar gas has been depleted). 
The \citetalias{Schulze:2015aa} Type~1 AGN ERDF agrees well with our Type~1 ERDF, while both the normalization and the shape of the \citetalias{Schulze:2015aa} Type~2 ERDF are significantly different from our Type~2 results. 
Specifically, the shapes of the two  \citetalias{Schulze:2015aa} ERDFs are consistent with each other, whereas the shapes of our Type~1 and Type~2 ERDFs are significantly different. 
Note that the predicted \citetalias{Schulze:2015aa} Type~2 estimates are based on several assumptions. As suggested in \citet{merloni2014}, even though the obscuration fraction is reported as redshift independent, some redshift dependence is still seen at high luminosities. Additionally, \citet{merloni2014} reported some issues that could lead to incorrect estimation of intrinsic luminosities (i.e., due to incorrect assumptions about the complex geometry of the obscurer). As this fraction is constrained using $\leq 10$ keV data, these uncertainties could be significant due to the degeneracy of AGN spectral parameters \citep{Gilli2007,Ricci2017_Xray_cat,ananna2020aa}. If the $z\simeq 1-2$ ERDF represents the underlying Type~2 population at that redshift, it would imply that AGN activity has decreased over time. The overall normalizations of the BHMFs at  $1 < z <2$ are also higher than the BHMFs of the local Universe, which supports the decreased activity scenario.

\subsection{AGN Duty Cycle}
\label{subsec:duty_cycle}
%
%

\begin{figure*}
	\includegraphics[width=.475\textwidth]{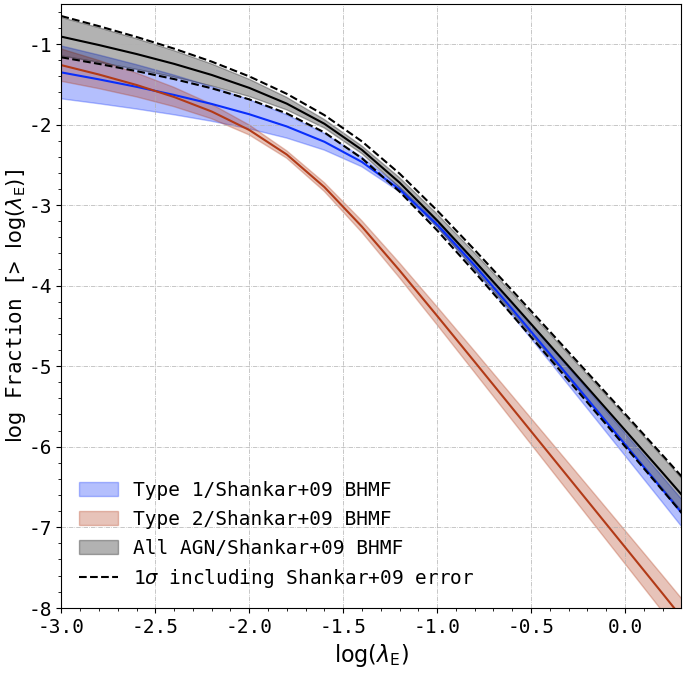} \hfill
	\includegraphics[width=.475\textwidth]{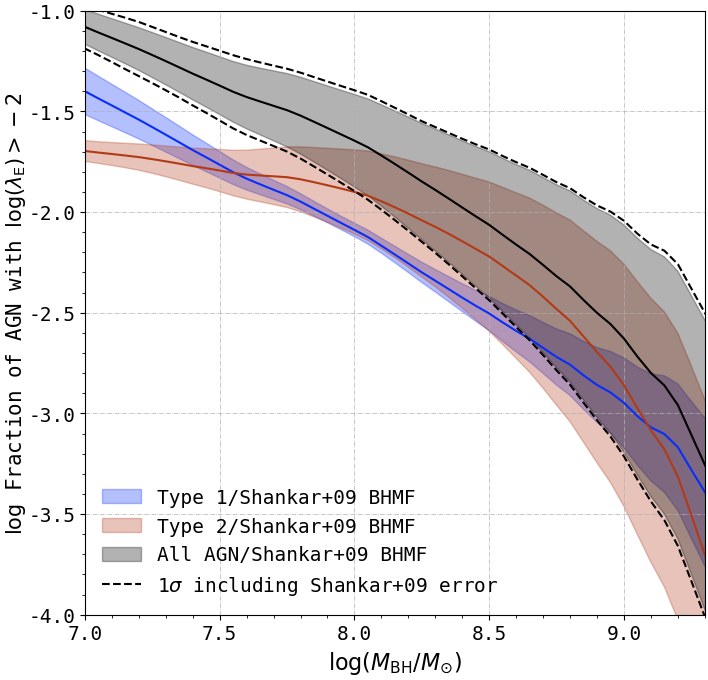}
	\caption{
	{\it Left panel:} The fraction of AGN that lie above $\log \lamEdd$ according to our model, relative to total number of SMBHs (from \citealt{Shankar:2009aa}). 
	{\it Right panel:} The fraction of AGN that lie above $\log \lamEdd > -2$ as a function of mass. 
	We divide our best fit Type~1 (\textit{blue line and shaded region}), Type~2 (\textit{red line and shaded region}) and overall (\textit{black line and shaded region}) functions by the local total BHMF (including inactive SMBH) from \citet{Shankar:2009aa}. For both panels, for the overall curve (\textit{black}) the shaded regions illustrate the uncertainty due to errors in our ERDFs and BHMFs, while the black dashed lines also include the uncertainty due to the ranges in the total BHMF, as shown in Figure 7 of \citet{Shankar:2009aa}. For the curves for Type 1 and Type 2 AGN only (blue and red), the shaded regions show the errors in ERDFs and BHMFs only, for ease of comparison between the two types. 
	The total SMBH mass density according to \citet{Shankar:2009aa} is \shankoldmassdensrange 
    $\Msol$ $h^{\rm 3}$ Mpc$^{-3}$, while the {\it active} SMBH mass density according to our analysis is \activemassdens\ $\Msol$ h$^{\rm 3}$ Mpc$^{-3}$ (i.e., \activemassdensrange\% of the total mass density).
	} 
	\label{fig:duty_cycle}
\end{figure*}

A highly useful observational constraint on theoretical models of SMBH evolution is the AGN duty cycle---that is, the fraction of all SMBHs (including inactive SMBHs) that are actively accreting, above a certain Eddington-ratio (and at any given cosmic epoch). 
The AGN duty cycle has been addressed by numerous observational and theoretical studies.
Some hydrodynamical galaxy simulations have tried to quantify the AGN duty cycle by tracing the gas inflow onto the central SMBHs.
\citet{Novak2011} simulated a single galaxy and found that the SMBH accretes at $\log \lamEdd > -3$ for $ \lesssim $ 30\% of the time span (12 Gyr) covered by the simulation. 
\citet{AnglesAlcazar2020} recently reported a duty cycle of $\sim$ 0.25 at $z < 1.1$. 
%
Phenomenological AGN population models can constrain and/or deduce the AGN duty cycle by linking the observed (redshift-resolved) AGN luminosity function with the local (active and inactive) BHMF, or indeed the (integrated) BH mass density, generally following the \citet{Soltan1982} argument (e.g., \citealt{cavaliere1989,Marconi:2004aa,Shankar:2009aa}).
For example, \citet{Shankar:2009aa} found that, in the local Universe, less than 1\% of all SMBHs should be considered as active (i.e., accreting at the fiducial accretion rate of their model; see their Figure 7). 
The active fraction or duty cycle is often defined as a ratio of luminosity function to mass function, independent of Eddington-ratio. 
%
Both simulations and population synthesis studies typically have to assume an AGN radiative efficiency \citep{Shankar:2009aa}, and often also have to assume 
an ERDF, or even a universal \lamEdd\ (e.g., \citealp{Shankar2013,Weigel:2017aa}). 

Highly-complete AGN surveys naturally provide the observational benchmark for the AGN duty cycle. For example, \cite{Goulding2010} reported an active fraction of $\approx$ 0.27, based on a volume-limited mid-IR selected sample of $D<15$ Mpc galaxies - although their definition of `active' includes AGN with Eddington-ratios as low as $\log \lamEdd \geq -5$. 
%

There are theoretical, phenomenological, and observational lines of argument for the AGN duty cycle to depend on galaxy and/or BH mass, and perhaps on other properties as well (e.g., galaxy environment, clustering; \citealp{haiman2001, martini2001, shen2007, whitemartini2008, shen2009, shankarweinberg2010, shankar2010c}). 
In Figure~\ref{fig:duty_cycle} we show the AGN duty cycle in the local Universe, based on our BASS/DR2 AGN sample. 
In the left panel of Fig.~\ref{fig:duty_cycle} we show the \lamEdd-dependent duty cycle, expressed as the cumulative probability of having $\log\lamEdd$ greater than a given value, $P(>\log\lamEdd)$. 
We calculate this probability by integrating over all BH mass bins and $\log\lamEdd$ bins above a given value, then dividing by the integrated total local BHMF (i.e., including inactive SMBHs), taken from \citet{Shankar:2009aa}.
The \citet{Shankar:2009aa} BHMF was compiled by taking into account the dispersions in all the local SMBHs that were available at the time \cite[e.g.,][]{mclure2002,MarconiHunt2003,tundo2007}. As discussed in \S\ref{sec:measurements}, 
\citep{Shankar2020_NatAst} reports a BHMF $\sim$4 lower than the \citet{Shankar:2009aa} BHMF (in terms of space densities at all masses), as it uses a recalibrated M--$\sigma$ relationship. As \citet{Koss_DR2_sigs} uses the canonical \citet{Kormendy:2013aa} prescription to estimate masses for BASS/DR2 objects, we compare our results to the \citet{Shankar:2009aa} BHMF in Figure~\ref{fig:duty_cycle}. 
%

The \Mbh--integrated AGN duty cycles for the entire BASS/DR2 sample relative to \cite{Shankar:2009aa} BHMF at $\log \lamEdd = -2, -1$ and 0 
 are about $P(\log \lamEdd > -2) \simeq $ \dutyshankoldlamnegtwo, $P(\log \lamEdd > -1) \simeq $ \dutyshankoldlamnegone, and  $P(\log \lamEdd > 0)\simeq $ \dutyshankoldlamzero, respectively. 
At the lowest \lamEdd\ threshold that is reasonable for radiatively-efficient SMBH accretion, we obtain  $P(\log\lamEdd > -3) \simeq $ \dutyshankoldrangelow--\dutyshankoldrangehigh.
According to the \citet{Shankar:2009aa} SMBH mass function, the total mass density of all SMBH in the local Universe is 
\shankoldmassdensrange $\Msol$ $h^{\rm 3}$ Mpc$^{-3}$, while the {\it active} SMBH mass density is \activemassdens\ $\Msol$ $h^{\rm 3}$ Mpc$^{-3}$ (i.e., \activemassdensrange\% of the total SMBH mass density).

Considering the Type~1 and Type~2 AGN in our samples, their duty cycles are essentially identical at the fiducial threshold corresponding to $P(\log~\lamEdd \simeq -2.5)$. 
For lower threshold Eddington-ratios ($\log \lamEdd < -2.5$) the cumulative duty cycle is slightly higher for Type~2 sources than it is for Type~1 sources (but within the $1 \sigma$ error budget), while for higher threshold \lamEdd, the duty cycle of Type~1 AGN is significantly higher. 
This means that a lower fraction of obscured AGN have such high \lamEdd, which is not surprising given the differences between the Type~1 and Type~2 ERDFs (Fig.~\ref{fig:bhmferdfdis}). 

The \textit{right} panel of Fig.~\ref{fig:duty_cycle} shows the fraction of active SMBHs (AGNs) with $\log \lamEdd\ > -2$ among the total SMBHs population (including inactive black holes), as a function of \Mbh. 
The general trend for all BASS AGNs is that the AGN fraction decreases with increasing \Mbh.  
This general trend is in agreement with what was found in several previous studies, including both direct observations \cite[e.g.,][]{Greene:2007aa,Goulding2010} and population models \cite[e.g.,][]{Marconi:2004aa,Shankar:2009aa,Shankar2013}.
Our AGN fraction  (\dutyshankoldrangelowperc--\dutyshankoldrangehighperc\%) 
is about an order of magnitude above what was found by \citet{Shankar:2009aa} using their fiducial model ($< 1$\% active), which assumes a constant radiative efficiency of 0.065. 
Our results are in slightly better agreement with the redshift-dependent Eddington-ratio model from the \citet{Shankar:2009aa} study. 
This demonstrates the importance of independent, observational determinations of the AGN ERDF, fractions, and duty cycles to (phenomenological) models of the cosmic evolution of SMBHs. 
We note that these trends may evolve with redshift.

Among the more nuanced trends in the right panel of Fig.~\ref{fig:duty_cycle}, we note that at lower masses, $\log(\Mbh/\Msun) < 9. $
, the fraction of Type~2 AGN with $\log\lamEdd > -2$ is higher than that of Type~1 AGNs. This is driven by the fact that the space density of Type~2 AGN is higher than that of Type~1 AGNs at $\log \lamEdd < -1.7$, as shown in Fig.~\ref{fig:bhmferdfdis}. This trend flips completely if the threshold is moved to $\log \lamEdd \geq -1$ (not shown in Figure), and Type~1 AGN dominates the fraction of AGN at all mass bins.

Assuming that the trends found here hold for other surveys (and out to higher redshifts), they highlight why different survey strategies may lead to ambiguous or contradictory conclusions about the obscured AGN fraction. Specifically, wide-field surveys that pick up the rarest, highest-\Mbh\ systems are expected to be biased towards unobscured systems; conversely, deeper (and narrower) surveys that uncover the more abundant lower-\Mbh\ population may be (mildly) biased towards obscured AGNs. \citet{mateos2017} reported that by studying the torus structure and covering factor of X-ray selected samples (at $< 10$ keV), they find that a significant population of obscured objects should exist at high luminosities, and are missed by X-ray surveys. These heavily obscured high-luminosity AGN have been identified in recent studies using IR, optical, and X-ray data \citep[e.g.,][]{yan2019,carroll2021}.  \citet{Treister2010} also reached similar conclusions by analyzing IR-selected sources. As discussed in \S\ref{sec:discussion_agn_obsc}, and as implied by Figure~\ref{fig:attenuation_curve} and these results, given Swift-BAT's current flux limits, many heavily obscured (i.e., $\log [N_{\rm H}/{\rm cm}^{-2}] > 25$), massive AGN may still be undetected by BAT. 

\subsection{Comparison between the Galaxy Stellar Mass Function and Active Black Hole Mass Function}
\label{ssec:gal_smbh_scaling}

\begin{figure}
\begin{centering}
	\includegraphics[width=.49\textwidth]{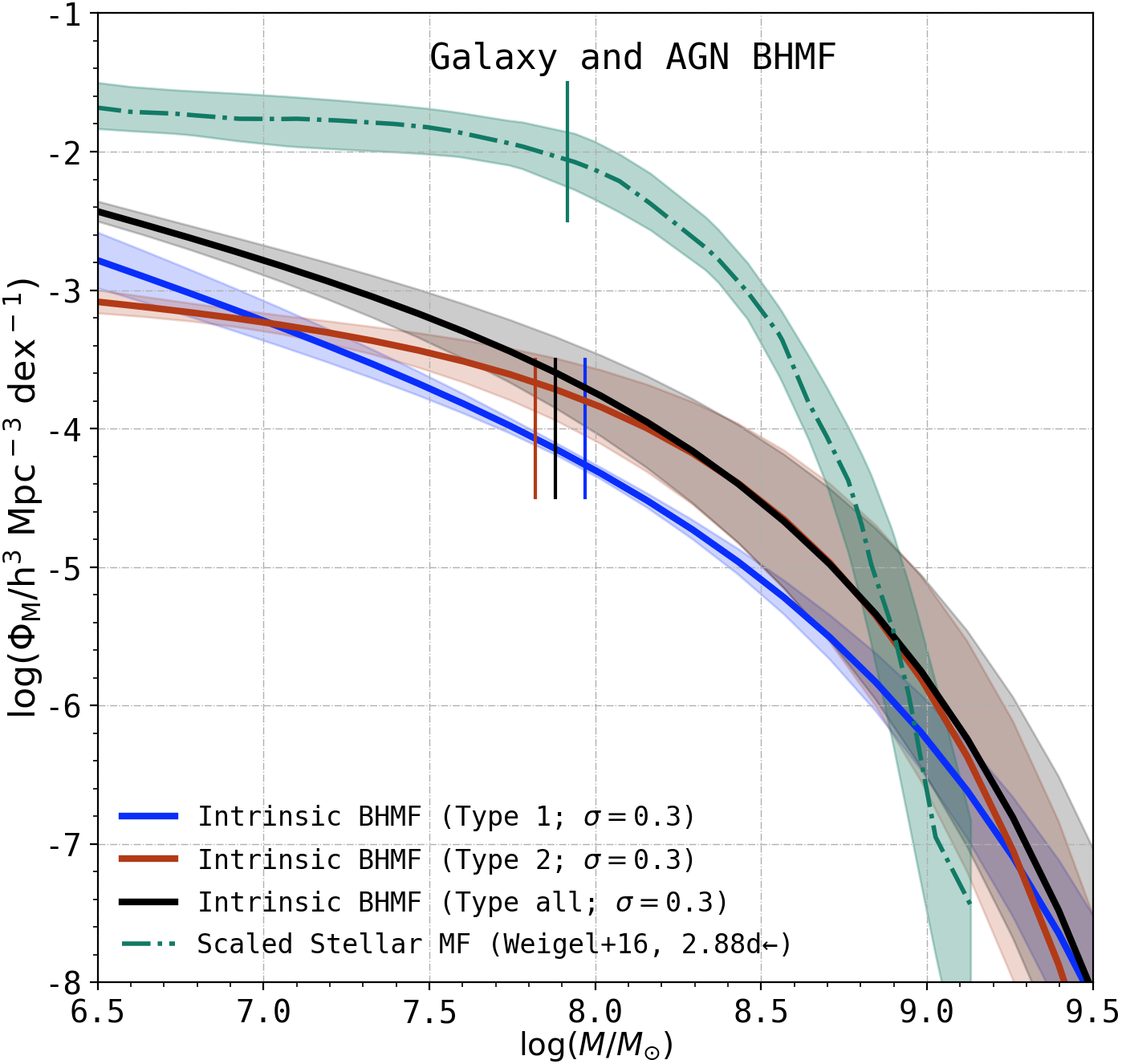}
	\caption{\label{fig:gal_bh_scaling} The BASS/DR2 black hole mass functions: Type~1 ({\it blue solid line}), Type~2 ({\it red solid line}) and all AGN ({\it black solid line}), compared to \citet{Weigel:2016aa} galaxy stellar mass function ({\it green dash-dotted line}), shifted left in mass by 2.88 dex to line up the breaks in galaxy Schechter function ({\it green vertical line}) and the modified Schechter function for all AGN ({\it black vertical line}). The dispersion in galaxy stellar mass function also includes the galaxy stellar mass function of \citet{peng2010}, \citet{Baldry:2012aa} and \citet{taylor2015}. All {\it shaded regions} show $\pm 1 \sigma$ errors for each function. }
\end{centering}
\end{figure}

\magenta{We finally use the BHMF we derive from the BASS DR2 data to speculate about the BH-to-stellar mass ratio.}
The local Universe provides ample evidence for a close relation between the mass of SMBHs and the stellar mass of their host galaxies (particularly their bulge components). 
The ratio of SMBH to stellar mass lies in the range $-3.55 < \log (\Mbh/\Mgal) < -2.31$ \cite[e.g.,][and references therein]{MarconiHunt2003,HaringRix2004, Sani2011_MM,Kormendy:2013aa,Marleau:2013aa,McConnell:2013aa,Reines:2015aa}.
If SMBH masses scale (roughly) linearly with host galaxy masses, then the corresponding mass functions should have similar shapes, with a horizontal shift that scales as $\Mbh/\Mgal$.\footnote{The vertical shift would scale with SMBH occupation fraction and AGN duty cycle.}
We explore the $\Mbh - \Mgal$ scaling relationship by comparing the break in our black hole mass function ($\Mstarbh$) with the break in galaxy stellar mass function ($\Mstargal$).

Some recent determinations of the galaxy stellar mass function, based on large optical low-redshift surveys, find breaks at $10.5 \lesssim \log(\Mstargal /\Msun) \lesssim 10.7$ \cite[e.g.,][]{MacLeod:2010aa,Baldry:2012aa,Weigel:2016aa}, with relatively limited variance between different galaxy types \cite[see discussion in, e.g.,][and references therein]{Moffett2016,Davidzon2017}. 
%
Comparing these galaxy stellar mass function breaks directly with the BHMF break we find for all BASS/DR2 AGNs [$\log (\Mstarbh/\Msun) =$\ \bhmflogmstar], we get \hbox{\galscalelow}~$ \lesssim \log( \Mstarbh / \Mstargal) \lesssim $\hbox{~\galscalehigh}. 
%
This agrees well with the range of $\log(\Mstarbh/\Mstargal)$ derived from direct measurements in individual systems.

\magenta{We caution that these results are highly speculative, as they inherently link the \textit{active} BHMF (i.e., the BHMF of AGN) to a quantity of stellar mass functions that are dominated by \textit{inactive} galaxies, thus assuming that the BASS-based \Mstarbh\ is representative of all SMBHs and/or that BASS AGN hosts are indistinguishable from the general galaxy population. 
The actual analysis of BASS AGN hosts and the measurement of their stellar masses is beyond the scope of this work, but is expected to be pursued in a future BASS study.}


Figure~\ref{fig:gal_bh_scaling} shows the galaxy mass function (shifted left by 2.88 dex for ease of comparison in the Figure) and AGN BHMF for our overall sample, as well as the Type~1 and Type~2 sub-samples. 
In this Figure, we use a modified Schechter function to represent the AGN samples, and a double Schechter function to represent the galaxy stellar mass function \cite[e.g.,][]{Weigel:2016aa, Weigel:2017aa}.
The shape of the galaxy stellar mass function may differ from the BH mass functions 
because of the different functional forms used to fit the data. However, this might also mean that the ratio of BH to galaxy mass, or even bulge to galaxy mass, varies with galaxy mass \cite[e.g.,][]{bell2017}. Fitting a modified Schechter function to galaxy stellar masses is beyond the scope of this work, but will also be explored in our future project.

We stress again that these simplistic galaxy-BH scaling relationships provide only a limited view into the relations between (BASS) AGNs and their hosts, as some studies suggest that close SMBH-host links should only be applicable to the (true) bulge or spheroidal components of galaxies and/or to certain types of galaxies \cite[see detailed discussion in, e.g.,][]{Graham2011,Kormendy:2013aa}.

\section{Summary and Conclusions}

We have determined the X-ray luminosity function (XLF), the black hole mass function (BHMF) and Eddington-ratio distribution function (ERDF) for the large and highly-complete ultra-hard X-ray-selected sample of BASS/DR2 AGN, covering the redshift range $0.01 \leq z \leq 0.3$. Our comprehensive methodology corrects these distributions for incompleteness at low masses and low Eddington-ratios, and also corrects for overestimation of space densities due to measurement uncertainty at high masses and high Eddington-ratios. 
We then convolved the bias-corrected BHMF and ERDF to verify that the observed XLF is reproduced self-consistently. 
We further calculated the XLF, BHMF and ERDF separately for Type~1 and Type~2 AGN. 
Indeed, thanks to the high sensitivity of \textit{Swift}-BAT to heavily obscured sources and the BASS spectroscopic follow-up, we are able to present a highly-complete determination of the BHMF and ERDF of Type~2 AGN.

We then use these key distribution functions to address several questions pertaining to the demographics of low-redshift AGNs. 

We summarize our inferences from this work as follows:

\begin{itemize}
    
    \item In the observed BASS/DR2 sample, the fraction of unabsorbed AGN increases with luminosity, the fraction of Compton-thin AGN decreases with luminosity and the fraction of Compton-thick objects stays constant with luminosity\footnote{\magenta{Note that some caution is required in this interpretation as the error bars for this trend in Compton-thick AGN is large due to small sample size.}} (as shown in Figure~\ref{fig:xlfznh}). 
    This result is consistent with the radiation-regulated unification model (proposed by \citealt{Ricci2017_Nat}).
    
    \item As shown in Figure~\ref{fig:bhmferdfdis}, the shape of the ERDF of Type~1 AGN is significantly different from that of Type~2 AGN, as the ERDF of Type~2 AGN is skewed towards low \lamEdd.
    The difference in the break in ERDF between Type~1 sample (\erdfloglamstartypeone) and Type~2 sample (\erdfloglamstartypetwo) is statistically significant. 
    \magenta{The increasing rarity of obscured AGNs above $\lamEdd\approx0.02$ is remarkably consistent} with the radiation-regulated unification model, and may indicate the role of blowout at high \lamEdd, while geometry and orientation dominate at low \lamEdd.
    
    \item As shown in Figure~\ref{fig:mass_indepen}, we demonstrate that the ERDF maintains its shape independent of BH mass, for two distinct mass regimes (and both Type~1 and Type~2 AGNs).
    
    
    \item Concerning the AGN duty cycles and mass density fraction, 
    we find that the fraction of Type~2 AGN is higher than Type~1 AGN at all masses (for $\log \lamEdd \geq -2$). 
    We find that the active fraction, defined as the fraction of AGNs with $\log \lamEdd > -3$ relative to the total BHMF (including relic systems) is \dutyshankoldrangelowperc--\dutyshankoldrangehighperc\%. In the local Universe, the percentage of mass in active SMBHs is \activemassdensrange\% of all SMBH mass.
    

\end{itemize}

Our extensive analysis opens the door for several potential follow-up investigations.
In the future, we will further explore the relationship between obscuring column density and Eddington ratio.  
With detailed host galaxy measurements for BASS AGNs, we may be able to study the key distributions functions (XLF, BHMF \& ERDF) split by host morphology, star formation state, environment (i.e., merger state), or other properties.
Combining our results with higher redshift samples, we expect that the present analysis of the BASS sample would serve as the low-redshift benchmark for studying the evolving population of accreting SMBHs, as probed by its key distribution functions. 


\section*{Acknowledgments}

We thank the anonymous reviewers for their constructive and detailed comments, which helped us improve the quality of this paper.
T.T.A. and R.C.H. acknowledge support from NASA through ADAP award 80NSSC19K0580, and the National Science Foundation through CAREER award 1554584. 
B.T. acknowledges support from the Israel Science Foundation (grant number 1849/19) and from the European Research Council (ERC) under the European Union's Horizon 2020 research and innovation program (grant agreement number 950533).
M.K. acknowledges support from NASA through ADAP award NNH16CT03C. 
C.M.U. acknowledges support from the National Science Foundation under Grant No. AST-1715512.  
C.R. acknowledges support from the Fondecyt Iniciacion grant 11190831. 
We acknowledge support from ANID-Chile Basal AFB-170002 and FB210003 (E.T., F.E.B.), FONDECYT Regular 1200495 and 1190818 (E.T., F.E.B.),  ANID Anillo ACT172033 (E.T.), Millennium Nucleus NCN19\_058 (TITANs; E.T.) and Millennium Science Initiative Program - ICN12\_009 (F.E.B.). %
K.O. acknowledges support from the National Research Foundation of Korea (NRF-2020R1C1C1005462). 
The work of K.I.\ is supported by the Japan Society for the Promotion of Science (JSPS) KAKENHI (18K13584, 20H01939). 
JdB acknowledges funding from the European Research Council (ERC) under the European Union's Horizon 2020 research and innovation programme (grant agreement No.726384/Empire).
This work was performed in part at Aspen Center for Physics, which is supported by National Science Foundation grant PHY-1607611.

This research made use of {\tt Astropy} 
\citep{astropy:2013, astropy:2018}, {\tt Matplotlib} \citep{Hunter:2007:90}, \textsc{Emcee} \citep{emcee}, {\tt NumPy} \citep{vanderWalt:2011:22}, {\tt Topcat} \citep{topcat}, \textsc{xpec} and \textsc{pyxpsec} \citep{xspec}, and {\tt ChainConsumer} \citep{chainconsumer}.




\bibliography{bib_lib_small,DR2_bib}{}
\bibliographystyle{aasjournal}



\appendix
\restartappendixnumbering
\section{Estimating random errors for the 1/\Vmax\ approach}
\label{sec:app_error_vmax}
To estimate the random errors on $\Phi_{\rm L}(\log L_{\rm X})$, $\Phi_{\rm M}(\log \Mbh)$, and $\xi(\log \lamEdd)$, we follow the approach by \cite{Weigel:2016aa} (see also \citealt{Gehrels:1986aa,Zhu:2009aa, Gilbank:2010aa}). The upper and lower errors on $\Phi_{\rm L}(\log L_{\rm X})$ in bin $j$ are given by:
\begin{equation}
\begin{aligned}
\sigma_{j, \rm up} =& -\Phi_{\mathrm{L}, j} + W_{\mathrm{eff}, j} \times \kappa_{\rm up}(N_{\mathrm{eff}, j})\\
\sigma_{j, \rm low} =& \Phi_{\mathrm{L}, j} - W_{\mathrm{eff}, j} \times \kappa_{\rm low}(N_{\mathrm{eff}, j}). 
\end{aligned}
\end{equation}
$W_{\rm eff}$ and $N_{\rm eff}$ correspond to the effective weight and the number, respectively, and are defined as:
\begin{equation}
\begin{aligned}
W_{\mathrm{eff}, j} =& \left(\sum_{i}^{N_{\rm bin}} \frac{1}{V_{\mathrm{max}, i}^2}\right) \times \left(\sum_{i}^{N_{\rm bin}} \frac{1}{V_{\mathrm{max}, i}}\right)^{-1}\\
N_{\mathrm{eff}, j} =& \left(\sum_{i}^{N_{\rm bin}} \frac{1}{V_{\mathrm{max}, i}} \right) \times \left(W_{\mathrm{eff}, j}\right)^{-1}.
\end{aligned}
\end{equation}
$\kappa_{\rm up}$ and $\kappa_{\rm low}$ represent the functions that allow us to compute the upper and lower limits on the effective number $N_{\rm eff}$ (see \citealt{Gehrels:1986aa}, equations 7 and 11).
To determine upper limits on $\Phi_{\rm L}(\log L_{\rm X})$, we compute $V_{\rm s}$, the comoving volume for the entire sky between $z_{\rm min, s}$ and $z_{\rm max, s}$. In bins with $N_{\rm bin} = 0$ the upper limit on $\Phi_{\mathrm{L}} (\log L_{\rm X})$ is then given by:
\begin{equation}
\sigma_{\rm limit} = -\frac{1}{V_{\rm s}} + \frac{1}{V_{\rm s}} \times \kappa_{\rm up}(N_{\rm eff} = 0),
\end{equation} 
with $\kappa_{\rm up}(N_{\rm eff} = 0)=1.841$ (\citealt{Gehrels:1986aa}). We compute the random errors on $\Phi_{\rm M}(\log \Mbh)$ and $\xi(\log \lamEdd)$ accordingly.

\section{Fitting the 1/\Vmax\ $\Phi$ values}
\label{sec:app_fit_vmax}
Below we outline how we find the best-fitting functional form for the XLF. The BHMF and the ERDF are fit accordingly. We fit all three distributions independently.

The errors on $\log \Phi_{\mathrm{L}}(\log L_{\rm X})$ are asymmetric. As we fit the values in log-space, we thus assume that $\log \Phi_{\rm L}(\log L_{\rm X})$ is distributed log-normally, rather than assuming a normal distribution. Following the method \cite{Weigel:2017aa}, we use an MCMC and the following probability density function for the fitting:
\begin{equation}
p(x, \mu, \sigma) = \frac{1}{x\sigma\sqrt{2\pi}} \times \exp\left(-\frac{(\ln x - \mu)^2}{2\sigma^2}\right).
\end{equation}
For the XLF, we determine the properties of $p$, i.e. $\mu$ and $\sigma$, in each $\log L_{\rm X}$ bin $j$. We use the following definitions:
\begin{equation}
\begin{aligned}
\bar{x}_j =& \log(\Phi_{\mathrm{L}, j}) + a\\
\mu_{j} =& \ln(\bar{x}_j)\\
\sigma_{j, 16} =& \frac{\ln(\bar{x}_j - \bar{\sigma}_{j, \rm low}) - \mu}{\rm{PPF(0.16)}}\\
\sigma_{j, 84} =& \frac{\ln(\bar{x}_j + \bar{\sigma}_{j, \rm up}) - \mu}{\rm{PPF(0.84)}}\\
\sigma_j =& \sqrt{\sigma_{j, 16}^2 + \sigma_{j, 84}^2}. 
\end{aligned}
\end{equation}
The constant, $a$, ensures that all $\bar{x}$ values are positive. $\rm{PPF}(0.16)$ and $\rm{PPF}(0.84)$ correspond to the value at which the integral over a normal distribution with $\mu = 0$, $\sigma = 1$ reaches 16 and 84 percent, respectively. We add $\sigma_{j, 16}$ and $\sigma_{j, 84}$ in quadrature to determine $\sigma_j$, since the log-normal distribution only represents an approximation for the distribution of $\log \Phi_{\rm L}(\log L_{\rm X})$ values.

For each functional XLF form that is proposed by the MCMC, we compute the predicted $\log \Phi_{\rm L, pred}$ values in all $\log L_{\rm X}$ bins. We then use $x_j = \log \Phi_{\mathrm{L, pred}, j}$ to compute the log-likelihood $\ln \mathcal{L}$:
\begin{equation}
\begin{aligned}
\mathcal{L} = & \prod_{j}^{\rm all\ bins} p(x_j, \mu_j, \sigma_j)\\
\ln \mathcal{L} \propto& - \sum_{j}^{\rm all\ bins} \ln (x_j + a) - \sum_j^{\rm all\ bins} \frac{(\ln(x_j + a) - \mu_j)^2}{2\sigma_j^2}.
\end{aligned}
\end{equation}

The MCMC allows us to maximize $\ln \mathcal{L}$ and to find the best-fitting parameters for the assumed functional form. To constrain the bright end slope of the XLF and the corresponding error we determine the sum of the $\gamma_1$ and $\epsilon_{\rm L}$ chains. We then determine the median $\gamma_2$ value and its credible intervals from this new chain. We proceed in the same way when fitting the ERDF with a double power-law. We do not include upper limits in the fitting procedure.

\section{Estimating the random error on the BHMF and the ERDF}
\label{sec:app_error_bhmf}
We estimate the random error on the bias-corrected $\Phi_{\rm M}(\log \Mbh)$ and $\xi(\log \lambda)$ by using the covariance matrix, which we derive from the MCMC chain. This approach is similar to the method used by \cite{Weigel:2016aa}.

We use the MCMC chain after burn-in to derive the covariance matrix $\Sigma$. On its main diagonal, $\Sigma$ contains the variance on the best-fitting $\Psi$ parameters. We also use the off-diagonal elements, which express the covariance. If we assume a modified Schechter function for the BHMF and a broken power law for the ERDF, $\Psi$ has six free parameters: the break and the two slopes of the BHMF ($M_{\rm BH}^*$, $\alpha$, $\beta$) and the break and the two slopes of the ERDF ($\lambda^*$, $\delta_1$, $\epsilon_{\delta}$). As discussed above, the normalization of $\Psi$ is kept constant in the MCMC. The $1\sigma$ random errors on the BHMF and the ERDF are given by:
\begin{equation}
\begin{aligned}
\sigma_{\Phi}^2(\log \Mbh) =& \left(\frac{\partial \Phi}{\partial M_{\rm BH}^*}\right)^2 \Sigma_{M M} + \left(\frac{\partial \Phi}{\partial \alpha}\right)^2 \Sigma_{\alpha \alpha} + \left(\frac{\partial \Phi}{\partial \beta}\right)^2  \Sigma_{\beta \beta} \\
+ & 2  \left(\frac{\partial \Phi}{\partial M_{\rm BH}^*}\right) \left(\frac{\partial \Phi}{\partial \alpha}\right) \Sigma_{M \alpha} + 2  \left(\frac{\partial \Phi}{\partial M_{\rm BH}^*}\right)  \left(\frac{\partial \Phi}{\partial \beta}\right) \Sigma_{M \beta} + 2  \left(\frac{\partial \Phi}{\partial \alpha}\right) \left(\frac{\partial \Phi}{\partial \beta}\right) \Sigma_{\alpha \beta}
\end{aligned}
\end{equation}
\begin{equation}
\begin{aligned}
\sigma_{\xi}^2(\log \lambda) =& \left(\frac{\partial \xi}{\partial \log \lambda^*}\right)^2 \Sigma_{\lambda \lambda} + \left(\frac{\partial\xi}{\partial \delta_1}\right)^2 \Sigma_{\delta \delta} + \left(\frac{\partial\xi}{\partial \epsilon_{\delta}}\right)^2 \Sigma_{\epsilon \epsilon}\\ 
+ & 2 \left(\frac{\partial\xi}{\partial \log \lambda^*}\right) \left(\frac{\partial\xi}{\partial \delta_1}\right) \Sigma_{\lambda \delta} + 2 \left(\frac{\partial\xi}{\partial \log \lambda^*}\right) \left(\frac{\partial\xi}{\partial \epsilon_{\delta}}\right) \Sigma_{\lambda \epsilon} + 2 \left(\frac{\partial\xi}{\partial \delta_1}\right) \left(\frac{\partial\xi}{\partial \epsilon_{\delta}}\right) \Sigma_{\delta \epsilon}.
\end{aligned}
\end{equation}
$\Sigma_{\rm{XX}}$ and $\Sigma_{\rm{XY}}$ correspond to the main and off-diagonal elements of the covariance matrix, respectively. As we assume the ERDF to be mass independent, we neglect the matrix elements that express the covariance between the BHMF and the ERDF.

\section{Testing the method}\label{sec:app_test}

We use two sets of parameters to create mock populations, and examine the robustness of our approach to recovering underlying parameter space using three different measurement uncertainties for each set of parameters.

\begin{figure*}[h]
    \centering
	\includegraphics[width=0.6\textwidth]{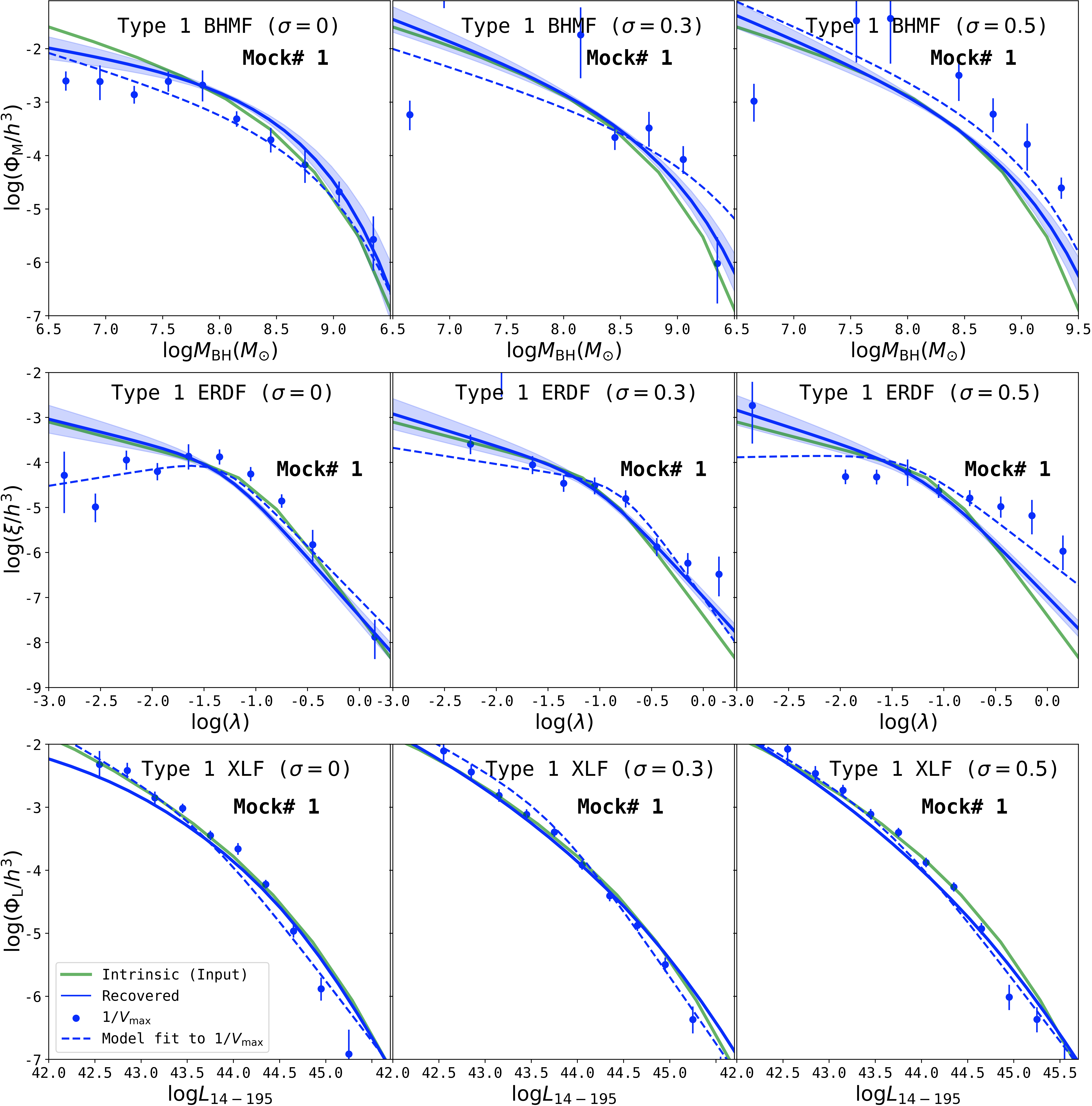}
	\includegraphics[width=0.6\textwidth]{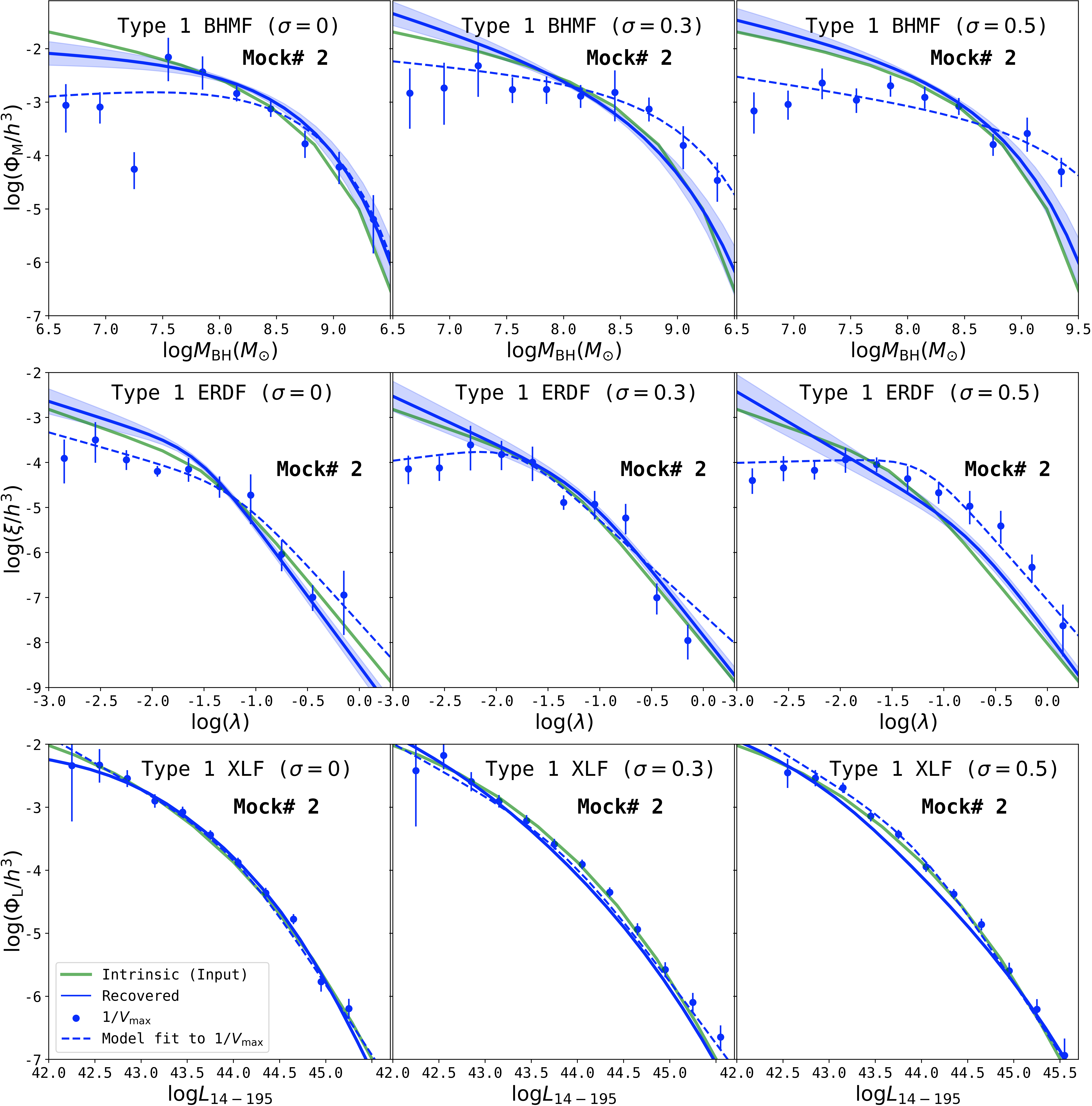}
	\caption{\label{fig:mock_catalog_type1} Results for Type~1 AGN 
	for two mock catalogs. For the top three panels (Mock\# 1), the parameters are fixed at $\log (\Phi^{*}/\rm{h^3\ Mpc^{-3}}) = -3.16$, $\log (M^*_{\rm BH}/\Msun) = 8$, $\alpha = -1.6$, $\beta = 0.6$, $\log (\xi^{*}/\rm{h^3\ Mpc^{-3}}) = -4.8$, $\log \lamEdd^* = -1$, $\delta_1 = 0.6$, $\epsilon_{\lamEdd} = 2.5$. For the bottom three panels (Mock\# 2), the parameters are fixed at $\log (\Phi^{*}/\rm{h^3\ Mpc^{-3}}) = -3.16$, $\log (M^*_{\rm BH}/\Msun) = 8.2$, $\alpha = -1.4$, $\beta = 0.7$, Each column represents a different dispersion in black hole mass and Eddington-ratio. From \textit{left} to \textit{right}, $\sigma_{\log \Mbh}$ and $\sigma_{\log \lamEdd}$ is increased from 0 to 0.3 to 0.5. For each plot, \textit{green lines} show the intrinsic function assumed for the mock catalog, the \textit{blue data points} show the results from 1/\Vmax\, the \textit{blue dashed lines} show the MCMC fit to these data points, and the \textit{blue solid lines} show our attempt to recover the underlying distributions according to the method outlined in \S\ref{sec:bias}.
	The bias-corrected intrinsic distributions are a much better match to the mock input catalog than the 1/\Vmax\ results.}
\end{figure*}

\begin{figure*}[h]
	\centering
	\includegraphics[width=0.6\textwidth]{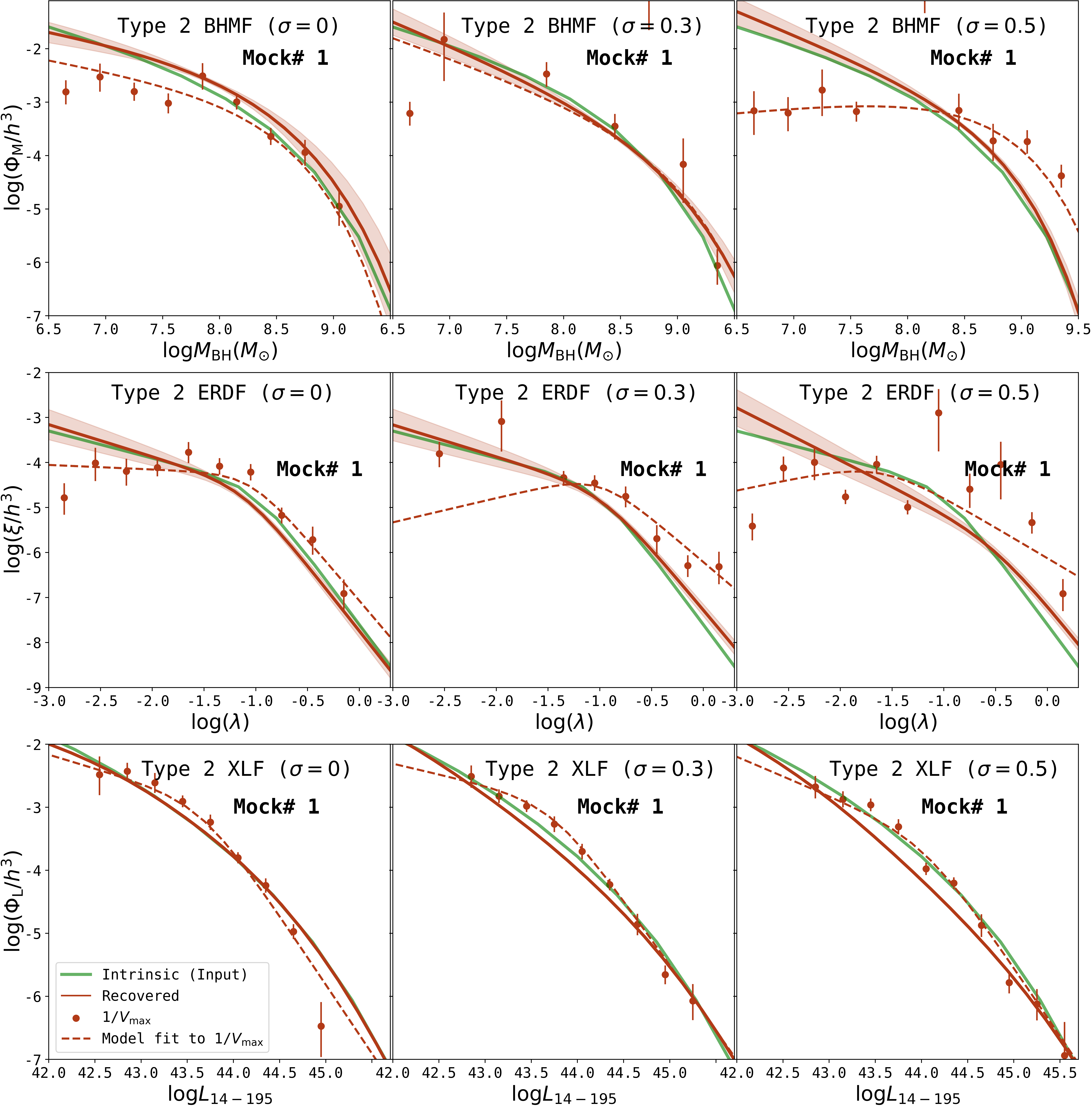} 
	\includegraphics[width=0.6\textwidth]{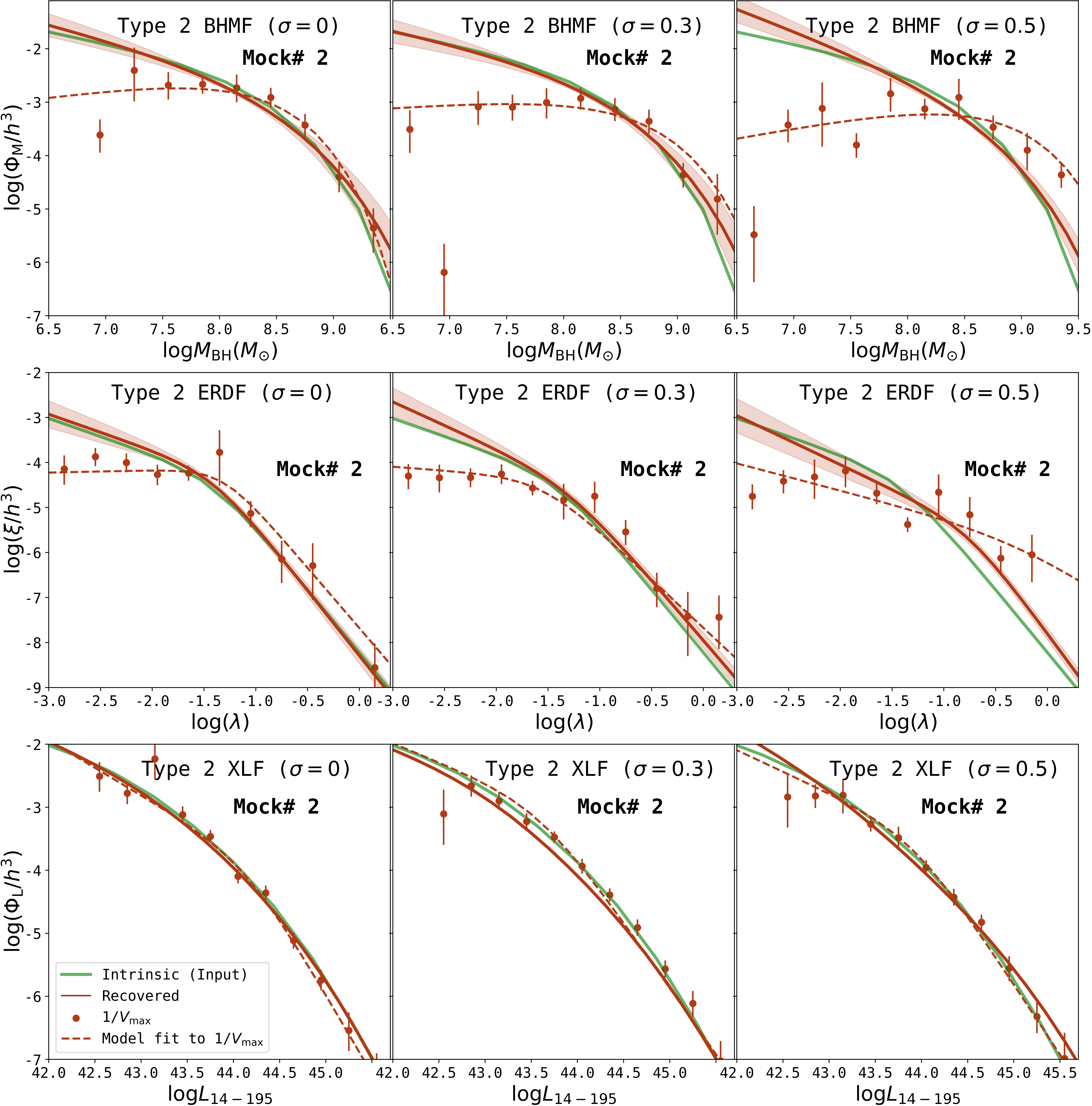}
	\caption{\label{fig:mock_catalog_type2} Results for Type~2 AGN 
	for two mock catalogs. For the top three panels (Mock\# 1), the parameters are fixed at $\log (\Phi^{*}/\rm{h^3\ Mpc^{-3}}) = -3.16$, $\log (M^*_{\rm BH}/\Msun) = 8$, $\alpha = -1.6$, $\beta = 0.6$, $\log (\xi^{*}/\rm{h^3\ Mpc^{-3}}) = -4.8$, $\log \lamEdd^* = -1$, $\delta_1 = 0.6$, $\epsilon_{\lamEdd} = 2.5$. For the bottom three panels (Mock\# 2), the parameters are fixed at $\log (\Phi^{*}/\rm{h^3\ Mpc^{-3}}) = -3.16$, $\log (M^*_{\rm BH}/\Msun) = 8.2$, $\alpha = -1.4$, $\beta = 0.7$, $\log (\xi^{*}/\rm{h^3\ Mpc^{-3}}) = -4.8$, $\log \lamEdd^* = -1.4$, $\delta_1 = 0.8$, $\epsilon_{\lamEdd} = 2$. Each column represents a different dispersion in black hole mass and Eddington-ratio; from \textit{left} to \textit{right}, $\sigma_{\log \Mbh}$ and $\sigma_{\log \lamEdd}$ are increased from 0 to 0.3 to 0.5. For each plot, \textit{green lines} show the intrinsic function assumed for the mock catalog, the \textit{red data points} show the results from 1/\Vmax\, the \textit{red dashed lines} show MCMC fit to these data points, and the \textit{red solid lines} show our attempt to recover the underlying distributions according to the method outlined in \S\ref{sec:bias}.
	As in the previous figure, the bias-corrected intrinsic distributions are a much better match to the mock input catalog than the 1/\Vmax\ results.}
\end{figure*}

\begin{figure*}[h]
    \centering
	\includegraphics[width=0.6\textwidth]{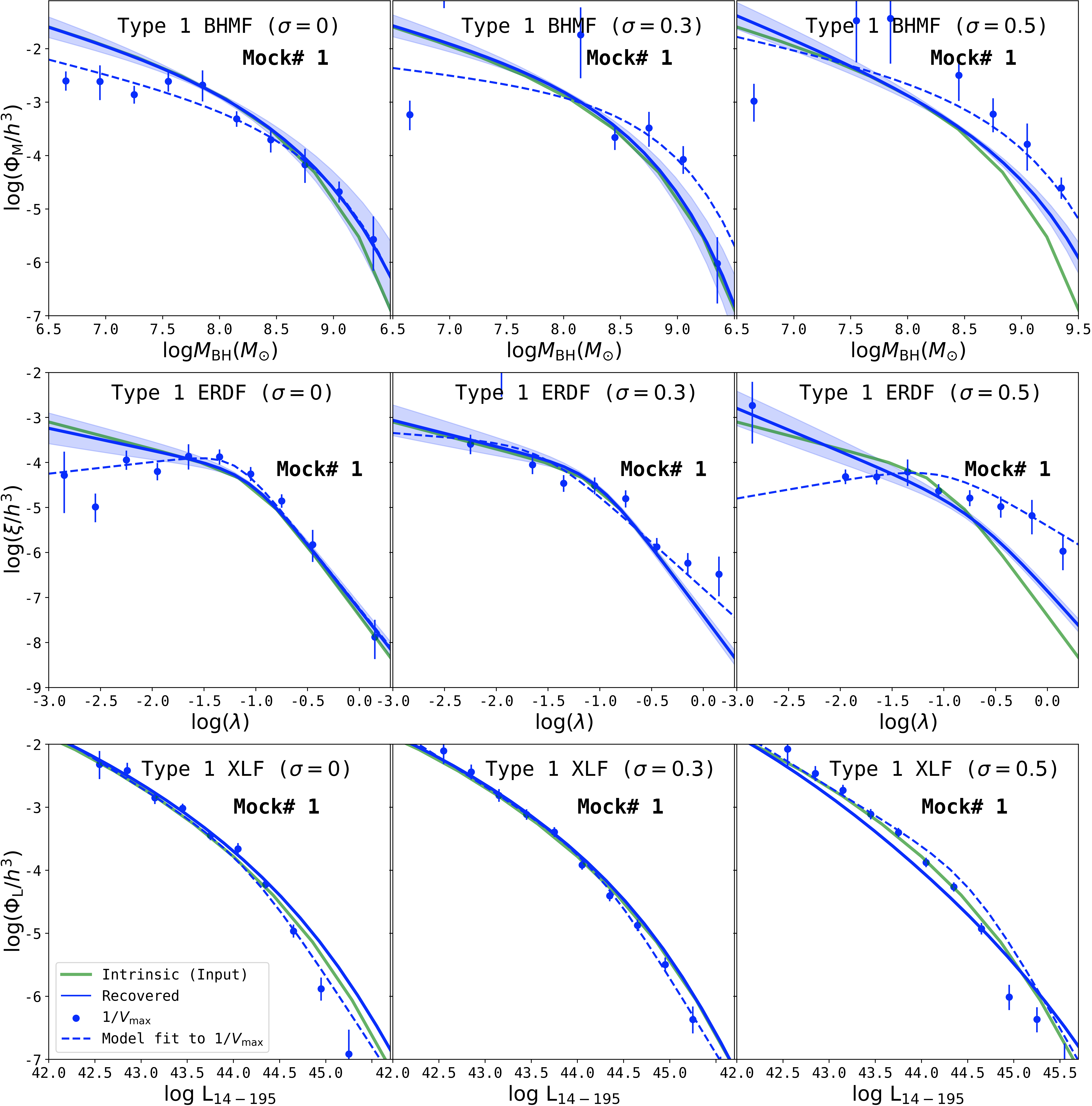}
	\includegraphics[width=0.6\textwidth]{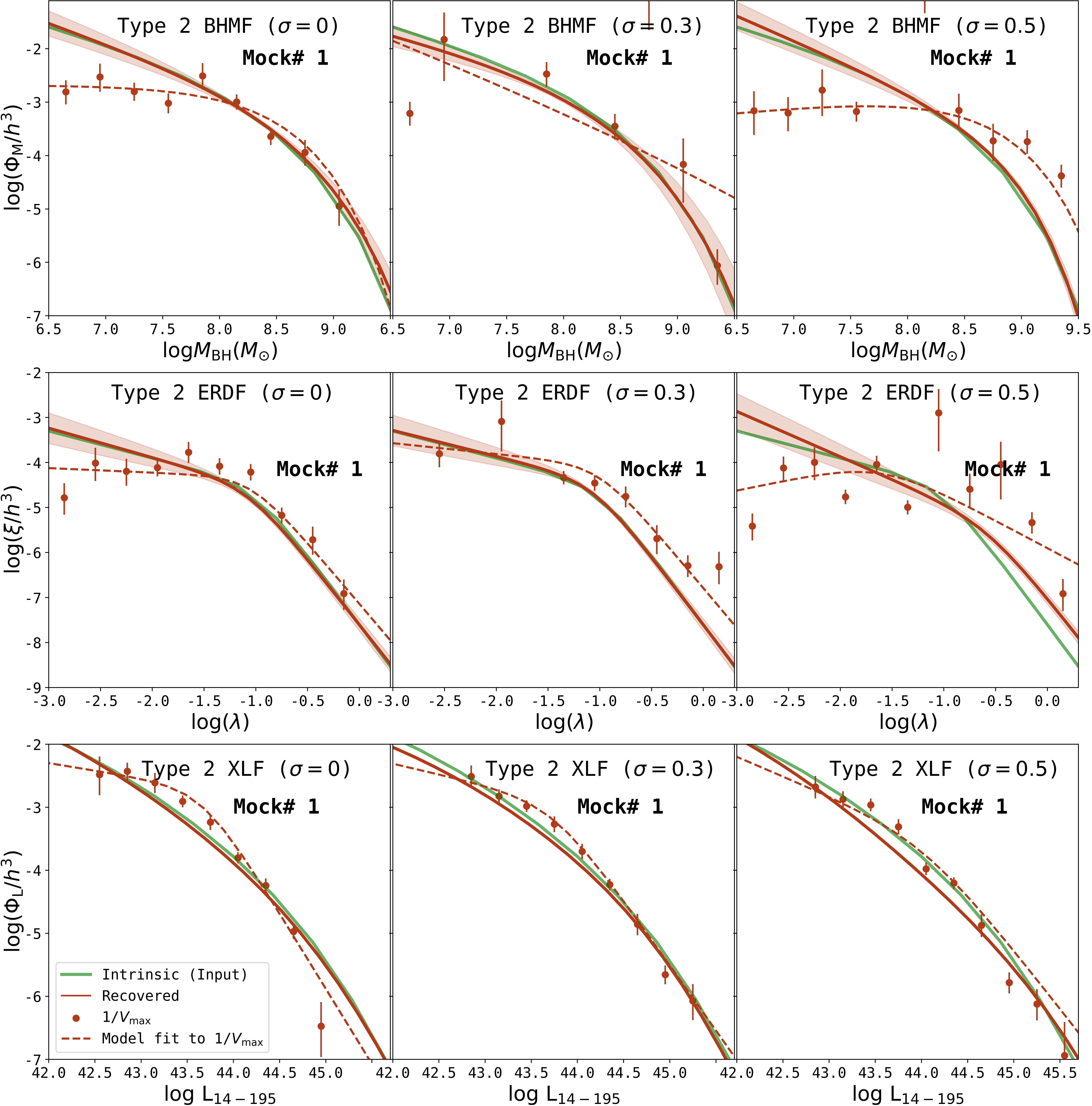}
	\caption{\label{fig:mock_catalog_variable_kbol} Results for Type~1 (top three rows) and Type~2 (bottom three rows) AGN 
	for mock catalog 1 with luminosity dependent bolometric correction \citep{Duras2020}. The intrinsic distributions are as described in Figures~\ref{fig:mock_catalog_type1} and \ref{fig:mock_catalog_type2} for mock 1. In this case, we also assume that the scattered intrinsic luminosity is scattered by $\sigma = 0.37$ and therefore the scatter on \lamEdd\ is $\sigma_{\log \lamEdd} = \sqrt{\sigma_{\log \Mbh}^2 + \sigma_{\rm \log L, scatt}^2} $. Each panel reports the assumed $\sigma = \sigma_{\log \Mbh} = 0, 0.3$ or $0.5$.	Even with these assumptions, bias-corrected intrinsic distributions are a much better match to the mock input catalog than the 1/\Vmax\ results.}
\end{figure*}

To test our approach to correct for sample truncation and to examine possible biases, we create mock catalogs. We assume the shapes of the BHMF and the ERDF. We randomly draw sources from the assumed distributions and subject them to selection effects. We then use this mock BASS survey as input, follow the steps outlined above, and test if our method allows us to recover the initial input distributions.

Besides the bias correction, this approach also allows us to test our ability to recover the XLF. The bolometric luminosity function is given by the convolution of the BHMF and the ERDF \cite[e.g.,][]{Weigel:2017aa}. We use equation \ref{eq:bhmflbol} and define the bolometric luminosity function in the following way:
\begin{equation}\label{eq:bhmfconvolution}
\Phi_{\rm L}(\log L_{\rm bol}) = \int_{\log \lambda_{\rm E, min, s}}^{\log \lambda_{\rm E, max, s}} \Phi_{\rm M}(\log L_{\rm bol} - \log \lamEdd -\lboltolam) \xi_{\rm norm}(\log \lamEdd) d\log \lamEdd .
\end{equation}
$\xi_{\rm norm}(\log \lamEdd)$ corresponds to the normalized ERDF, given by:
\begin{equation}
\xi_{\rm norm}(\log \lamEdd) = \frac{\xi(\log \lamEdd)}{\int_{\log \lambda_{\rm E, min, s}}^{\log \lambda_{\rm E, max, s}} \xi(\log \lamEdd) d\log \lamEdd} .
\end{equation}
After applying the bolometric correction (Equation~\ref{eq:l14195_to_lbol}), we are able to compare the XLF that we recover with the 1/\Vmax\ method to our prediction. 

To create a mock sample, we proceed according to the following steps:
\begin{itemize}
	\item $N_{\rm draw}$: To determine the sample size of our mock catalog, $N_{\rm draw}$, we integrate the assumed BHMF from $\log M_{\rm BH, min, s}$ to $\log M_{\rm BH, max, s}$ and multiply this space density with the comoving volume for the entire sky between $z_{\rm min, s}$ and $z_{\rm max, s}$. Note that to increase the sample size the simulated volume can be increased. 

	\item $\log \Mbh$: We construct the cumulative distribution function (CDF) of the assumed black hole mass function to randomly draw $\log \Mbh$ values. We draw $N_{\rm draw}$ values between 0 and 1 from a uniform distribution and invert the CDF to determine $\log \Mbh$. 

	\item $\log \lamEdd$: We use the CDF of the assumed ERDF to assign $N_{\rm draw}$ Eddington-ratio values. By drawing from the CDFs of the BHMF and the ERDF, the information on their assumed normalizations, $\Phi^{*}$ and $\xi^{*}$, is lost. Only $N_{\rm draw}$ affects the resulting normalizations. As we determine $N_{\rm draw}$ by integrating over the BHMF, the assumed $\xi^{*}$ is irrelevant. By construction, integrating over the predicted BHMF and ERDF will result in the same space densities. 

	\item $z$: We assign a redshift between $z_{\rm min, s}$ and $z_{\rm max, s}$ to each of the $N_{\rm draw}$ entries in our mock catalog. We assume that, within the considered redshift range, $\Phi_{\rm M}(\log \Mbh)$ and $\xi(\log \lamEdd)$ remain constant. Following \cite{Herbel:2017aa}, we draw $z$ values from a non-uniform distribution to account for the evolution of the comoving volume with $z$.

	\item $\log L_{\rm X}$: Once we have estimated $\log \Mbh$ and $\log \lamEdd$, we are able to compute bolometric luminosities, according to equation \ref{eq:bhmflbol}. By assuming a constant bolometric correction, \magenta{or alternatively following the luminosity dependent bolometric correction presented in \citet[][see Appendix~\ref{sec:app_variable_kbol} below]{Duras2020}}, we are able to translate these bolometric luminosities to (ultra-hard) X-ray luminosities $\log L_{\rm X}$. 
	\magenta{In the latter scenario, we assign a scatter of $\siglscatt = 0.37$ dex to the intrinsic luminosity \cite[see][]{Duras2020}, to account for the scatter in the bolometric to X-ray luminosity conversion, and calculate a scattered `intrinsic' luminosity $\log L_{\rm X,~scatt}$.}

	\item $\log \NH$: For Type~2 AGN, we draw a sample of $\log N_H$ with intrinsic distribution described in \citet{Ricci:2015aa}, as described in detail in \S\ref{sec:pi}, \magenta{based on scattered X-ray luminosity $\log L_{\rm X,~scatt}$.}
	
	\item \magenta{Mass estimation uncertainty effects: We add an additional (log-normal) scatter term, with a standard deviation between 0 and 0.5 dex (see Table \ref{tab:bhmfpara}) to the simulated $\log \Mbh$ 
	values, to account for the uncertainties related to determining \Mbh\ from observations. Note that these uncertainties are in fact dominated by systematic uncertainties inherent to the mass estimation methods.
	We then recalculate observed $\lambda_{\rm E,~obs}$ using the scattered mass and $\log L_{\rm X,~scatt}$ values. These values are used for computing the unbiased ERDF and BHMF.}
	
	\item Selection effects: Finally, to create a realistic mock catalog, we expose the simulated sources to selection effects. The flux limit of the BASS survey represents the most prominent bias. Using $\log L_{\rm X}$, $z$ and $\log N_{\rm H}$, we compute the hard X-ray flux of each source in our mock catalog. For each object $i$, we randomly draw a value between 0 and 1 from a uniform distribution. If this random value lies below $\Omega_{\rm sel}(\log F_{\mathrm{X}, i}$), the source remains in the sample. We eliminate all other sources which, according to the flux-area curve, are too faint to be included in the sample.

\end{itemize}

In Figures~\ref{fig:mock_catalog_type1} and \ref{fig:mock_catalog_type2} we show two examples for our random draw test. We assume the following initial BHMF and ERDF parameters: 
for Mock\# 1, $\log (\Phi^{*}/\rm{h^3\ Mpc^{-3}}) = -3.16$, $\log (M^*_{\rm BH}/\Msun) = 8$, $\alpha = -1.6$, $\beta = 0.6$, $\log (\xi^{*}/\rm{h^3\ Mpc^{-3}}) = -4.8$, $\log \lamEdd^* = -1$, $\delta_1 = 0.6$, $\epsilon_{\lamEdd} = 2.5$. For Mock\# 2, $\log (\Phi^{*}/\rm{h^3\ Mpc^{-3}}) = -3.16$, $\log (M^*_{\rm BH}/\Msun) = 8.2$, $\alpha = -1.4$, $\beta = 0.7$, $\log (\xi^{*}/\rm{h^3\ Mpc^{-3}}) = -4.8$, $\log \lamEdd^* = -1.4$, $\delta_1 = 0.8$, $\epsilon_{\lamEdd} = 2$. 
As mentioned above, the normalization of the ERDF ($\xi^*$) cannot be constrained with our random draw method. By construction, the integral over the ERDF corresponds to the integral over the BHMF. In all panels of the figure, we illustrate the marginalized probability distribution functions and give the recovered, best-fitting, bias-corrected BHMF and ERDF parameters.


For both mock catalogs, the effect of adding uncertainty is evident at the high mass end of the BHMF and the high \lamEdd\ end of the ERDF: both distributions are steep and objects are scattered into higher $\Mbh$ and \lamEdd\ bins as dispersions are increased (left to right in Figures~\ref{fig:mock_catalog_type1} and \ref{fig:mock_catalog_type2}).
The Figures show that our method allows us to recover the initial input functions for $\sigma_{\log \Mbh}$ and $\sigma_{\log \lambda_{\rm E}}$ between $0-0.5$. 
\magenta{In Figure~\ref{fig:mock_catalog_variable_kbol}, we use a luminosity dependent bolometric correction from \citet{Duras2020}, and assume a scatter in the bolometric correction of $\siglscatt = 0.37$. 
Therefore, the scatter on $\log \lamEdd$ is $\sigma_{\log \lamEdd} = \sqrt{\sigma_{\log \Mbh}^2 + \siglscatt^2} \simeq 0.37{-}0.62$ in that case. This high value of \siglscatt (and resulting $\sigma_{\log \lamEdd}$) is somewhat extreme, but serves to demonstrate the effect of having a large uncertainty in luminosity due to measurement uncertainty and bolometric correction.} 
The 1/\Vmax\ values of the XLF are consistent with what we expect from the convolution. For the BHMF and the ERDF, the 1/\Vmax\ values are consistent with the assumed input functions at high masses and Eddington-ratios. At low and high \Mbh\ and \lamEdd\ values, the effect of sample truncation is evident. The bias-corrected intrinsic function (shown with red and blue solid lines) is better at recovering the true underlying function than the 1/\Vmax\ data points and fits at this end as well.

\magenta{
\section{Luminosity Dependent Bolometric Correction}\label{sec:app_variable_kbol}

In the main part of the paper, we use Equation~\ref{eq:l14195_to_lbol} to straightforwardly convert between X-ray and bolometric luminosities (and vice versa) during the forward modeling. 
We also use this conversion to calculate the Eddington ratios of our AGNs from their BH masses and X-ray luminosities. 
Here we explore the effects of using, instead, a luminosity-dependent bolometric correction, as supported by several studies \cite[e.g.,][and references therein]{Marconi:2004aa,Hopkins2007_QLF_obs,Lusso2012,Duras2020}. 
Specifically, we choose to use the recently published prescriptions of \citet{Duras2020}, which are based on a large sample of AGNs covering a broad range in redshift and luminosity (including the Swift-BAT AGNs at low redshifts). 

To convert $\log L_{\rm bol}$ to $\log L_{\rm 2{-}10\,kev}$ we apply a bolometric correction $\log \kappa_{\rm 2{-}10\,kev} \equiv L_{\rm bol} / L_{\rm 2{-}10\,kev}$, which following \cite{Duras2020} is derived using the functional form
\begin{equation}\label{eq:durasKbol}
\kappa_{\rm 2{-}10\,kev} (L_{\rm bol}) = a \left[ 1+\left( \frac{\log(L_{\rm bol}/L_{\odot})}{b} \right)^c \right] \, .
\end{equation}
Similarly, to convert from $\log L_{\rm 2{-}10\,kev}$ to $\log L_{\rm bol}$ (i.e., to calculate \lamEdd) we use the functional form:
\begin{equation}\label{eq:durasK210}
\kappa_{\rm 2{-}10\,kev} (L_{X}) = a \left[ 1+\left( \frac{\log(L_{X}/L_{\odot})}{b} \right)^c \right] \, .
\end{equation} 
For both forms of conversion, the parameters $a$, $b$ and $c$ are taken from Table~1 of \citet{Duras2020}. In Equation~\ref{eq:durasKbol}, we use the prescribed variables for each AGN type (i.e., Type~1 and Type~2). 
As our main analysis relies on {\it ultra-hard} X-ray luminosities relevant for the BAT survey, we use the conversion from $\log L_{\rm 2{-}10\,kev}$ to $\log L_{\rm 14{-}195\,kev}$:
\begin{equation}
    \log L_{\rm 14{-}195\,kev} = \log L_{\rm 2{-}10\,kev} + 0.39
\end{equation}
This conversion factor is the median difference between intrinsic 2--10 keV and 14--195 keV luminosity of BAT sources within our main sample. A model dependent approach would consider the distribution of X-ray spectral shapes and/or the measured $2{-}10$ keV luminosities of BASS/DR2 AGNs (\citealt{Ricci2017_Xray_cat}; see also \citealt{ananna2020aa}). However, exploring a more complex conversion in this step is beyond the scope of this work, and would make a comparison to our main analysis, which uses a constant bolometric correction, rather ambiguous.

We next demonstrate the key results of our analysis using this luminosity dependent bolometric correction. We present one set of results assuming $\siglscatt = 0$ for comparison with our main results, and additional results assuming a scatter in luminosity of 0.37 dex - which reflects the scatter in $\kappa_{\rm 2{-}10\,kev}$ found by \citet[][see their Table~1]{Duras2020}. As this systematic uncertainty in luminosity is rather high, we consider it as the total uncertainty in luminosity (i.e., the measurement uncertainty is assumed to be negligible). As noted in Appendix \ref{sec:app_test} above, this scatter in luminosity adds in quadrature to the uncertainty in BH masses, and translates into a higher total uncertainty on \lamEdd\ (due to the direct dependence on \Lbol):  $\sigma_{\log \lamEdd} (= \sqrt{\sigma_{\log \Mbh}^2 + \siglscatt^2}) = 0.48{-}0.62$.

\begin{figure*}
	\centering
	\includegraphics[width=0.5\textwidth]{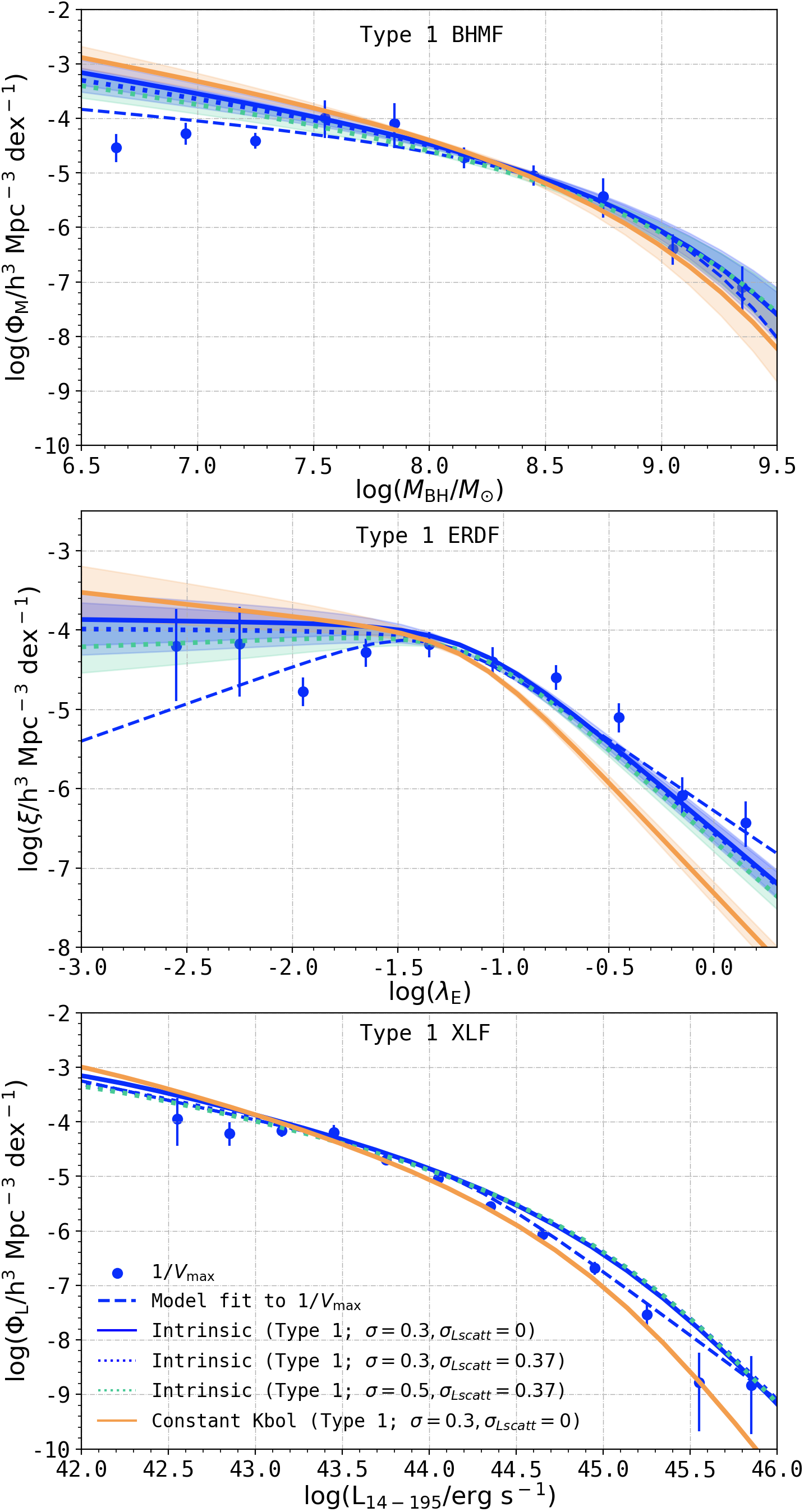}~
	\includegraphics[width=0.5\textwidth]{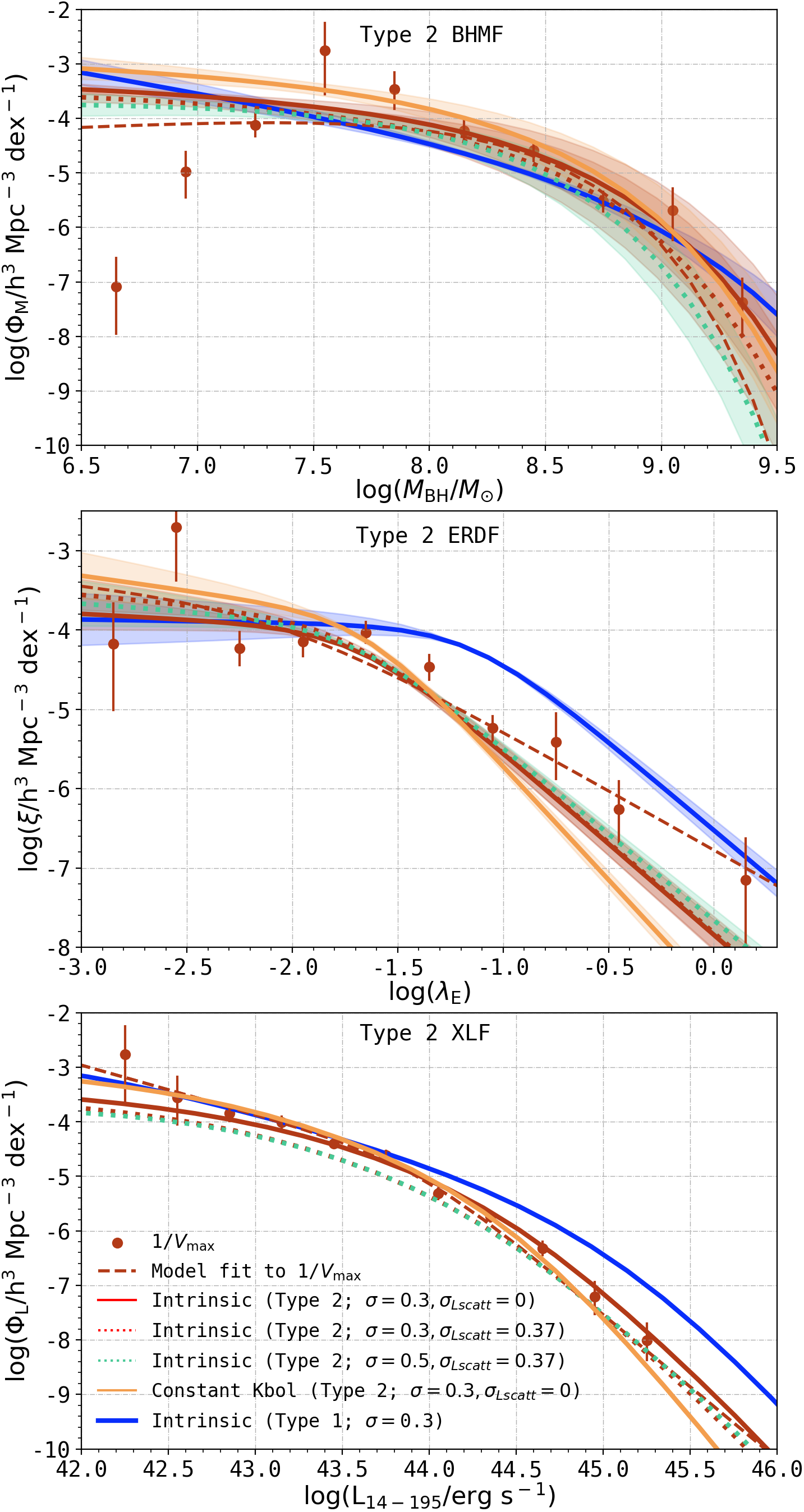}
	\caption{\label{fig:duras_type12_results} The BHMF, ERDF and reconstructed XLF of BASS/DR2 AGN, derived using a luminosity-dependent bolometric correction.
	The distributions of BH mass, Eddington ratio, and luminosity are shown in the {\it top}, {\it middle} and {\it bottom} panels, respectively, for both Type~1 \& Type~2 AGNs in BASS/DR2 ({\it left} and {\it right}).
	These are calculated using the \magenta{luminosity dependent bolometric corrections of  \cite{Duras2020}}.
	For each type of AGNs, combining the BHMF ({\it top} row) and ERDF ({\it middle} row) reproduces the observed XLF ({\it bottom} row). 
	{\it Data points} come from the 1/\Vmax\ analysis and {\it dashed lines} represent fits to these {\it data points}. 
	The {\it solid lines} and {\it shaded areas} show the final, bias-corrected intrinsic distribution functions. 
	\magenta{We illustrate intrinsic distributions derived assuming uncertainties of $\sigma_{\log \Mbh} = \sigma_{\lamEdd} 0.3$ and $\sigma_{\log \Mbh} = 0.3,~\siglscatt = 0.37$ (i.e., $\sigma_{\log \lamEdd} = 0.47$ dex)  with \textit{blue solid} and \textit{dotted} lines for Type 1 AGN, respectively (and red solid and dotted line for Type 2 AGN). The \textit{green dotted lines} in left (right) panels are calculated assuming $\sigma_{\log \Mbh} = 0.5,~\siglscatt = 0.37$ (i.e., $\sigma_{\log \lamEdd} = 0.62$ dex) for Type~1 (Type~2) AGN.
	The best-fit distributions of Type~1 AGNs (for $\sigma_{\log \Mbh} = \sigma_{\log \lamEdd} = 0.3$ case) are shown also in the {\it right} panels, to highlight differences between Type~1 and 2 AGNs. Both panels are also overplotted with constant bolometric correction results (\textit{orange solid line}) from the main part of the text ($\sigma_{\log \Mbh} = \sigma_{\log \lamEdd} = 0.3$ case) for comparison.}
	The 1/\Vmax\ values (split by AGN type) are reported in Tables~\ref{tab:bhmfxlfvmaxxlf}, \ref{tab:bhmfvmax} and \ref{tab:erdfvmax}.}
\end{figure*}

\begin{figure}
	\centering
	\includegraphics[width=0.5\textwidth]{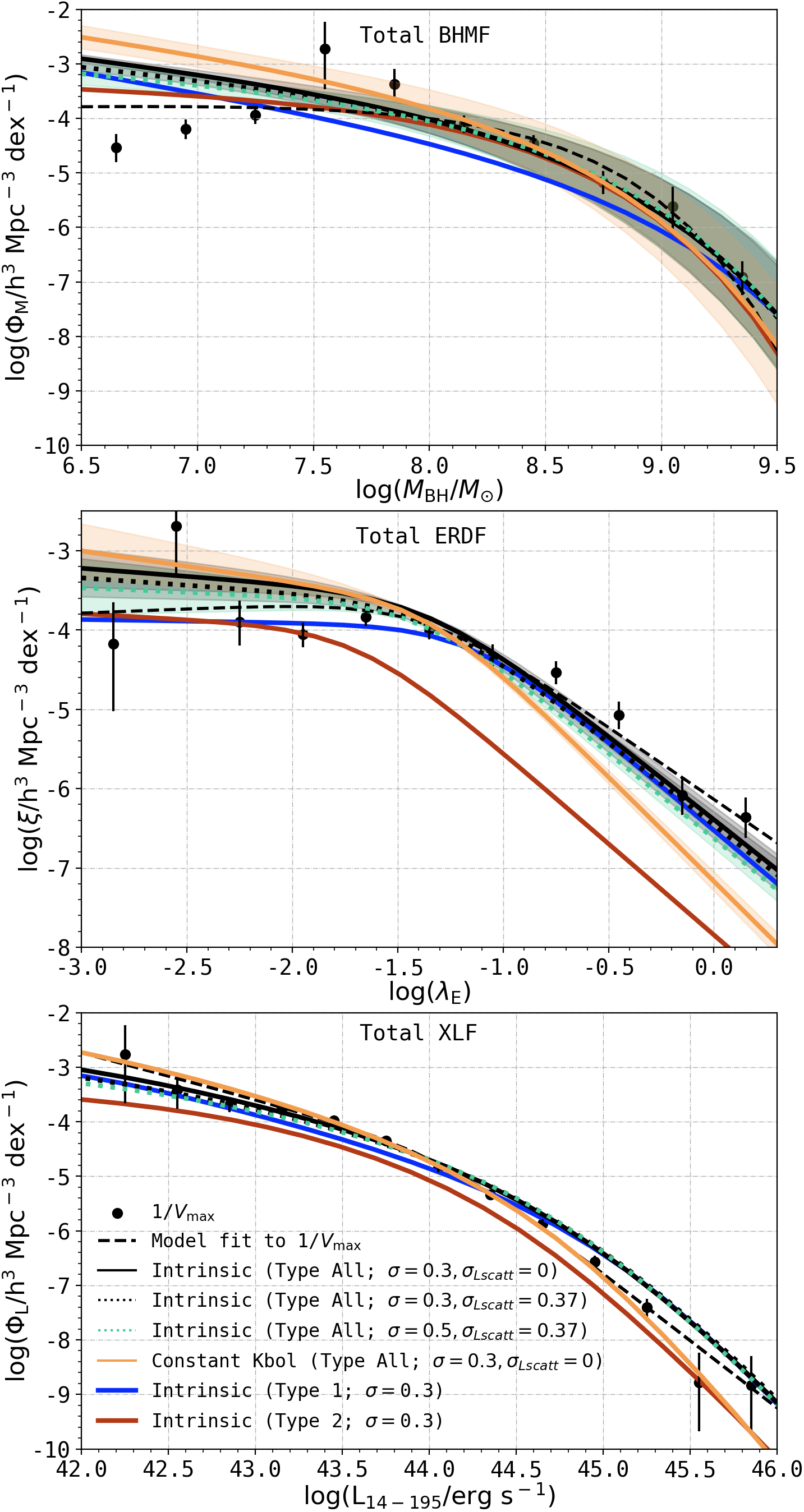}
	\caption{\label{fig:duras_typeall_results} As in Fig.~\ref{fig:duras_type12_results}, the BHMFs, ERDFs and reconstructed XLFs are shown in the {\it top}, {\it middle} and {\it bottom} panels, respectively, for both Type~1 \& Type~2 AGNs in BASS/DR2 ({\it left} and {\it right}). These functions are calculated assuming a luminosity dependent bolometric correction \citep{Duras2020}.  
	We show the intrinsic distributions derived for all AGNs assuming uncertainties of either $\sigma_{\log \Mbh} = 0.3$ and $\sigma_{\log \lamEdd} = 0.47$ dex or $\sigma_{\log \Mbh} = 0.5$ and $\sigma_{\log \lamEdd} = 0.62$ dex in ({\it black} and {\it green} lines, respectively).
	The {\it black data points} in the bottom panel show the direct 1/\Vmax\ XLF estimates, and the {\it black dashed line} is the fit to those points. 
	{\it Solid black lines} show the final, bias-corrected, intrinsic distribution functions for the complete AGN sample.}
\end{figure}


In Tables~\ref{tab:bhmfbhmferdffitBHMF_app} and \ref{tab:bhmfbhmferdffitERDF_app}, we tabulate the results derived using the \citet{Duras2020} luminosity-dependent bolometric correction. 
In Figures~\ref{fig:duras_type12_results} and \ref{fig:duras_typeall_results}, we plot the results along with $\sigma_{\log \Mbh} = \sigma_{\log \lamEdd} = 0.3$ results from the main analysis for comparison. These figures show that the \citet{Duras2020} bolometric correction prescription produces significantly different ERDF shapes compared to those derived using a constant bolometric correction (shown using orange solid line in these plots), especially at high $\lamEdd$ values. 
However, we stress that our key conclusion concerning the difference in characteristic break Eddington ratio (\lamEddstar) between Type 1 and Type 2 AGNs still holds for this bolometric correction prescription. 
Specifically, with the \citet{Duras2020} bolometric corrections we obtain $\log\lamEddstar = $ \erdfloglamstartypeonedurassigLzero\ and \erdfloglamstartypetwodurassigLzero\ for Type 1 and Type 2 AGNs, respectively.
The latter value is, again, remarkably consistent with what is expected from the radiation-driven unification scenario, as discussed in Section~\ref{sec:discussion_agn_obsc} (see \citealp{Fabian2009_feedback, Ricci2017_Nat}).
On the other hand, when experimenting with luminosity-dependent bolometric corrections, we note that varying $\sigma_{\log \Mbh}$ and $\siglscatt$ does not significantly change the results. 

Given the wide range of bolometric corrections investigated in the literature, and the complexity of applying such prescriptions when deriving the {\it intrinsic} distributions of key AGN properties, we prefer not to recommend one set of bolometric corrections in this work. Instead, we present the constant bolometric correction results in the main part of the text due to its simplicity, and leave it to the discretion of the reader to choose to use either those results, or those based on the \cite{Duras2020} prescription, presented in this Appendix. 


\begin{deluxetable*}{lclll}
\label{tab:bhmfbhmferdffitBHMF_app}
\tablecaption{Sample truncation corrected BHMF$^{\rm a}$, and fit to /\Vmax\ values for the BHMF for Type~1, Type~2 AGN, and both samples together. All results were calculated assuming \citet{Duras2020} bolometric correction.} 
\tablewidth{0pt}
\tablehead{
\colhead{} & \colhead{$\log (M_{\rm BH}^{*}/\Msun)$} & \colhead{$\log (\Psi^{*}/\rm{h^{3}\ Mpc^{-3}})$} & \colhead{$\alpha$} & \colhead{$\beta$} }
\startdata
        All & & & & \\
        Intrinsic (Variable $\kappa_{\rm bol}$; $\sigma = 0.3,~\sigma_{\rm Lscatt} = 0$) & \bhmflogmstardurassigLzero & \bhmflogphistardurassigLzero & \bhmfalphadurassigLzero & \bhmfbetadurassigLzero \\
		Intrinsic (Variable $\kappa_{\rm bol}$; $\sigma = 0.3,~\sigma_{\rm Lscatt} = 0.37$) & \bhmflogmstarduras & \bhmflogphistarduras & \bhmfalphaduras & \bhmfbetaduras \\
		Intrinsic (Variable $\kappa_{\rm bol}$; $\sigma = 0.5,~\sigma_{\rm Lscatt} = 0.37$) & \bhmflogmstarsigfiveduras & \bhmflogphistarsigfiveduras & \bhmfalphasigfiveduras & \bhmfbetasigfiveduras \\
		\hline
		\hline
		Type~1 & & & & \\
		Intrinsic (Variable $\kappa_{\rm bol}$; $\sigma = 0.3,~\sigma_{\rm Lscatt} = 0$) & \bhmflogmstartypeonedurassigLzero & \bhmflogphistartypeonedurassigLzero & \bhmfalphatypeonedurassigLzero & \bhmfbetatypeonedurassigLzero \\
		Intrinsic (Variable $\kappa_{\rm bol}$; $\sigma = 0.3,~\sigma_{\rm Lscatt} = 0.37$) & \bhmflogmstartypeoneduras & \bhmflogphistartypeoneduras & \bhmfalphatypeoneduras & \bhmfbetatypeoneduras \\
		Intrinsic (Variable $\kappa_{\rm bol}$; $\sigma = 0.5,~\sigma_{\rm Lscatt} = 0.37$) & \bhmflogmstartypeonesigfiveduras & \bhmflogphistartypeonesigfiveduras & \bhmfalphatypeonesigfiveduras & \bhmfbetatypeonesigfiveduras \\
		1/\Vmax\ & {\bhmfvmaxtyponelogMbh} & {\bhmfvmaxtyponephi} & {\bhmfvmaxtyponealpha} & {\bhmfvmaxtyponebeta}\\
		\hline
		Type~2 & & & & \\
		Intrinsic (Variable $\kappa_{\rm bol}$; $\sigma = 0.3,~\sigma_{\rm Lscatt} = 0$) & \bhmflogmstartypetwodurassigLzero & \bhmflogphistartypetwodurassigLzero & \bhmfalphatypetwodurassigLzero & \bhmfbetatypetwodurassigLzero \\
		Intrinsic (Variable $\kappa_{\rm bol}$; $\sigma = 0.3,~\sigma_{\rm Lscatt} = 0.37$) & \bhmflogmstartypetwoduras & \bhmflogphistartypetwoduras & \bhmfalphatypetwoduras & \bhmfbetatypetwoduras \\
		Intrinsic (Variable $\kappa_{\rm bol}$; $\sigma = 0.5,~\sigma_{\rm Lscatt} = 0.37$) & \bhmflogmstartypetwosigfiveduras & \bhmflogphistartypetwosigfiveduras & \bhmfalphatypetwosigfiveduras & \bhmfbetatypetwosigfiveduras \\
		1/\Vmax\ & {\bhmfvmaxtyptwologMbh} & {\bhmfvmaxtyptwophi} & {\bhmfvmaxtyptwoalpha} & {\bhmfvmaxtyptwobeta}\\
\enddata
\tablenotetext{a}{We assume a modified Schechter function (see equation \ref{eq:bhmf_msch_def}) for the BHMF.}
\end{deluxetable*}

\begin{deluxetable*}{lclll}
\label{tab:bhmfbhmferdffitERDF_app}
\tablecaption{Sample truncation-corrected ERDF$^{\rm a}$, and fit to /\Vmax\ values for the ERDF for Type~1 AGN, Type~2 AGN and the full AGN sample. All results were calculated assuming \citet{Duras2020} bolometric correction.}
\tablewidth{0pt}
\tablehead{
\colhead{} & \colhead{$\log \lamEdd^{*}$} & \colhead{$\log (\xi^{*}/\rm{h^3\ Mpc^{-3}})$} & \colhead{$\delta_1$} & \colhead{$\epsilon_{\lambda}$} }
\startdata
		All & & & & \\
		Intrinsic (Variable $\kappa_{\rm bol}$; $\sigma = 0.3,~\sigma_{\rm Lscatt} = 0$) &{\erdfloglamstardurassigLzero} & {\erdflogxistardurassigLzero} & {\erdfdeltaadurassigLzero} & \erdfepislonlamdurassigLzero \\
		Intrinsic (Variable $\kappa_{\rm bol}$; $\sigma = 0.3,~\sigma_{\rm Lscatt} = 0.37$) &{\erdfloglamstarduras} & {\erdflogxistarduras} & {\erdfdeltaaduras} & {\erdfepislonlamduras} \\
		Intrinsic (Variable $\kappa_{\rm bol}$; $\sigma = 0.5,~\sigma_{\rm Lscatt} = 0.37$) & {\erdfloglamstarsigfiveduras} & {\erdflogxistarsigfiveduras} & {\erdfdeltaasigfiveduras} & {\erdfepislonlamsigfiveduras} \\
		1/\Vmax\ & {\erdfvmaxalllam} & {\erdfvmaxallphi} & {\erdfvmaxallgamma} & {\erdfvmaxalleps} \\
		\hline
		Type~1 & & & & \\
		Intrinsic (Variable $\kappa_{\rm bol}$; $\sigma = 0.3,~\sigma_{\rm Lscatt} = 0$) &{\erdfloglamstartypeonedurassigLzero} & {\erdflogxistartypeonedurassigLzero} & {\erdfdeltaatypeonedurassigLzero} & {\erdfepislonlamtypeonedurassigLzero} \\
		Intrinsic (Variable $\kappa_{\rm bol}$; $\sigma = 0.3,~\sigma_{\rm Lscatt} = 0.37$) &{\erdfloglamstartypeoneduras} & {\erdflogxistartypeoneduras} & {\erdfdeltaatypeoneduras} & {\erdfepislonlamtypeoneduras} \\
		Intrinsic (Variable $\kappa_{\rm bol}$; $\sigma = 0.5,~\sigma_{\rm Lscatt} = 0.37$) & {\erdfloglamstartypeonesigfiveduras} & {\erdflogxistartypeonesigfiveduras} & {\erdfdeltaatypeonesigfiveduras} & {\erdfepislonlamtypeonesigfiveduras} \\
		1/\Vmax & {\erdfvmaxtyponelam} & {\erdfvmaxtyponephi} & {\erdfvmaxtyponegamma} & {\erdfvmaxtyponeeps}  \\
		\hline
		Type~2 & & & & \\
		Intrinsic (Variable $\kappa_{\rm bol}$; $\sigma = 0.3,~\sigma_{\rm Lscatt} = 0$) &{\erdfloglamstartypetwodurassigLzero} & {\erdflogxistartypetwodurassigLzero} & {\erdfdeltaatypetwodurassigLzero} & {\erdfepislonlamtypetwodurassigLzero} \\
		Intrinsic (Variable $\kappa_{\rm bol}$; $\sigma = 0.3,~\sigma_{\rm Lscatt} = 0.37$) &{\erdfloglamstartypetwoduras} & {\erdflogxistartypetwoduras} & {\erdfdeltaatypetwoduras} & {\erdfepislonlamtypetwoduras} \\
		Intrinsic (Variable $\kappa_{\rm bol}$; $\sigma = 0.5,~\sigma_{\rm Lscatt} = 0.37$) & {\erdfloglamstartypetwosigfiveduras} & {\erdflogxistartypetwosigfiveduras} & {\erdfdeltaatypetwosigfiveduras} & {\erdfepislonlamtypetwosigfiveduras} \\
		1/\Vmax\ & {\erdfvmaxtyptwolam} & {\erdfvmaxtyptwophi} & {\erdfvmaxtyptwogamma} & {\erdfvmaxtyptwoeps} \\
\enddata
\tablenotetext{a}{We assume a double power-law shape for the ERDF (see equation \ref{eq:erdf_dbpl_def}).}
\end{deluxetable*}

} 

\section{Figures and Tables of Key Results}\label{sec:app_tables}



\begin{figure*}[h]
	\centering
	\includegraphics[width=0.9\textwidth]{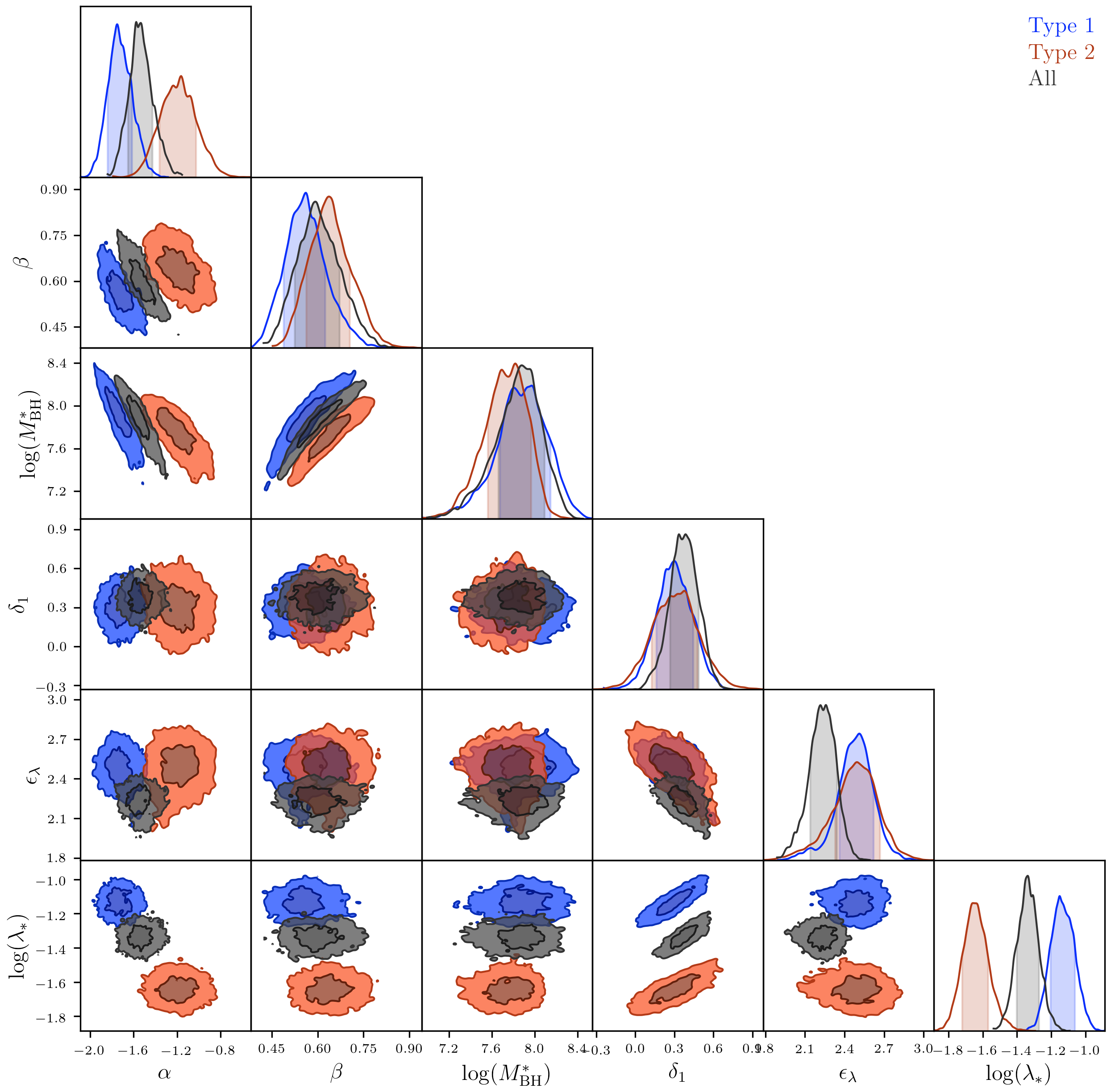}
	\caption{\label{fig:chain_overlap} The three probability chains (Type~1 in \textit{blue}, Type~2 in \textit{red}, overall in \textit{black}) plotted together.}
\end{figure*}

\begin{deluxetable*}{lclll}
\label{tab:bhmfbhmferdffitERDFmassdep}
\tablecaption{Sample truncation-corrected ERDF for Type~1 AGN and Type~2 AGN in low ($\log \Mbh \leq 7.8$) and high mass ($\log \Mbh \geq 8.2$) bins.}
\tablewidth{0pt}
\tablehead{
\colhead{} & \colhead{$\log \lamEdd^{*}$} & \colhead{$\log (\xi^{*}/\rm{h^3\ Mpc^{-3}})$} & \colhead{$\delta_1$} & \colhead{$\epsilon_{\lambda}$} }
\startdata
		Type~1 & & & & \\
		Intrinsic Low Mass & {\erdfloglamstarlowmasstypeone} & {\erdflogxistarlowmasstypeone} & {\erdfdeltaalowmasstypeone} & {\erdfepislonlamlowmasstypeone} \\
		Intrinsic High Mass & {\erdfloglamstarhighmasstypeone} & {\erdflogxistarhighmasstypeone} & {\erdfdeltaahighmasstypeone} & {\erdfepislonlamhighmasstypeone} \\
		\hline
		Type~2 & & & & \\
		Intrinsic Low Mass & {\erdfloglamstarlowmasstypetwo} & {\erdflogxistartypetwo} & {\erdfdeltaalowmasstypetwo} & {\erdfepislonlamlowmasstypetwo} \\
		Intrinsic High Mass & {\erdfloglamstarhighmasstypetwo} & {\erdflogxistarhighmasstypetwo} & {\erdfdeltaahighmasstypetwo} & {\erdfepislonlamhighmasstypetwo} \\
\enddata
\end{deluxetable*}

\begin{deluxetable*}{lllll|llll|llll} 
\label{tab:bhmfxlfvmaxxlf}
\tablecaption{1/\Vmax values of the XLFs for all AGN selected by the cuts described in Table~\ref{tab:bhmfpara}. Here we also present the 1/\Vmax for the Type~1 and Type~2 samples separately, which together make the overall sample. Note that to plot these data points, the bin size $d\log L_{\rm X} = \binxlf$ dex has to be subtracted from the $\log \Phi_{\rm L}$ values.}
\tablewidth{0pt}
\tablehead{\colhead{\textbf{All}} & \colhead{} & \colhead{} & \colhead{} & \colhead{} & \colhead{\textbf{Type~1}} & \colhead{} & \colhead{} & \colhead{} & \colhead{\textbf{Type~2}} & \colhead{} & \colhead{} & \colhead{} \\[-0.3cm]
\colhead{$\log (L_{\rm X}/\rm erg\ s^{-1})$} & \colhead{{$N$}} & \colhead{$\log (\Phi_{\rm L})$} & \colhead{$\sigma_{\rm{up}}$} & \colhead{$\sigma_{\rm{low}}$} & \colhead{$N$} & \colhead{$\log (\Phi_{\rm L})$} & \colhead{$\sigma_{\rm{up}}$} & \colhead{$\sigma_{\rm{low}}$} & \colhead{$N$} & \colhead{$\log (\Phi_{\rm L})$} & \colhead{$\sigma_{\rm{up}}$} & \colhead{$\sigma_{\rm{low}}$}}
\startdata
42.25 & 1 & -2.77 & 0.52 & -0.87 & & & & & 1 & -2.77 & 0.52 & -0.87 \\
42.55 & 5 & -3.41 & 0.29 & -0.34 & 2 & -3.95 & 0.37 & -0.47 & 3 & -3.55 & 0.38 & -0.5 \\
42.85 & 22 & -3.69 & 0.11 & -0.11 & 8 & -4.22 & 0.19 & -0.2 & 14 & -3.85 & 0.14 & -0.14 \\
43.15 & 59 & -3.78 & 0.07 & -0.07 & 28 & -4.16 & 0.1 & -0.1 & 31 & -4.01 & 0.1 & -0.1 \\
43.45 & 89 & -3.99 & 0.08 & -0.08 & 53 & -4.2 & 0.12 & -0.12 & 36 & -4.41 & 0.09 & -0.09 \\
43.75 & 111 & -4.35 & 0.05 & -0.05 & 62 & -4.7 & 0.06 & -0.06 & 49 & -4.61 & 0.08 & -0.08 \\
44.05 & 104 & -4.86 & 0.05 & -0.05 & 75 & -5.04 & 0.06 & -0.06 & 29 & -5.32 & 0.09 & -0.09 \\
44.35 & 86 & -5.34 & 0.05 & -0.05 & 56 & -5.56 & 0.06 & -0.06 & 30 & -5.74 & 0.09 & -0.09 \\
44.65 & 63 & -5.87 & 0.07 & -0.07 & 45 & -6.07 & 0.08 & -0.08 & 18 & -6.32 & 0.13 & -0.13 \\
44.95 & 32 & -6.57 & 0.1 & -0.1 & 26 & -6.69 & 0.1 & -0.1 & 6 & -7.21 & 0.27 & -0.31 \\
45.25 & 12 & -7.41 & 0.14 & -0.15 & 9 & -7.53 & 0.17 & -0.18 & 3 & -8.01 & 0.31 & -0.36 \\
45.55 & 1 & -8.78 & 0.52 & -0.87 & 1 & -8.78 & 0.52 & -0.87 & & & & \\
45.85 & 1 & -8.84 & 0.52 & -0.87 & 1 & -8.84 & 0.52 & -0.87 & & & & \\
\hline
{} & {586} & {} & {} & {} & {366} & {} & {} & {} & {220} & {} & {} & {} \\
\hline
\enddata
\end{deluxetable*}

\begin{deluxetable*}{lllll|llll|llll}
\label{tab:bhmfvmax}
\tablecaption{1/\Vmax\ values for the BHMF of all Table~\ref{tab:bhmfpara} selected AGN, Type~1 AGN and Type~2 AGN (uncorrected for sample truncation). The quantities in bracket shows the 1/\Vmax\ values after excluding one object that lies at the low flux threshold of the attenuation curve (discussed in \S\ref{sec:Vmax}).}
\tablewidth{0pt}
\tablehead{\colhead{\textbf{All}} & \colhead{} & \colhead{} & \colhead{} & \colhead{} & \colhead{\textbf{Type~1}} & \colhead{} & \colhead{} & \colhead{} & \colhead{\textbf{Type~2}} & \colhead{} & \colhead{} & \colhead{} \\[-0.3cm]
\colhead{$\log (\Mbh /\Msun)$} & \colhead{$N$} & \colhead{$\log (\Psi)$} & \colhead{$\sigma_{\rm{up}}$} & \colhead{$\sigma_{\rm{low}}$} & \colhead{$N$} & \colhead{$\log (\Psi)$} & \colhead{$\sigma_{\rm{up}}$} & \colhead{$\sigma_{\rm{low}}$} & \colhead{$N$} & \colhead{$\log (\Psi)$} & \colhead{$\sigma_{\rm{up}}$} & \colhead{$\sigma_{\rm{low}}$}}
\startdata
6.65 & 16 & -4.53 & 0.22 & -0.25 & 15 & -4.53 & 0.22 & -0.25 & 1 & -7.08 & 0.52 & -0.87 \\
6.95 & 27 & -4.2 & 0.16 & -0.17 & 23 & -4.28 & 0.18 & -0.19 & 4 & -4.98 & 0.37 & -0.47 \\
7.25 & 56 & -3.94 & 0.13 & -0.14 & 42 & -4.41 & 0.12 & -0.13 & 14 & -4.12 & 0.2 & -0.22 \\
7.55 & 76 (75) & -2.73 (-3.76) & 0.47 (0.19) & -0.72 (-0.2) & 48 & -3.99 & 0.3 & -0.35 & 28 (27) & -2.75 (-4.16) & 0.5 (0.17) & -0.8 (-0.18) \\
7.85 & 112 & -3.37 & 0.25 & -0.29 & 65 & -4.09 & 0.35 & -0.43 & 47 & -3.46 & 0.31 & -0.36 \\
8.15 & 116 & -4.1 & 0.14 & -0.14 & 68 & -4.72 & 0.17 & -0.17 & 48 & -4.22 & 0.18 & -0.19 \\
8.45 & 99 & -4.46 & 0.14 & -0.14 & 56 & -5.05 & 0.16 & -0.17 & 43 & -4.59 & 0.18 & -0.19 \\
8.75 & 52 & -5.17 & 0.18 & -0.19 & 28 & -5.43 & 0.31 & -0.37 & 24 & -5.52 & 0.17 & -0.18 \\
9.05 & 23 & -6.05 & 0.18 & -0.2 & 15 & -6.4 & 0.24 & -0.27 & 8 & -6.31 & 0.28 & -0.33 \\
9.35 & 8 & -6.91 & 0.27 & -0.31 & 5 & -7.1 & 0.36 & -0.45 & 3 & -7.37 & 0.43 & -0.6 \\
\hline
{} & {586} & {} & {} & {} & {366} & {} & {} & {} & {220} & {} & {} & {} \\
\enddata
\end{deluxetable*}

\begin{deluxetable*}{lllll|llll|llll}
\label{tab:erdfvmax}
\tablecaption{1/\Vmax\ values for the ERDF of all Table~\ref{tab:bhmfpara} selected AGN, Type~1 AGN and Type~2 AGN (uncorrected for sample truncation). The quantities in bracket shows the 1/\Vmax\ values after excluding one object that lies at the low flux threshold of the attenuation curve (discussed in \S\ref{sec:Vmax}).}
\tablewidth{0pt}
\tablehead{\colhead{\textbf{All}} & \colhead{} & \colhead{} & \colhead{} & \colhead{} & \colhead{\textbf{Type~1}} & \colhead{} & \colhead{} & \colhead{} & \colhead{\textbf{Type~2}} & \colhead{} & \colhead{} & \colhead{} \\[-0.3cm]
\colhead{$\log \lamEdd$} & \colhead{$N$} & \colhead{$\log (\xi)$} & \colhead{$\sigma_{\rm{up}}$} & \colhead{$\sigma_{\rm{low}}$} & \colhead{$N$} & \colhead{$\log (\xi)$} & \colhead{$\sigma_{\rm{up}}$} & \colhead{$\sigma_{\rm{low}}$} & \colhead{$N$} & \colhead{$\log (\xi)$} & \colhead{$\sigma_{\rm{up}}$} & \colhead{$\sigma_{\rm{low}}$}}
\startdata
-2.85 & 3 & -4.07 & 0.41 & -0.55 & & & & & 3 & -4.07 & 0.41 & -0.55 \\
-2.55 & 13 (12) & -2.7 (-3.52) & 0.45 (0.35) & -0.64 (-0.43) & 4 & -4.19 & 0.45 & -0.64 & 9 (8) & -2.71 (-3.62) & 0.46 (0.43) & -0.68 (-0.59) \\
-2.25 & 29 & -3.89 & 0.25 & -0.28 & 9 & -4.18 & 0.46 & -0.67 & 20 & -4.2 & 0.19 & -0.2 \\
-1.95 & 71 & -4.0 & 0.13 & -0.14 & 26 & -4.74 & 0.15 & -0.16 & 45 & -4.09 & 0.16 & -0.17 \\
-1.65 & 118 & -3.89 & 0.1 & -0.1 & 57 & -4.29 & 0.16 & -0.17 & 61 & -4.11 & 0.13 & -0.14 \\
-1.35 & 133 & -3.97 & 0.1 & -0.1 & 89 & -4.14 & 0.13 & -0.14 & 44 & -4.46 & 0.15 & -0.16 \\
-1.05 & 105 & -4.32 & 0.14 & -0.15 & 79 & -4.39 & 0.16 & -0.17 & 26 & -5.18 & 0.22 & -0.24 \\
-0.75 & 62 & -4.74 & 0.11 & -0.11 & 53 & -4.78 & 0.12 & -0.12 & 9 & -5.78 & 0.26 & -0.29 \\
-0.45 & 34 & -5.1 & 0.16 & -0.17 & 33 & -5.1 & 0.17 & -0.17 & 1 & -7.46 & 0.52 & -0.87 \\
-0.15 & 11 & -6.23 & 0.23 & -0.25 & 10 & -6.29 & 0.25 & -0.28 & 1 & -7.15 & 0.52 & -0.87 \\
0.15 & 6 & -6.53 & 0.28 & -0.32 & 6 & -6.53 & 0.28 & -0.32 & & & & \\
0.45 & 1 & -7.08 & 0.52 & -0.87 & & & & & 1 & -7.08 & 0.52 & -0.87 \\
\hline
{} & {586} & {} & {} & {} & {366} & {} & {} & {} & {220} & {} & {} & {} \\
\enddata
\end{deluxetable*}


\begin{deluxetable*}{lllll|llll|llll}
\label{tab:xlfvmaxallsamples}
\tablecaption{1/\Vmax values for XLFs shown in Figure~\ref{fig:xlfobs}. The ``All BASS AGN'' sample falls in the $z \leq 0.3$ range, and the other two samples fall within the $0.01 \leq z \leq 0.3$ range.}
\tablewidth{0pt}
\tablehead{\colhead{\textbf{All BASS AGN}} & \colhead{} & \colhead{} & \colhead{} & \colhead{} & \colhead{} & \colhead{\textbf{No Obs Corr}} & \colhead{} & \colhead{} & \colhead{} & \colhead{\textbf{w/ Obs Corr}} & \colhead{} & \colhead{} \\[-0.3cm]
\colhead{$\log (L_{\rm X}/\rm erg\ s^{-1})$} & \colhead{{$N$}} & \colhead{$\log (\Phi_{\rm L}/\rm{h^3\ Mpc^{-3}})$} & \colhead{$\sigma_{\rm{up}}$} & \colhead{$\sigma_{\rm{low}}$} & \colhead{$N$} & \colhead{$\log (\Phi_{\rm L}/\rm{h^3\ Mpc^{-3}})$} & \colhead{$\sigma_{\rm{up}}$} & \colhead{$\sigma_{\rm{low}}$} & \colhead{$N$} & \colhead{$\log (\Phi_{\rm L}/\rm{h^3\ Mpc^{-3}})$} & \colhead{$\sigma_{\rm{up}}$} & \colhead{$\sigma_{\rm{low}}$}}
\startdata
40.75 & 1 & -2.21 & 0.52 & -0.87 & 0 &  &  &  & 0 &  &  &  \\
41.05 & 1 & -2.62 & 0.52 & -0.87 & 0 &  &  &  & 0 &  &  &  \\
41.35 & 1 & -3.16 & 0.52 & -0.87 & 0 &  &  &  & 0 &  &  &  \\
41.65 & 3 & -2.82 & 0.3 & -0.36 & 0 &  &  &  & 0 &  &  &  \\
41.95 & 8 & -2.94 & 0.18 & -0.19 & 0 &  &  &  & 0 &  &  &  \\
42.25 & 9 & -3.43 & 0.17 & -0.19 & 1 & -3.12 & 0.52 & -0.87 & 1 & -2.77 & 0.52 & -0.87 \\
42.55 & 16 & -3.52 & 0.12 & -0.13 & 6 & -3.56 & 0.22 & -0.24 & 6 & -3.37 & 0.27 & -0.3 \\
42.85 & 33 & -3.68 & 0.09 & -0.09 & 28 & -3.67 & 0.1 & -0.1 & 28 & -3.6 & 0.1 & -0.1 \\
43.15 & 68 & -3.84 & 0.06 & -0.06 & 61 & -3.85 & 0.06 & -0.06 & 61 & -3.77 & 0.07 & -0.07 \\
43.45 & 95 & -4.09 & 0.05 & -0.05 & 90 & -4.11 & 0.05 & -0.05 & 90 & -3.98 & 0.08 & -0.08 \\
43.75 & 115 & -4.43 & 0.04 & -0.04 & 113 & -4.43 & 0.04 & -0.04 & 113 & -4.34 & 0.05 & -0.05 \\
44.05 & 106 & -4.9 & 0.05 & -0.05 & 106 & -4.89 & 0.05 & -0.05 & 106 & -4.85 & 0.05 & -0.05 \\
44.35 & 90 & -5.35 & 0.05 & -0.05 & 89 & -5.36 & 0.05 & -0.05 & 89 & -5.33 & 0.05 & -0.05 \\
44.65 & 70 & -5.9 & 0.06 & -0.06 & 70 & -5.9 & 0.06 & -0.06 & 70 & -5.84 & 0.06 & -0.06 \\
44.95 & 39 & -6.53 & 0.08 & -0.08 & 39 & -6.53 & 0.08 & -0.08 & 39 & -6.47 & 0.08 & -0.09 \\
45.25 & 13 & -7.39 & 0.14 & -0.14 & 13 & -7.39 & 0.14 & -0.14 & 13 & -7.37 & 0.14 & -0.14 \\
45.55 & 2 & -8.5 & 0.37 & -0.47 & 2 & -8.5 & 0.37 & -0.47 & 2 & -8.49 & 0.37 & -0.47 \\
45.85 & 1 & -8.84 & 0.52 & -0.87 & 1 & -8.84 & 0.52 & -0.87 & 1 & -8.84 & 0.52 & -0.87 \\
\hline
{} & {671$^{\rm a}$} & {} & {} & {} & {619} & {} & {} & {} & {619} & {} & {} & {} \\
\enddata
\tablenotetext{a}{The lowest luminosity object has a luminosity of $\log$ L$_{\rm 14-195} = 38.56$, and a \Vmax\ very close to zero (i.e., 1/\Vmax\ = -$\infty$). Including that object will bring the total to \nnonblazarhighredshiftcut.}
\end{deluxetable*}

\begin{deluxetable*}{lllll|llll|llll}
\label{tab:xlfvmaxabsbins}
\tablecaption{1/\Vmax values of XLFs in different obscuration bins, shown in Figure~\ref{fig:xlfznh}. All samples fall within the $0.01 \leq z \leq 0.3$ range.}
\tablewidth{0pt}
\tablehead{\colhead{\textbf{$\log {\rm N}_{\rm H} \leq 22$}} & \colhead{} & \colhead{} & \colhead{} & \colhead{} & \colhead{} & \colhead{\textbf{$22 \leq \log {\rm N}_{\rm H} \leq 24$}} & \colhead{} & \colhead{} & \colhead{} & \colhead{\textbf{$\log {\rm N}_{\rm H} \geq 24$}} & \colhead{}  \\[-0.3cm]
\colhead{$\log (L_{\rm X}/\rm erg\ s^{-1})$} & \colhead{{$N$}} & \colhead{$\log (\Phi_{\rm L}/\rm{h^3\ Mpc^{-3}})$} & \colhead{$\sigma_{\rm{up}}$} & \colhead{$\sigma_{\rm{low}}$} & \colhead{$N$} & \colhead{$\log (\Phi_{\rm L}/\rm{h^3\ Mpc^{-3}})$} & \colhead{$\sigma_{\rm{up}}$} & \colhead{$\sigma_{\rm{low}}$} & \colhead{$N$} & \colhead{$\log (\Phi_{\rm L}/\rm{h^3\ Mpc^{-3}})$} & \colhead{$\sigma_{\rm{up}}$} & \colhead{$\sigma_{\rm{low}}$}}
\startdata
42.25 & & & & & 1 & -2.77 & 0.52 & -0.87 & & & & \\
42.55 & 1 & -4.45 & 0.52 & -0.87 & 5 & -3.41 & 0.29 & -0.34 & & & & \\
42.85 & 12 & -4.06 & 0.15 & -0.15 & 14 & -3.87 & 0.14 & -0.15 & 2 & -4.57 & 0.37 & -0.47 \\
43.15 & 24 & -4.25 & 0.1 & -0.1 & 34 & -4.04 & 0.09 & -0.09 & 3 & -4.63 & 0.31 & -0.38 \\
43.45 & 39 & -4.46 & 0.08 & -0.08 & 42 & -4.41 & 0.08 & -0.08 & 9 & -4.52 & 0.25 & -0.28 \\
43.75 & 48 & -4.83 & 0.07 & -0.07 & 54 & -4.68 & 0.07 & -0.07 & 11 & -5.02 & 0.16 & -0.17 \\
44.05 & 67 & -5.1 & 0.06 & -0.06 & 32 & -5.33 & 0.09 & -0.09 & 7 & -5.77 & 0.19 & -0.21 \\
44.35 & 44 & -5.66 & 0.07 & -0.07 & 44 & -5.62 & 0.07 & -0.07 & 1 & -7.05 & 0.52 & -0.87 \\
44.65 & 42 & -6.14 & 0.07 & -0.08 & 25 & -6.26 & 0.1 & -0.1 & 3 & -6.74 & 0.31 & -0.37 \\
44.95 & 28 & -6.69 & 0.09 & -0.09 & 8 & -7.16 & 0.18 & -0.19 & 3 & -7.19 & 0.31 & -0.37 \\
45.25 & 9 & -7.53 & 0.17 & -0.18 & 3 & -8.02 & 0.3 & -0.36 & 1 & -8.38 & 0.52 & -0.87 \\
45.55 & 1 & -8.78 & 0.52 & -0.87 & 1 & -8.81 & 0.52 & -0.87 & & & & \\
45.85 & 1 & -8.84 & 0.52 & -0.87 & & & & & & & & \\
\hline
{} & {316} & {} & {} & {} & {263} & {} & {} & {} & {40} & {} & {} & {} \\
\enddata
\end{deluxetable*}

\begin{figure*}
	\includegraphics[width=0.5\textwidth]{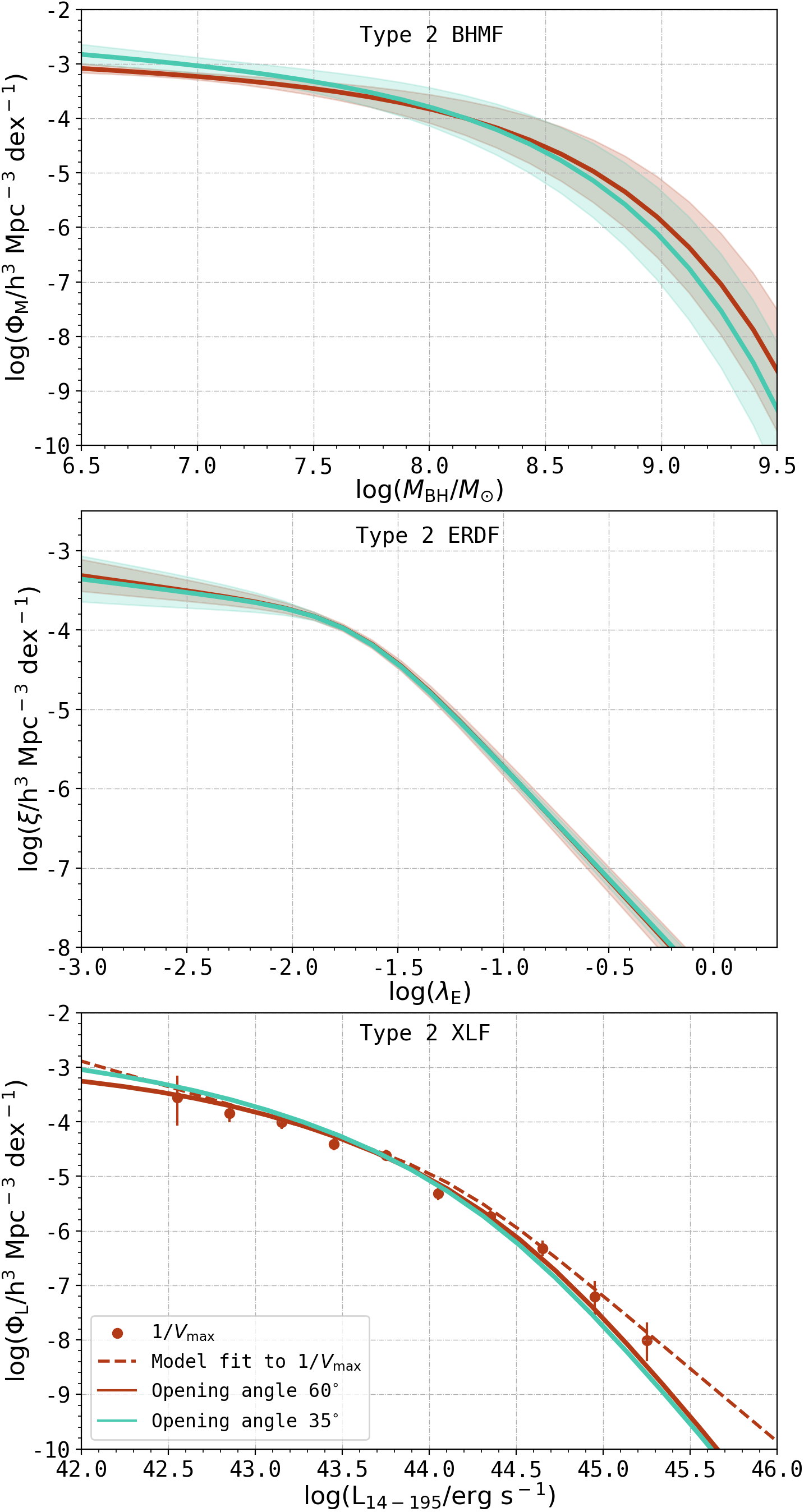}~
	\includegraphics[width=.5\textwidth]{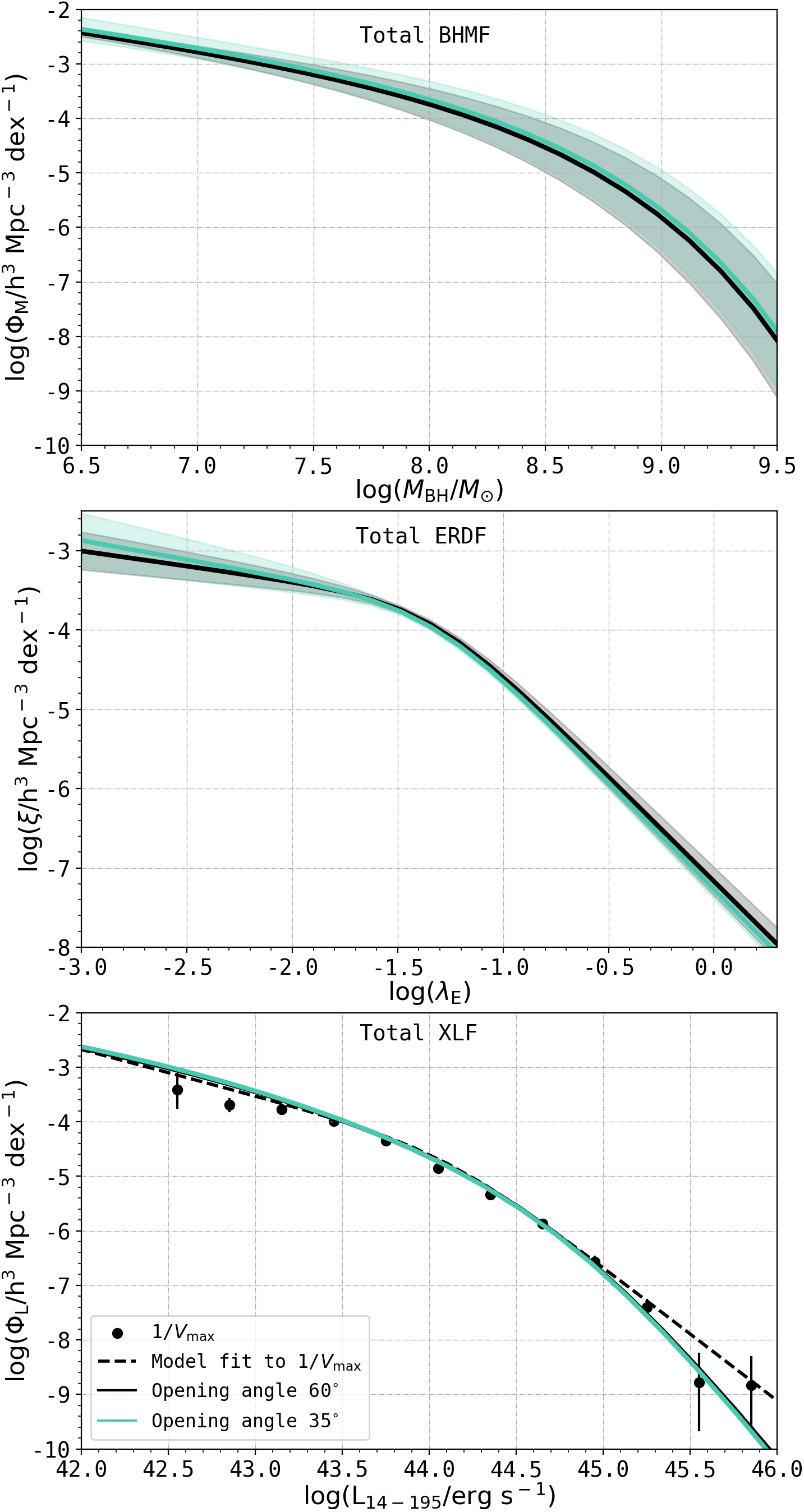}
	\caption{\label{fig:bhmferdfdisoa35}Number densities 
	of BASS 
	Type~2 AGN ({\it left panels, red points/lines}) and all AGN ({\it right panels; black points/lines}) assuming an absorption function and attenuation curve for a template spectra with torus opening angle of 60$^{\circ}$. Overplotted {\it green lines} show results for a template spectra with a torus opening angle of 35$^{\circ}$.}
\end{figure*}

\begin{figure*}[h]
\label{fig:pesky_include}
	\centering
	\includegraphics[width=0.6\textwidth]{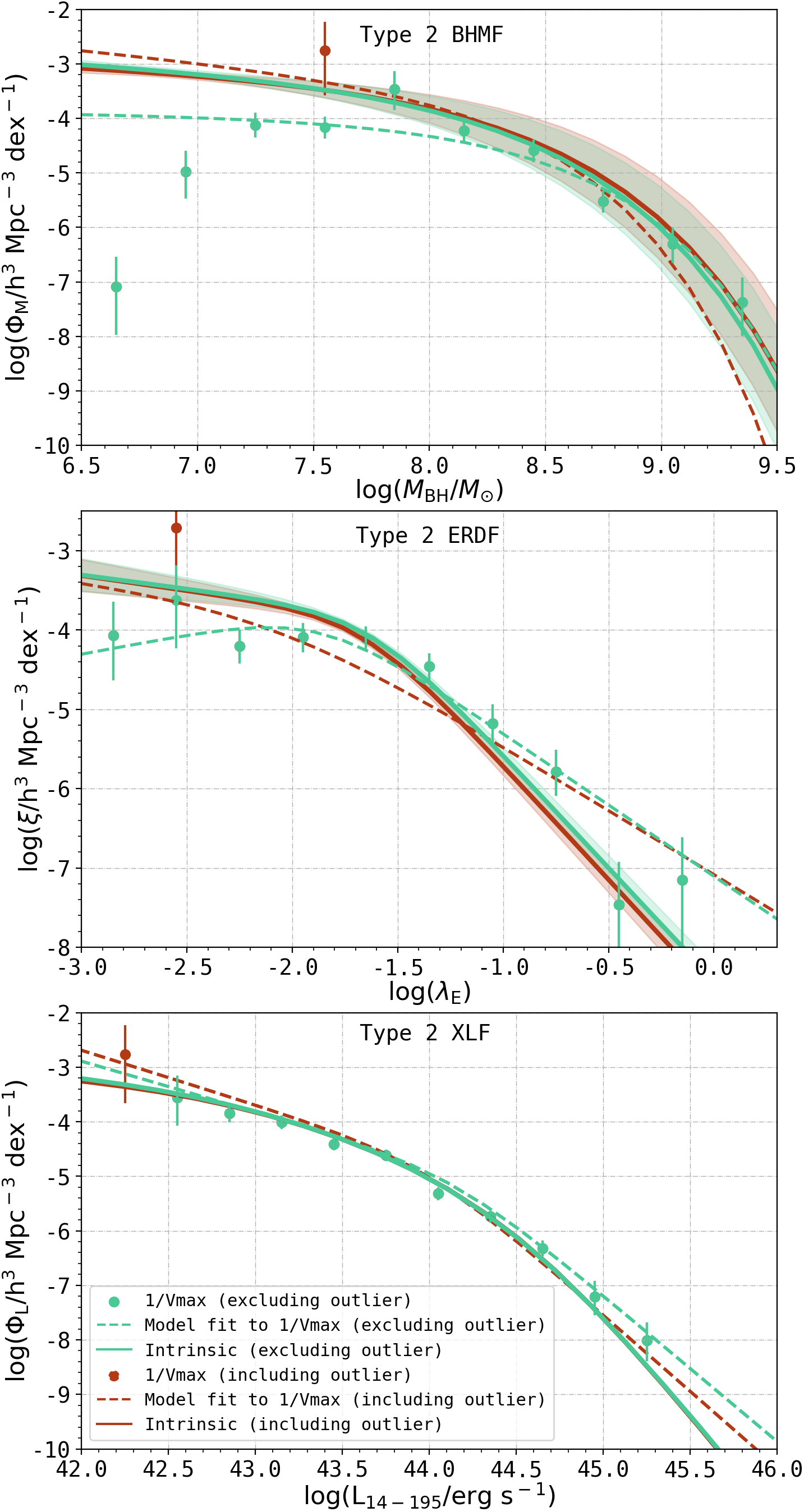}
	\caption{The impact on our analysis due to the lowest luminosity and lowest redshift object within our redshift restricted sample (NGC 5283, or BAT ID 684). 
	\textit{Green points} show the 1/\Vmax\ measurements excluding this single source, as is done throughout our main analysis, while the \textit{red point} shows how these change when the source is included in the sample (affecting only one bin in luminosity, \Mbh, and \lamEdd). 
	This object falls at the flux threshold of the overall survey (according to the curve shown in Figure~\ref{fig:attenuation_curve}), but given the variation in the sky-sensitivity (i.e. flux-area) across the surveyed area, the corresponding  1/\Vmax\ value is likely to be inaccurate. 
	The different lines demonstrate that --- unlike some of the fits to 1/\Vmax\ measurements --- the {\it intrinsic} distribution functions are robust to the choice of including or excluding this single source; for all three distribution functions, the solid lines of the two cases (almost) completely overlap, while the dashed lines do not.
	We provide all 1/\Vmax\ measurements, both with and without this single source, in Tables~\ref{tab:bhmfxlfvmaxxlf}, \ref{tab:bhmfvmax} and \ref{tab:erdfvmax}.
	}
\end{figure*}

\end{document}

%% file: parameter_shortcuts.tex
\newcommand{\nxraydetected}{838} 
\newcommand{\nfaintdual}{37} 
\newcommand{\ntotremoved}{181}

\newcommand{\nnonblazar}{713} 

\newcommand{\nnonblazarnonfaint}{678} 
\newcommand{\nnonblazarlowredshiftcut}{659} 
\newcommand{\nlowzexclude}{53} 
\newcommand{\nnonblazarredshiftcut}{619} 
\newcommand{\nhighzexclude}{6} 
\newcommand{\ntotallcuts}{586} 
\newcommand{\nnonblazarhighredshiftcut}{672} 

\newcommand{\ntotallcutstypeone}{366} 
\newcommand{\ntotallcutstypetwo}{220} 

\newcommand{\galscalehigh}{-2.62}
\newcommand{\galscalelow}{-2.82}
\newcommand{\bhmflogphistar}{-3.52}
\newcommand{\bhmfalpha}{$-1.576^{+0.147}_{-0.078}$}
\newcommand{\bhmfbeta}{$0.593^{+0.078}_{-0.069}$}
\newcommand{\bhmflogmstar}{$7.88^{+0.21}_{-0.22}$}
\newcommand{\erdflogxistar}{-3.64}
\newcommand{\erdfdeltaa}{$0.38^{+0.10}_{-0.12}$}
\newcommand{\erdfepislonlam}{$2.260^{+0.082}_{-0.121}$}
\newcommand{\erdfloglamstar}{$-1.338\pm 0.065$}
\newcommand{\bhmflogphistarsigfive}{-3.37}
\newcommand{\bhmfalphasigfive}{$-1.26^{+0.19}_{-0.11}$}
\newcommand{\bhmfbetasigfive}{$0.630^{+0.065}_{-0.086}$}
\newcommand{\bhmflogmstarsigfive}{$7.67^{+0.25}_{-0.20}$}
\newcommand{\erdflogxistarsigfive}{-3.8}
\newcommand{\erdfdeltaasigfive}{$0.28^{+0.13}_{-0.12}$}
\newcommand{\erdfepislonlamsigfive}{$2.72^{+0.13}_{-0.14}$}
\newcommand{\erdfloglamstarsigfive}{$-1.249^{+0.059}_{-0.061}$}
\newcommand{\bhmflogphistarsigLtwo}{-3.67}
\newcommand{\bhmfalphasigLtwo}{$-1.530^{+0.114}_{-0.094}$}
\newcommand{\bhmfbetasigLtwo}{$0.612^{+0.053}_{-0.063}$}
\newcommand{\bhmflogmstarsigLtwo}{$7.92^{+0.13}_{-0.22}$}
\newcommand{\erdflogxistarsigLtwo}{-3.76}
\newcommand{\erdfdeltaasigLtwo}{$0.40\pm 0.12$}
\newcommand{\erdfepislonlamsigLtwo}{$2.322^{+0.095}_{-0.105}$}
\newcommand{\erdfloglamstarsigLtwo}{$-1.286^{+0.059}_{-0.076}$}
\newcommand{\bhmflogphistaroathreefive}{-3.49}
\newcommand{\bhmfalphaoathreefive}{$-1.576^{+0.057}_{-0.157}$}
\newcommand{\bhmfbetaoathreefive}{$0.600^{+0.066}_{-0.082}$}
\newcommand{\bhmflogmstaroathreefive}{$7.92^{+0.18}_{-0.27}$}
\newcommand{\erdflogxistaroathreefive}{-3.68}
\newcommand{\erdfdeltaaoathreefive}{$0.484^{+0.091}_{-0.133}$}
\newcommand{\erdfepislonlamoathreefive}{$2.210^{+0.066}_{-0.106}$}
\newcommand{\erdfloglamstaroathreefive}{$-1.332^{+0.077}_{-0.068}$}
\newcommand{\bhmflogphistartypeone}{-4.19}
\newcommand{\bhmfalphatypeone}{$-1.753^{+0.137}_{-0.088}$}
\newcommand{\bhmfbetatypeone}{$0.561^{+0.062}_{-0.073}$}
\newcommand{\bhmflogmstartypeone}{$7.97^{+0.17}_{-0.30}$}
\newcommand{\erdflogxistartypeone}{-4.08}
\newcommand{\erdfdeltaatypeone}{$0.30\pm 0.14$}
\newcommand{\erdfepislonlamtypeone}{$2.51^{+0.11}_{-0.15}$}
\newcommand{\erdfloglamstartypeone}{$-1.152^{+0.089}_{-0.053}$}
\newcommand{\bhmflogphistartypeonesigfive}{-4.27}
\newcommand{\bhmfalphatypeonesigfive}{$-1.56\pm 0.13$}
\newcommand{\bhmfbetatypeonesigfive}{$0.590^{+0.104}_{-0.050}$}
\newcommand{\bhmflogmstartypeonesigfive}{$7.91^{+0.16}_{-0.27}$}
\newcommand{\erdflogxistartypeonesigfive}{-4.23}
\newcommand{\erdfdeltaatypeonesigfive}{$0.13^{+0.15}_{-0.14}$}
\newcommand{\erdfepislonlamtypeonesigfive}{$2.97^{+0.18}_{-0.19}$}
\newcommand{\erdfloglamstartypeonesigfive}{$-1.103^{+0.065}_{-0.054}$}
\newcommand{\bhmflogphistartypeonesigLtwo}{-4.27}
\newcommand{\bhmfalphatypeonesigLtwo}{$-1.73\pm 0.11$}
\newcommand{\bhmfbetatypeonesigLtwo}{$0.566^{+0.085}_{-0.071}$}
\newcommand{\bhmflogmstartypeonesigLtwo}{$7.93^{+0.25}_{-0.22}$}
\newcommand{\erdflogxistartypeonesigLtwo}{-4.09}
\newcommand{\erdfdeltaatypeonesigLtwo}{$0.27^{+0.13}_{-0.12}$}
\newcommand{\erdfepislonlamtypeonesigLtwo}{$2.57^{+0.12}_{-0.16}$}
\newcommand{\erdfloglamstartypeonesigLtwo}{$-1.138^{+0.070}_{-0.058}$}
\newcommand{\bhmflogphistartypetwo}{-3.6}
\newcommand{\bhmfalphatypetwo}{$-1.16^{+0.14}_{-0.20}$}
\newcommand{\bhmfbetatypetwo}{$0.637^{+0.067}_{-0.075}$}
\newcommand{\bhmflogmstartypetwo}{$7.82^{+0.15}_{-0.26}$}
\newcommand{\erdflogxistartypetwo}{-3.82}
\newcommand{\erdfdeltaatypetwo}{$0.376^{+0.099}_{-0.253}$}
\newcommand{\erdfepislonlamtypetwo}{$2.50\pm 0.17$}
\newcommand{\erdfloglamstartypetwo}{$-1.657^{+0.087}_{-0.064}$}
\newcommand{\bhmflogphistartypetwosigfive}{-3.6}
\newcommand{\bhmfalphatypetwosigfive}{$-0.99^{+0.21}_{-0.19}$}
\newcommand{\bhmfbetatypetwosigfive}{$0.703^{+0.086}_{-0.069}$}
\newcommand{\bhmflogmstartypetwosigfive}{$7.76^{+0.19}_{-0.15}$}
\newcommand{\erdflogxistartypetwosigfive}{-3.92}
\newcommand{\erdfdeltaatypetwosigfive}{$0.30^{+0.21}_{-0.20}$}
\newcommand{\erdfepislonlamtypetwosigfive}{$2.53^{+0.26}_{-0.11}$}
\newcommand{\erdfloglamstartypetwosigfive}{$-1.593^{+0.080}_{-0.096}$}
\newcommand{\bhmflogphistartypetwosigLtwo}{-3.64}
\newcommand{\bhmfalphatypetwosigLtwo}{$-1.18^{+0.16}_{-0.17}$}
\newcommand{\bhmfbetatypetwosigLtwo}{$0.617^{+0.100}_{-0.044}$}
\newcommand{\bhmflogmstartypetwosigLtwo}{$7.79^{+0.17}_{-0.22}$}
\newcommand{\erdflogxistartypetwosigLtwo}{-3.84}
\newcommand{\erdfdeltaatypetwosigLtwo}{$0.32^{+0.18}_{-0.20}$}
\newcommand{\erdfepislonlamtypetwosigLtwo}{$2.50\pm 0.17$}
\newcommand{\erdfloglamstartypetwosigLtwo}{$-1.628^{+0.079}_{-0.084}$}
\newcommand{\bhmflogphistartypetwooathreefive}{-3.44}
\newcommand{\bhmfalphatypetwooathreefive}{$-1.26^{+0.17}_{-0.16}$}
\newcommand{\bhmfbetatypetwooathreefive}{$0.635^{+0.085}_{-0.053}$}
\newcommand{\bhmflogmstartypetwooathreefive}{$7.73^{+0.25}_{-0.12}$}
\newcommand{\erdflogxistartypetwooathreefive}{-3.8}
\newcommand{\erdfdeltaatypetwooathreefive}{$0.33^{+0.15}_{-0.20}$}
\newcommand{\erdfepislonlamtypetwooathreefive}{$2.51^{+0.14}_{-0.16}$}
\newcommand{\erdfloglamstartypetwooathreefive}{$-1.675^{+0.091}_{-0.067}$}
\newcommand{\bhmfvmaxallphi}{$-4.33^{+0.22}_{-0.37}$}
\newcommand{\bhmfvmaxallalpha}{$-1.06^{+0.12}_{-0.30}$}
\newcommand{\bhmfvmaxallbeta}{$0.574^{+0.112}_{-0.040}$}
\newcommand{\bhmfvmaxalllogMbh}{$8.12^{+0.14}_{-0.11}$}
\newcommand{\erdfvmaxallphi}{$-3.76^{+0.18}_{-0.30}$}
\newcommand{\erdfvmaxallgamma}{$-0.02^{+0.29}_{-0.39}$}
\newcommand{\erdfvmaxalleps}{$2.06^{+0.30}_{-0.27}$}
\newcommand{\erdfvmaxalllam}{$-1.19^{+0.23}_{-0.21}$}
\newcommand{\bhmfvmaxtyponephi}{$-5.10^{+0.26}_{-0.50}$}
\newcommand{\bhmfvmaxtyponealpha}{$-1.35^{+0.18}_{-0.23}$}
\newcommand{\bhmfvmaxtyponebeta}{$0.681^{+0.087}_{-0.114}$}
\newcommand{\bhmfvmaxtyponelogMbh}{$8.73^{+0.26}_{-0.31}$}
\newcommand{\erdfvmaxtyponephi}{$-4.02^{+0.22}_{-0.32}$}
\newcommand{\erdfvmaxtyponegamma}{$-0.51^{+0.53}_{-0.41}$}
\newcommand{\erdfvmaxtyponeeps}{$2.57^{+0.33}_{-0.45}$}
\newcommand{\erdfvmaxtyponelam}{$-1.06^{+0.28}_{-0.25}$}
\newcommand{\bhmfvmaxtyptwophi}{$-4.33^{+0.19}_{-0.23}$}
\newcommand{\bhmfvmaxtyptwoalpha}{$-1.04^{+0.30}_{-0.29}$}
\newcommand{\bhmfvmaxtyptwobeta}{$0.732^{+0.074}_{-0.050}$}
\newcommand{\bhmfvmaxtyptwologMbh}{$8.102^{+0.172}_{-0.095}$}
\newcommand{\erdfvmaxtyptwophi}{$-3.74^{+0.35}_{-0.43}$}
\newcommand{\erdfvmaxtyptwogamma}{$-0.50^{+1.08}_{-0.56}$}
\newcommand{\erdfvmaxtyptwoeps}{$2.30^{+0.81}_{-0.69}$}
\newcommand{\erdfvmaxtyptwolam}{$-1.87^{+0.38}_{-0.40}$}

\newcommand{\xlfvmaxtypeonephi}{$-4.75^{+0.27}_{-0.36}$}
\newcommand{\xlfvmaxtypeonegamma}{$0.69^{+0.22}_{-0.26}$}
\newcommand{\xlfvmaxtypeoneeps}{$1.60^{+0.22}_{-0.14}$}
\newcommand{\xlfvmaxtypeoneL}{$44.12^{+0.21}_{-0.22}$}
\newcommand{\xlfvmaxtypetwophi}{$-4.88^{+0.29}_{-0.49}$}
\newcommand{\xlfvmaxtypetwogamma}{$0.93^{+0.21}_{-0.26}$}
\newcommand{\xlfvmaxtypetwoeps}{$1.75^{+0.20}_{-0.21}$}
\newcommand{\xlfvmaxtypetwoL}{$44.14^{+0.13}_{-0.33}$}
\newcommand{\xlfvmaxtypeallphi}{$-4.46^{+0.19}_{-0.44}$}
\newcommand{\xlfvmaxtypeallgamma}{$0.85^{+0.17}_{-0.25}$}
\newcommand{\xlfvmaxtypealleps}{$1.60^{+0.19}_{-0.12}$}
\newcommand{\xlfvmaxtypeallL}{$44.10^{+0.20}_{-0.19}$}

\newcommand{\xlfvmaxallnoobscorrphi}{$-4.71^{+0.25}_{-0.19}$}
\newcommand{\xlfvmaxallnoobscorrgamma}{$0.75^{-0.12}_{+0.11}$}
\newcommand{\xlfvmaxallnoobscorreps}{$1.64^{+0.11}_{-0.10}$}
\newcommand{\xlfvmaxallnoobscorrL}{$44.17^{+0.13}_{-0.18}$}
\newcommand{\xlfvmaxlowzcutobscorrphi}{$-4.78^{+0.35}_{-0.39}$}
\newcommand{\xlfvmaxlowzcutobscorrgamma}{$0.92^{+0.21}_{-0.25}$}
\newcommand{\xlfvmaxlowzcutobscorreps}{$1.63^{+0.10}_{-0.19}$}
\newcommand{\xlfvmaxlowzcutobscorrL}{$44.24^{+0.24}_{-0.31}$}
\newcommand{\xlfvmaxlowzcutnoobscorrphi}{$-4.79^{+0.34}_{-0.30}$}
\newcommand{\xlfvmaxlowzcutnoobscorrgamma}{$0.87^{+0.19}_{-0.21}$}
\newcommand{\xlfvmaxlowzcutnoobscorreps}{$1.59^{+0.14}_{-0.13}$}
\newcommand{\xlfvmaxlowzcutnoobscorrL}{$44.23^{+0.18}_{-0.24}$}
\newcommand{\xlfvmaxunabsphi}{$-5.04^{+0.36}_{-0.31}$}
\newcommand{\xlfvmaxunabsgamma}{$0.77^{+0.16}_{-0.26}$}
\newcommand{\xlfvmaxunabseps}{$1.53^{+0.21}_{-0.13}$}
\newcommand{\xlfvmaxunabsL}{$44.17^{+0.19}_{-0.20}$}
\newcommand{\xlfvmaxctnphi}{$-5.35^{+0.40}_{-0.04}$}
\newcommand{\xlfvmaxctngamma}{$0.99^{+0.13}_{-0.16}$}
\newcommand{\xlfvmaxctneps}{$1.81^{+0.17}_{-0.18}$}
\newcommand{\xlfvmaxctnL}{$44.30^{+0.06}_{-0.19}$}
\newcommand{\xlfvmaxctkphi}{$-4.51^{+0.50}_{-0.44}$}
\newcommand{\xlfvmaxctkgamma}{$0.33^{+0.44}_{-0.40}$}
\newcommand{\xlfvmaxctkeps}{$1.81^{+0.13}_{-0.57}$}
\newcommand{\xlfvmaxctkL}{$43.59^{+0.38}_{-0.23}$}

\newcommand{\nlowlumagn}{15} 

\newcommand{\nblazars}{105} 

\newcommand{\lboltolam}{38.18}

\newcommand{\dutyshankoldrangehigh}{0.16}
\newcommand{\dutyshankoldrangelow}{0.1}
\newcommand{\dutyshankoldrangehighperc}{16}
\newcommand{\dutyshankoldrangelowperc}{10}
\newcommand{\dutyshankoldlamnegtwo}{$2.85 \times 10^{-2}$}
\newcommand{\dutyshankoldlamnegone}{$6.45 \times 10^{-4}$}
\newcommand{\dutyshankoldlamzero}{$1.61 \times 10^{-6}$}

\newcommand{\erdfdeltaalowmasstypetwo}{$0.64^{+0.26}_{-0.28}$}
\newcommand{\erdfepislonlamlowmasstypetwo}{$2.05^{+0.16}_{-0.26}$}
\newcommand{\erdfloglamstarlowmasstypetwo}{$-1.59^{+0.15}_{-0.19}$}
\newcommand{\erdflogxistarhighmasstypetwo}{-5.18}
\newcommand{\erdfdeltaahighmasstypetwo}{$0.37^{+0.17}_{-0.26}$}
\newcommand{\erdfepislonlamhighmasstypetwo}{$2.43^{+0.18}_{-0.20}$}
\newcommand{\erdfloglamstarhighmasstypetwo}{$-1.69^{+0.11}_{-0.13}$}

\newcommand{\erdflogxistarlowmasstypeone}{-4.15}
\newcommand{\erdfdeltaalowmasstypeone}{$0.36^{+0.19}_{-0.15}$}
\newcommand{\erdfepislonlamlowmasstypeone}{$2.67^{+0.17}_{-0.23}$}
\newcommand{\erdfloglamstarlowmasstypeone}{$-0.898^{+0.100}_{-0.094}$}
\newcommand{\erdflogxistarhighmasstypeone}{-5.29}
\newcommand{\erdfdeltaahighmasstypeone}{$0.32^{+0.19}_{-0.14}$}
\newcommand{\erdfepislonlamhighmasstypeone}{$2.46^{+0.22}_{-0.21}$}
\newcommand{\erdfloglamstarhighmasstypeone}{$-1.313^{+0.101}_{-0.087}$}

\newcommand{\activemassdens}{$3.58^{+0.41}_{-0.23} \times 10^{-4}$}

\newcommand{\shankoldmassdensrange}{${\approx}(3{-}5) \times 10^{-5}$}
\newcommand{\activemassdensrange}{$6{-}10$}

\newcommand{\bhmflogphistardurassigLzero}{-3.91}
\newcommand{\bhmfalphadurassigLzero}{$-1.481^{+0.122}_{-0.079}$}
\newcommand{\bhmfbetadurassigLzero}{$0.572^{+0.068}_{-0.075}$}
\newcommand{\bhmflogmstardurassigLzero}{$7.96^{+0.25}_{-0.21}$}
\newcommand{\erdflogxistardurassigLzero}{-3.55}
\newcommand{\erdfdeltaadurassigLzero}{$0.20^{+0.11}_{-0.12}$}
\newcommand{\erdfepislonlamdurassigLzero}{$1.908^{+0.086}_{-0.074}$}
\newcommand{\erdfloglamstardurassigLzero}{$-1.346^{+0.098}_{-0.086}$}
\newcommand{\bhmflogphistarduras}{-3.97}
\newcommand{\bhmfalphaduras}{$-1.401^{+0.109}_{-0.085}$}
\newcommand{\bhmfbetaduras}{$0.598^{+0.061}_{-0.075}$}
\newcommand{\bhmflogmstarduras}{$8.00^{+0.14}_{-0.27}$}
\newcommand{\erdflogxistarduras}{-3.64}
\newcommand{\erdfdeltaaduras}{$0.18^{+0.17}_{-0.10}$}
\newcommand{\erdfepislonlamduras}{$1.908^{+0.093}_{-0.096}$}
\newcommand{\erdfloglamstarduras}{$-1.35^{+0.11}_{-0.10}$}
\newcommand{\bhmflogphistarsigfiveduras}{-3.96}
\newcommand{\bhmfalphasigfiveduras}{$-1.32^{+0.14}_{-0.10}$}
\newcommand{\bhmfbetasigfiveduras}{$0.597^{+0.095}_{-0.063}$}
\newcommand{\bhmflogmstarsigfiveduras}{$7.98^{+0.15}_{-0.29}$}
\newcommand{\erdflogxistarsigfiveduras}{-3.65}
\newcommand{\erdfdeltaasigfiveduras}{$0.11^{+0.15}_{-0.18}$}
\newcommand{\erdfepislonlamsigfiveduras}{$2.05^{+0.11}_{-0.14}$}
\newcommand{\erdfloglamstarsigfiveduras}{$-1.380^{+0.104}_{-0.097}$}
\newcommand{\bhmflogphistartypeonedurassigLzero}{-4.7}
\newcommand{\bhmfalphatypeonedurassigLzero}{$-1.693^{+0.115}_{-0.073}$}
\newcommand{\bhmfbetatypeonedurassigLzero}{$0.592^{+0.056}_{-0.107}$}
\newcommand{\bhmflogmstartypeonedurassigLzero}{$8.25^{+0.21}_{-0.28}$}
\newcommand{\erdflogxistartypeonedurassigLzero}{-3.94}
\newcommand{\erdfdeltaatypeonedurassigLzero}{$0.04^{+0.15}_{-0.13}$}
\newcommand{\erdfepislonlamtypeonedurassigLzero}{$2.20^{+0.10}_{-0.12}$}
\newcommand{\erdfloglamstartypeonedurassigLzero}{$-1.153^{+0.109}_{-0.066}$}
\newcommand{\bhmflogphistartypeoneduras}{-4.57}
\newcommand{\bhmfalphatypeoneduras}{$-1.589^{+0.077}_{-0.113}$}
\newcommand{\bhmfbetatypeoneduras}{$0.564^{+0.076}_{-0.070}$}
\newcommand{\bhmflogmstartypeoneduras}{$8.13^{+0.22}_{-0.23}$}
\newcommand{\erdflogxistartypeoneduras}{-4.03}
\newcommand{\erdfdeltaatypeoneduras}{$0.02^{+0.15}_{-0.16}$}
\newcommand{\erdfepislonlamtypeoneduras}{$2.19^{+0.14}_{-0.12}$}
\newcommand{\erdfloglamstartypeoneduras}{$-1.145^{+0.067}_{-0.109}$}
\newcommand{\bhmflogphistartypeonesigfiveduras}{-4.77}
\newcommand{\bhmfalphatypeonesigfiveduras}{$-1.61^{+0.10}_{-0.12}$}
\newcommand{\bhmfbetatypeonesigfiveduras}{$0.575^{+0.090}_{-0.081}$}
\newcommand{\bhmflogmstartypeonesigfiveduras}{$8.22^{+0.20}_{-0.30}$}
\newcommand{\erdflogxistartypeonesigfiveduras}{-4.03}
\newcommand{\erdfdeltaatypeonesigfiveduras}{$-0.10^{+0.23}_{-0.15}$}
\newcommand{\erdfepislonlamtypeonesigfiveduras}{$2.43^{+0.18}_{-0.17}$}
\newcommand{\erdfloglamstartypeonesigfiveduras}{$-1.13\pm 0.11$}
\newcommand{\bhmflogphistartypetwodurassigLzero}{-3.95}
\newcommand{\bhmfalphatypetwodurassigLzero}{$-1.13^{+0.16}_{-0.17}$}
\newcommand{\bhmfbetatypetwodurassigLzero}{$0.630^{+0.082}_{-0.071}$}
\newcommand{\bhmflogmstartypetwodurassigLzero}{$7.89^{+0.15}_{-0.28}$}
\newcommand{\erdflogxistartypetwodurassigLzero}{-3.98}
\newcommand{\erdfdeltaatypetwodurassigLzero}{$0.14^{+0.21}_{-0.17}$}
\newcommand{\erdfepislonlamtypetwodurassigLzero}{$2.16^{+0.16}_{-0.17}$}
\newcommand{\erdfloglamstartypetwodurassigLzero}{$-1.68\pm 0.11$}
\newcommand{\bhmflogphistartypetwoduras}{-4.07}
\newcommand{\bhmfalphatypetwoduras}{$-1.11^{+0.16}_{-0.13}$}
\newcommand{\bhmfbetatypetwoduras}{$0.650^{+0.066}_{-0.082}$}
\newcommand{\bhmflogmstartypetwoduras}{$7.85^{+0.17}_{-0.22}$}
\newcommand{\erdflogxistartypetwoduras}{-3.87}
\newcommand{\erdfdeltaatypetwoduras}{$0.25^{+0.20}_{-0.23}$}
\newcommand{\erdfepislonlamtypetwoduras}{$2.03^{+0.18}_{-0.16}$}
\newcommand{\erdfloglamstartypetwoduras}{$-1.73\pm 0.12$}
\newcommand{\bhmflogphistartypetwosigfiveduras}{-4.05}
\newcommand{\bhmfalphatypetwosigfiveduras}{$-0.99^{+0.25}_{-0.15}$}
\newcommand{\bhmfbetatypetwosigfiveduras}{$0.708^{+0.054}_{-0.096}$}
\newcommand{\bhmflogmstartypetwosigfiveduras}{$7.81^{+0.18}_{-0.20}$}
\newcommand{\erdflogxistartypetwosigfiveduras}{-3.91}
\newcommand{\erdfdeltaatypetwosigfiveduras}{$0.19\pm 0.24$}
\newcommand{\erdfepislonlamtypetwosigfiveduras}{$1.99^{+0.24}_{-0.11}$}
\newcommand{\erdfloglamstartypetwosigfiveduras}{$-1.72^{+0.12}_{-0.15}$}